\documentclass[article]{aa}

\usepackage{natbib}
\usepackage{amsmath}
\usepackage{graphicx}
\usepackage{txfonts}
\usepackage{lscape}
\usepackage{gensymb}
\usepackage{color}
\usepackage[version=3]{mhchem}

\usepackage{array}
\newcolumntype{H}{>{\iffalse}c<{\fi}@{}}

\bibpunct{(}{)}{;}{a}{}{,}

 \renewenvironment{thebibliography}[1]{%
  \begin{oldthebibliography}{#1}%
   \setlength{\parskip}{0ex}%
   \setlength{\itemsep}{0ex}%
 }%
 {%
  \end{oldthebibliography}%
 }

\graphicspath{{./figures/}}

\makeatletter
\def\input@path{{./tables/}}
\makeatother

\DeclareRobustCommand{\ion}[2]{\textup{#1\,\textsc{\lowercase{#2}}}}

\DeclareRobustCommand{\kms}{km\,${\rm s}^{-1}$}
\DeclareRobustCommand{\jyb}{Jy\,beam${}^{-1}$}

\DeclareRobustCommand{\us}{\,}

\begin{document} 

  \title{OH absorption in the first quadrant of the Milky Way as seen by THOR\thanks{This project is based on observations made with the VLA telescope under the program IDs: 12A-161, 13A-120, 14B-148. The observations were conducted as part of the THOR survey (The \ion{H}{i}, OH, Recombination line survey of the Milky Way; http://www.mpia.de/thor/)}}

  \author{
  	M. R. Rugel\inst{1}\and
  	H. Beuther\inst{1}\and
  	S. Bihr\inst{1}\and
  	Y. Wang\inst{1}\and
  	J. Ott\inst{2}\and
  	A. Brunthaler\inst{3}\and
  	A. Walsh\inst{4}\and
	S. C. O. Glover\inst{5}\and
  	P. F. Goldsmith\inst{6}\and
	L. D. Anderson\inst{7,8,9}\and
	N. Schneider\inst{10}\and
	K. M. Menten\inst{3}\and
	S. E. Ragan\inst{11}\and
	J. S. Urquhart\inst{12}\and 
        R. S. Klessen\inst{5,13}\and
  	J. D. Soler\inst{1}\and
        N. Roy\inst{14}\and
        J. Kainulainen\inst{1} \and
        T. Henning\inst{1} \and
        F. Bigiel\inst{5} \and
        R. J. Smith\inst{15} \and
        F. Wyrowski\inst{3} \and
  	S. N. Longmore\inst{16}
	}
  \institute{
  	Max Planck Institute for Astronomy, K\"onigstuhl 17, 69117 Heidelberg, Germany\\ \email{rugel@mpia.de} \and
  	National Radio Astronomy Observatory, PO Box O, 1003 Lopezville Road, Socorro, NM 87801, USA \and
	Max Planck Institute for Radioastronomy, Auf dem H\"ugel 69, 53121 Bonn, Germany \and 
	International Centre for Radio Astronomy Research, Curtin University, GPO Box U1987, Perth WA 6845, Australia \and
	Universit\"at Heidelberg, Zentrum f\"ur Astronomie, Institut f\"ur
        Theoretische Astrophysik, Albert-Ueberle-Str. 2, 69120
        Heidelberg, Germany \and
	Jet Propulsion Laboratory, California Institute of Technology, 4800 Oak Grove Drive, Pasadena, CA 91109, USA \and
	Department of Physics and Astronomy, West Virginia University, Morgantown WV 26506, USA \and
	Adjunct Astronomer at the Green Bank Observatory, P.O. Box 2, Green Bank WV 24944, USA \and
	Center for Gravitational Waves and Cosmology, West Virginia University, Chestnut Ridge Research Building, Morgantown, WV 26505, USA \and
	I. Physikalisches Institut, University of Cologne, Z\"ulpicher Str. 77, 50937 K\"oln, Germany \and
	School of Physics and Astronomy, Cardiff University, Queen's Buildings, The Parade, Cardiff, CF24 3AA, UK\and
	School of Physical Sciences, University of Kent, Ingram Building, Canterbury, Kent CT2 7NH, UK \and
	Universit\"{a}t Heidelberg, Interdisziplin\"{a}res Zentrum f\"{u}r Wissenschaftliches Rechnen, INF 205, 69120 Heidelberg, Germany\and
        Department of Physics, Indian Institute of Science, Bangalore 560012, India\and
        Jodrell Bank Centre for Astrophysics, School of Physics and Astronomy, 
        The University of Manchester, Oxford Road, Manchester, M13 9PL, UK\and
	Astrophysics Research Institute, Liverpool John Moores University, 146 Brownlow Hill, Liverpool L3 5RF, UK
	}
	
 \abstract
  {The hydroxyl radical (OH) is present in the diffuse molecular and partially atomic 
   phases of the interstellar medium (ISM), but its abundance relative to hydrogen is not
   clear.}
  {We aim to evaluate the abundance of OH with respect to molecular
    hydrogen using OH absorption against cm-continuum sources over the first
    Galactic quadrant.}
  {This OH study is part of the \ion{H}{i}/OH/Recombination line survey of the
    inner Milky Way (THOR). THOR is a Karl G. Jansky Very Large Array (VLA)
    large program of atomic, molecular and ionized gas in the range $15\degree \leq l
    \leq 67\degree$ and $|b| \leq 1\degree$. It is the highest-resolution unbiased OH absorption survey to
    date towards this region. We combine the optical depths derived from these observations
    with literature \ce{^{13}CO}(1-0) and \ion{H}{i} observations
    to determine the OH abundance.}
   {We detect absorption in the 1665 and 1667\,MHz transitions, that is, the ``main'' hyperfine structure 
   lines, for continuum sources stronger than $F_{\rm cont} \geq 0.1$\us\jyb. 
   OH absorption is found against approximately 15\% of these continuum sources with increasing fractions for stronger sources. 
   Most of the absorption occurs in molecular clouds that are
   associated with Galactic \ion{H}{ii} regions. We find \ce{OH} and
   \ce{^{13}CO} gas to have similar kinematic properties. 
   The data indicate that the \ce{OH} abundance decreases with increasing hydrogen column density. 
   The derived OH abundance with respect to the total hydrogen nuclei column density (atomic and molecular phase) is 
   in agreement with a constant abundance for $A_V < 10-20$. Towards the lowest column densities, we find sources that 
   exhibit OH absorption but no \ce{^{13}CO} emission, indicating that OH is a well suited tracer of the 
   low column density molecular gas. We also present spatially resolved OH absorption towards the 
   prominent extended \ion{H}{ii}-region W43.}
   {The unbiased nature of the THOR survey opens a new window onto the gas properties of the 
   interstellar medium. The characterization of the OH abundance over a large range of hydrogen gas column densities 
   contributes to the understanding of OH as a molecular gas tracer and provides a starting point for future investigations.}

\date{Received XXX; accepted XXX}
\maketitle

\section{Introduction}	   \label{sec:introduction}
Molecular clouds are the hosts of star formation. Studying their physical and chemical properties, 
their formation, and their evolution, is crucial for understanding key characteristics of the Milky Way Galaxy, e.g.,
the mass of stars that can be formed out of the gas reservoir
in the Milky Way \citep[e.g.,][]{McKeeOstriker:2007aa,DobbsKrumholz:2014aa,HeyerDame:2015aa,KlessenGlover:2016aa}. 
In particular, the formation of molecular clouds from
diffuse atomic gas is of central concern. 

\smallskip
\noindent
Most molecular gas is in the form of molecular hydrogen,
\ce{H_2}, which is difficult to observe directly in the cold
environments of molecular clouds. While \ce{CO} is frequently used as a tracer of 
\ce{H_2} in the Milky Way \citep[e.g.,][]{Miville-DeschenesMurray:2017aa}, observational
and theoretical studies suggest that a significant fraction of the molecular gas is 
not traced by CO \citep[e.g.,][]{GrenierCasandjian:2005aa,Planck-CollaborationAde:2011aa,PinedaLanger:2013aa,SmithGlover:2014aa}. 
Therefore, a search for alternative molecular gas tracers is necessary.

\smallskip
\noindent
\ce{OH} is a potential tracer for molecular gas in transition regions. 
It was first detected by \citet{WeinrebBarrett:1963aa} and was
one of the earliest molecules studied in detail in many regions of the 
Galactic plane \citep[e.g.,][]{Goss:1968aa,Turner:1979aa,DawsonWalsh:2014aa}, 
as it has easily-accessible ground state transitions at cm-wavelengths. 
Recent high sensitivity studies found OH emission that extends beyond the molecular cloud envelope traced by \ce{CO}
surveys \citep[e.g.,][using the GBT with $7\farcm6$ and the Arecibo telescope with $3\arcmin$
resolution, respectively]{AllenHogg:2015aa,XuLi:2016aa}. A detailed comparison of the ``CO-dark'' gas fraction and OH 
across a molecular cloud boundary in Taurus found OH to be present in ``CO-dark' regions with $A_V< 1.5\,{\rm mag}$ \citep{XuLi:2016aa}. 
Complementary studies show that OH is present in ``partially atomic, partially molecular'', warm ($\sim100\,{\rm K}$) \ion{H}{i} halos \citep{WannierAndersson:1993aa},
and show that its column density increases with increasing $N_\ion{H}{i}$ for $N_{\ion{H}{i}}<1.0\times10^{21}\,{\rm cm}^{-2}$ \citep{TangLi:2017ab}. 
Additionally, OH has also been found to be correlated with visual extinction in diffuse clouds \citep[][observed
at $22\arcmin$ resolution with the 37m telescope of the Vermilion River
Observatory]{Crutcher:1979aa}. These results strongly suggest its presence in transition regions between atomic and molecular gas. 

\smallskip
\noindent
The \ce{OH} abundance towards higher extinction regions is on the other hand of 
interest for the determination of magnetic field strengths from Zeeman splitting of OH absorption 
lines. To understand the gas densities at which \ce{OH} traces 
the magnetic fields, precise knowledge of the \ce{OH} abundance 
at different densities is indispensable. In particular, the behavior of the 
\ce{OH} abundance in regions of higher density is not yet well understood, neither theoretically nor observationally
\citep[e.g.,][]{HeilesGoodman:1993aa}. 

\smallskip
\noindent
There are three different types of chemical reactions in molecular clouds that can influence the abundance
of \ce{OH} \citep[e.g.,][]{van-DishoeckHerbst:2013aa}: Gas phase ion-neutral chemistry,
important in diffuse and cold environments (``diffuse'' chemistry), neutral-neutral chemistry, 
important for warm regions (>200K), and grain surface chemistry, which depends on the
strength of the radiation field and the temperature. The fractional abundance
of \ce{OH} is closely related to that of \ce{H2O} if diffuse chemistry or photodesorption of water from grains is dominant \citep{HollenbachKaufman:2012aa}.
In high temperature environments, e.g., in shocks, this changes, favoring the formation of \ce{H2O} in the 
case of very high temperatures, unless strong ultraviolet radiation is present
to photo-dissociate \ce{H2O} and thus to increase the amount of OH in the gas phase 
\citep[e.g.,][]{NeufeldKaufman:2002aa,van-DishoeckHerbst:2013aa}. 

\smallskip
\noindent
The fractional OH abundance has been found to be constant for {$A_V< 7\, {\rm mag}$} and hydrogen nuclei number densities of $n\lesssim2500$\us${\rm cm}^{-3}$ \citep[e.g.,][]{Crutcher:1979aa}. Typical values for the OH abundance with respect to total hydrogen nuclei column density are $X_{\rm OH} \sim 4.0\times10^{-8}$ \citep{Goss:1968aa,Crutcher:1979aa,HeilesGoodman:1993aa}, and with respect to molecular hydrogen column density $X_{\rm OH} \sim 1.0\times10^{-7}$ \citep[e.g.,][]{LisztLucas:2002aa}. Other studies exist, however, which also found higher values for the OH abundance, i.e. of a few $\times 10^{-7}$ in molecular cloud boundaries, with a decreasing trend towards $X_{\rm OH} \sim 1.5\times10^{-7}$ at visual extinctions $A_V \geq 2.5 {\rm mag}$ \citep{XuLi:2016aa}. Once molecular cloud regions fall into the line-of-sight where UV radiation is attenuated, the OH abundance is no longer expected to be constant \citep[][and references therein]{HeilesGoodman:1993aa}. Models predict the depletion of oxygen bearing species from the gas phase in the absence of photodesorption of water ice, which occurs at $A_V \sim 6\,{\rm mag}$, depending on the strength of the radiation field \citep{HollenbachKaufman:2012aa}.  

\smallskip
\noindent
The transitions investigated in this paper are the $\Lambda$ doubling
transitions of the OH ground state, the ${\rm {}^{2}\Pi_{3/2};J=3/2}$ state. The
transitions at 1665{\us}MHz and 1667{\us}MHz (``main lines'') are 5 and 9 times stronger
than the satellite transitions at 1612{\us}MHz and 1720{\us}MHz \citep[``satellite lines''; e.g.,][]{Elitzur:1992aa}. 
While the satellite lines are easily anomalously excited, e.g., through ambient infrared radiation
(that is, are subject to population inversion and show non-thermal, maser emission), 
it requires higher densities to anomalously excite the main lines, 
which are mostly also found to be optically thin \citep[e.g.,][]{Goss:1968aa,Crutcher:1979aa,HeilesGoodman:1993aa}. 
Observations of OH transitions at 1665{\us}MHz and 1667{\us}MHz in
absorption against strong cm-continuum sources therefore provide a
possibility to determine the optical depth of the OH ground state transitions 
directly \citep[e.g.,][]{Goss:1968aa, StanimirovicWeisberg:2003aa}.

\smallskip
\noindent
Strong maser emission from OH 1665 and 1667 MHz has also been found, predominantly towards high mass young stellar objects, but also towards evolved stars \citep[e.g.,][]{ArgonReid:2000aa}. They are pumped by the strong far infrared field emitted by the warm ($T \sim 150\us{\rm K}$) dust in their host stars' dense ($\sim 10^7${\us}cm$^{-3}$) envelopes \citep[e.g.,][]{CesaroniWalmsley:1991aa}. In the course of the THOR survey (The \ion{H}{i}, OH, Recombination line survey of the Milky Way;
\citealt{BeutherBihr:2016aa}) many such OH masers have been detected \citep[see, e.g.,][]{WalshBeuther:2016aa}, but are not the topic of the present paper.

\smallskip
\noindent
The determination of \ce{OH} column densities from hyperfine ground state absorption observations
requires an assumption regarding the excitation temperature of the transitions. 
The \ce{OH} excitation temperature of the main lines depends on the volume density and ionization fraction, 
and only weakly on the kinetic temperature \citep[e.g.,][]{GuibertRieu:1978aa}. The 
critical density ($n_{\rm crit} = {A_{\rm ul}}/{\gamma_{\rm ul}}$; $A_{\rm ul}$ is the Einstein coefficient for 
spontaneous emission and ${\gamma_{\rm ul}}$ the collisional deexcitation rate coefficient), a measure of when collisional processes 
dominate the deexcitation of the upper energy levels of a transition, is typically found around $n_{\rm crit}\sim0.5\,{\rm cm^{-3}}$
for the OH transitions at 1665\,MHz and 1667\,MHz. The transitions are typically found to be subthermally excited, with excitation temperatures
of $T_{\rm ex} = 5-10\,{\rm K}$ \citep[e.g.,][]{ColganSalpeter:1989aa}. 
The reason for this is that densities much higher than $n_{\rm crit}$ are needed for thermalization. These densities exceed those typical of boundary regions of molecular clouds ($n\leq10^{3}\,{\rm cm}^{-3}$). 
Firstly, once stimulated emission and absorption of the cosmic microwave background are included, the effective critical density required for the collisional and radiative deexcitation rates to balance is $n\gtrsim10^{3}\,{\rm cm}^{-3}$ (e.g., \citealt{WannierAndersson:1993aa}).
Secondly, the small energy separation of the OH lines ($E_u/k \sim 0.1\,{\rm K}$) makes the lines harder to thermalize for any given $T_{\rm kin}$, such that $n\gtrsim10^{3}\,{\rm cm}^{-3}$ are required to thermalize the lines even if stimulated emission and absorption are not taken into account. 

\smallskip
\noindent
Within the THOR survey, we observed the ground state \ce{OH} transitions
at a high angular resolution of 20\arcsec\ and compared our results to those obtained from tracers of
atomic and molecular gas at comparable angular resolution across the first quadrant of
the Milky Way. The present paper addresses two aspects of the OH data: The detection
statistics of OH main line absorption and the utility of the OH ground state
transitions as molecular and atomic gas tracers based on comparisons of column
densities and kinematic properties. 

\smallskip
\noindent
The paper is structured as follows: In Section~2, we present the observations and delineate 
the use of ancillary data. Section~3 gives the results that
are discussed in Section~4. The conclusions are provided in Section~5 and
the Appendix gives additional information about the \ce{OH} detections. 

\section{Observations and data reduction} \label{sec:observations}
We have mapped the four OH ground state transitions in the first quadrant of the Milky Way with the Karl G. Jansky Very Large Array (VLA) in C-configuration. The observations are part of the large program THOR, with data taken over several observational periods, mapping between $l$ = 14.5\degree\ and $l$ = 67.25\degree, $|b|\leq$1.1\degree. Here we present \ce{OH} observations in absorption for the entire survey region and include the \ce{OH} absorption data in the pilot study of 4 square degrees around the star-forming region W43, which have already been presented in \citet{WalshBeuther:2016aa}. As the observing strategy was discussed in \citet{BeutherBihr:2016aa}, we will restrict the discussion of the THOR data in this paper to the \ce{OH} absorption observations. 

\smallskip
\noindent
The OH satellite line transitions, located at 1.612231 and 1.720530 GHz \citep{Schoiervan-der-Tak:2005aa,Offervan-Hermet:1994aa}, were observed with two 2-MHz-wide spectral windows. The two main line transitions at 1.665402 and 1.667359 GHz were observed in one 4-MHz-wide spectral window for $l=29.2\degree-31.5\degree$, $l=37.9\degree-47.1\degree$ and $l=51.2\degree-67.0\degree$. The rest of the survey coverage was mapped in the 1.665-GHz transition alone, using a 2-MHz-wide spectral window. Channel widths for all transitions are 3.9{\us}kHz ($\sim$ 0.7\us\kms) in the pilot study 
($l=29.2\degree-31.5\degree$, $|b|\leq1.1\degree$) and 7.8{\us}kHz ($\sim$ 1.4\us\kms) for the rest of the survey. The channel width of the \ce{OH} transitions was chosen to be equivalent to the simultaneously-conducted \ion{H}{i} observations, which in turn were following the spectral resolution of the existing \ion{H}{i} Very Large Array Galactic Plane Survey (VGPS; \citealt{StilTaylor:2006aa}) for comparability. All data have been taken at a total integration time per pointing of 5$-$6 min, split into 3 observations of equal time to improve the $uv$-coverage.

\smallskip
\noindent
The data were calibrated with the CASA calibration pipeline, and the solutions were iterated after removing data of individual baselines and antennas in time ranges in which these contain artifacts. Using CASA\footnote{\url{http://casa.nrao.edu}; version 4.2.2}, data were continuum-subtracted and gridded on a common velocity grid of 1.5\us\kms\ resolution, and subsequently inverted and deconvolved with the CASA task \texttt{clean}. The line free channels were cleaned separately to obtain the continuum at 1666{\us}MHz for the 1665/1667{\us}MHz transitions. These continuum data were used for the later analysis for consistency in calibration. 

\smallskip
\noindent
The angular resolution of the data is between ${12\farcs7\times12\farcs4}$ and ${18\farcs7\times12\farcs5}$, depending on the transition and on the elevation of the source at the time of observation. We regrid all data to the Galactic coordinate system and smooth all data to a resolution of ${20\arcsec\times20\arcsec}$. The noise is typically about 10~m\jyb\ at a velocity resolution of 1.5{\us}\kms, except for areas around strong emission sources.

\smallskip
\noindent
The \ce{^{13}CO}(1-0) observations employed as tracer for the molecular gas are taken from GRS \citep[Galactic Ring Survey,][]{JacksonRathborne:2006aa} and for two sources  (G60.882$-$0.132 and G61.475+0.092) that lie beyond $l= 60\degree$ from the Exeter FCRAO \ce{CO} survey \citep{MottramBrunt:2010aa}. Both datasets were taken with a single dish telescope having a $46\arcsec$ FWHM beamsize. All data have been converted to main-beam temperature ($T_{\rm mb}$) using a beam efficiency of $\eta = 0.48$ \citep{JacksonRathborne:2006aa}. 
The \ion{H}{i} 21 cm absorption gives column densities of the atomic gas and the data are also from THOR \citep{BihrBeuther:2015aa,BeutherBihr:2016aa}. As the \ion{H}{i} spectral cubes (${20\arcsec\times20\arcsec}$ resolution) were imaged without continuum subtraction, the continuum is extracted from the line-free channels in the provided spectral cubes and used later to derive the line-to-continuum ratio. All datasets are gridded on the same coordinate system as the 1666{\us}MHz continuum image. 

\smallskip
\noindent
A continuum catalog was extracted from the narrow band continuum maps at 1666{\us}MHz with a spatial resolution of ${20\arcsec\times20\arcsec}$ with the source finding algorithm
\texttt{blobcat} \citep[][]{HalesMurphy:2012aa}. The noise maps were created using the residual maps that have been produced during deconvolution 
\citep[see also][]{BihrJohnston:2016aa}. To verify the completeness of the catalog, it was matched to the continuum source catalog of the THOR survey for sources with $F_{\rm cont}\ge0.1${\us}\jyb, which was derived from \mbox{128-MHz-wide} spectral windows and therefore has higher sensitivity (\citealt{BihrJohnston:2016aa}, Wang et al., in prep.). One continuum source, G30.854+0.151, which was not detected in the narrow-band catalog due to strong sidelobes from a nearby continuum source, was subsequently added to the detections, and the flux from the broadband continuum catalog was used for quantitative analysis in the following. 

\smallskip
\noindent
For the quantitative comparison of OH and \ion{H}{i} absorption to \ce{^{13}CO}(1-0) emission in Section~\ref{sec:ohabundance}, the spatial resolution of the OH, \ion{H}{i} and continuum datacubes is degraded to match the resolution of \ce{^{13}CO}(1-0) data at 46''. For simplicity, the deconvolved images are smoothed with a Gaussian kernel in the image plane. The relevant quantity for absorption measurements in Sect.~\ref{sec:ohabundance} is the ratio of absorption line to continuum. While the baselines of the VLA in C-configuration sample angular scales of 46'', the actual scales probed depend intrinsically on the emission structure of the continuum source. Scales of 46'' are only probed for continuum emission extending at least 46'' in angular size. For inhomogeneous emission, patches of stronger continuum emission will have a larger contribution to the line-to-continuum ratio. With this general consideration regarding absorption measurements, smoothing the data to 46'' resolution gets closest to the scales probed by \ce{^{13}CO} emission. An example of an extended OH absorption map at ${20\arcsec\times20\arcsec}$ resolution is provided for the star forming complex W43 in Sect.~\ref{sec:analysisw43}. 

\smallskip
\noindent
To minimize the introduction of systematic errors, the continuum is derived from line-free channels and thus has the same uv-coverage and calibration as the spectral line data. The median noise in the smoothed spectral line cubes is at 0.013{\us}\jyb, with variations between 0.008{\us}\jyb\ and 0.020{\us}\jyb\ (see also gray lines in Fig.~\ref{fig:plot_oh1665_vs_contemp_with_tau_min}). An example spectrum of all transitions is shown in Fig.~\ref{fig:plot_positions_single}. 

\smallskip
\noindent
To address extended \ce{OH} absorption in W43 \citep[e.g.,][]{SmithBiermann:1978aa,MotteNguyen-Luong:2014aa}, we employ the full resolution OH 1667{\us}MHz data (${20\arcsec\times20\arcsec}$). It is compared to APEX observations of 870\,$\mu$m dust emission from the ATLASGAL\footnote{\url{http://atlasgal.mpifr-bonn.mpg.de/cgi-bin/ATLASGAL_DATABASE.cgi}} survey \citep{SchullerMenten:2009aa,ContrerasSchuller:2013aa,UrquhartCsengeri:2014aa} and IRAM \ce{C^{18}O}(2-1) emission from the literature \citep{CarlhoffNguyen-Luong:2013aa}. Both datasets were smoothed to the same spatial resolution and the \ce{C^{18}O}(2-1) emission to the same spectral resolution as the \ce{OH} data. 

\section{Results}     \label{sec:results}
\subsection{Detection statistics}
The OH main line transitions at 1665{\us}MHz and 1667{\us}MHz are searched 
for absorption features at the locations of peaks in the continuum maps at 1666{\us}MHz.
Absorption lines which are detected at a signal-to-noise level larger than 4
are classified as detections. In total, significant OH absorption is found against 42 continuum sources (Fig.~\ref{fig:continuum_image}).

\begin{figure}
 \centering
 \includegraphics[width=0.99\columnwidth]{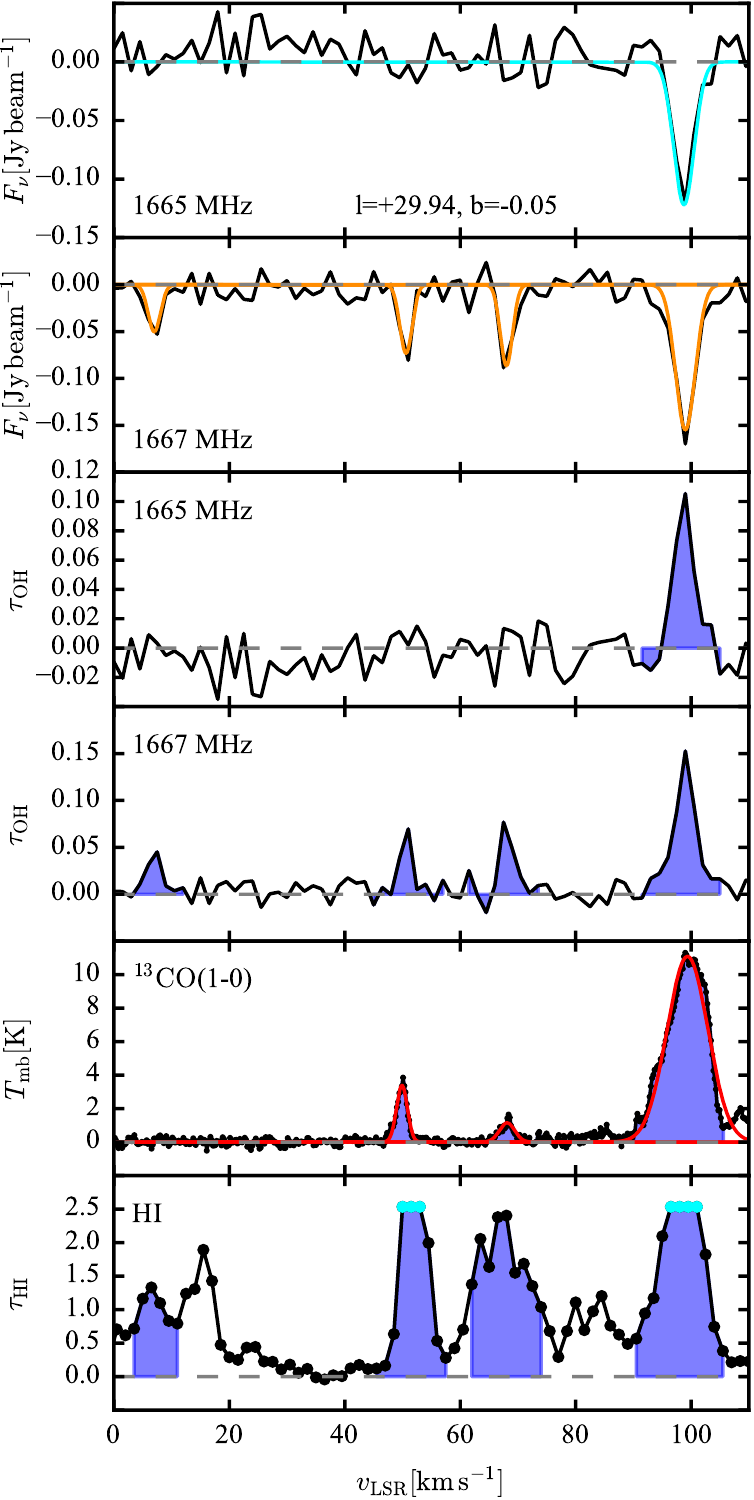} 
 \caption{Example spectra and optical depths at $l$=+29.935\degree, $b$=$-$0.053\degree\ (at 46'' resolution). The topmost two panels show 1665{\us}MHz and 1667{\us}MHz absorption features. The fitted Gaussian profiles for the 1665{\us}MHz~line~(cyan) and 1667{\us}MHz~line~(orange) are overlaid. The two middle panels show the optical depth of the 1665{\us}MHz and 1667{\us}MHz transitions. The second panel from the bottom shows emission of ${}^{13}{\rm CO}(1-0)$ in main-beam temperature ($T_{\rm mb}$), overlaid with a fitted Gaussian profile (red). The lowermost panel shows \ion{H}{i} optical depth as measured from the absorption spectra. Lower limits (cyan dots) are given for saturated bins. The blue shaded area in the lower four panels denotes the area of the transitions, from which the column densities are determined.}
 \label{fig:plot_positions_single}
\end{figure}

\begin{figure*}
 \centering
 \includegraphics[width=0.99\textwidth]{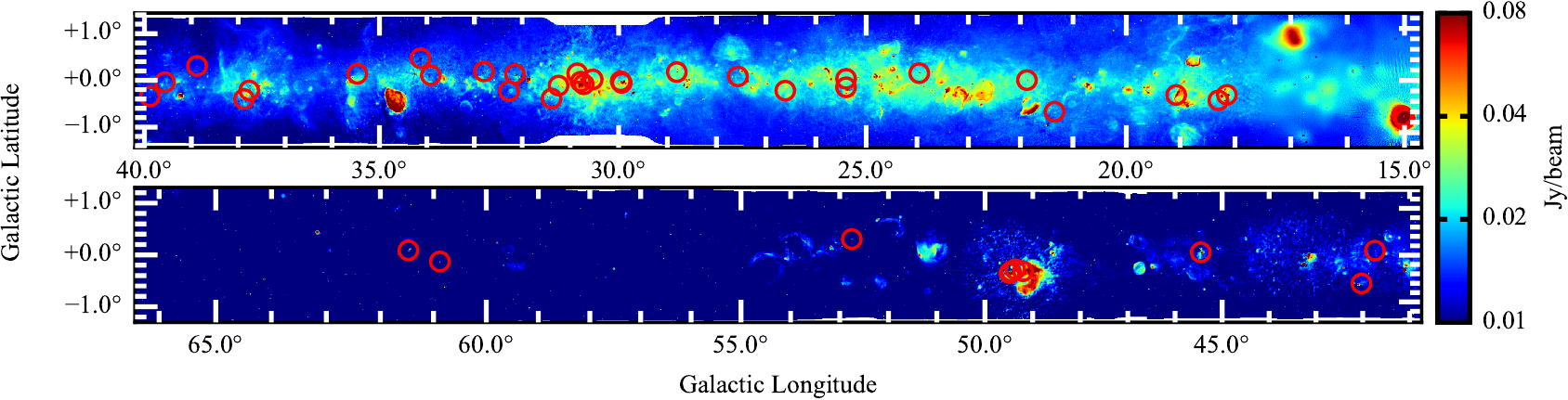} 
 \caption{Detections of OH absorption at 1665 and 1667\,MHz (red circles) overplotted on continuum emission at 1.4\,GHz from the combined THOR and VGPS data (\citealt{BeutherBihr:2016aa}, Wang et al., submitted).}
 \label{fig:continuum_image}
\end{figure*}

\smallskip
\noindent
As \ce{OH}~absorption can occur at multiple velocities due to
clouds along the line of sight, in total we find absorption in both main lines
in 59~velocity components. Of these, we find 30~and~17~velocity components
exclusively in the 1665{\us}MHz and 1667{\us}MHz \ce{OH}~ground~state~transitions,
respectively, and 12 in both lines, matching in velocity. We detect a higher
number of OH~1665{\us}MHz transitions, because the 1667{\us}MHz~transition was
observed only around W43, during the pilot study, and during the second part
of the THOR~survey (see section~\ref{sec:observations}). 
Conversely, we sometimes find OH features in the 
1667{\us}MHz~transition that have no counterparts in the 1665{\us}MHz~transition. 
This is expected, as the statistical weight of this
transition is roughly twice as large, hence at a given sensitivity, optical 
depths and column densities twice as low can be probed. Examples of these are absorption at 7.0, 51.0 and 67.5\us\kms\ towards 
G29.935$-$0.053 (Fig.~\ref{fig:plot_positions_single}). 

\smallskip
\noindent
The continuum sources with detections are listed in Table~\ref{tbl:detections}. 
The spectra of the detected absorption lines are displayed
in Figs.~\ref{fig:plot_positions_00}~-~\ref{fig:plot_positions_12}.

\begin{figure}
 \centering
 \includegraphics[width=0.99\columnwidth]{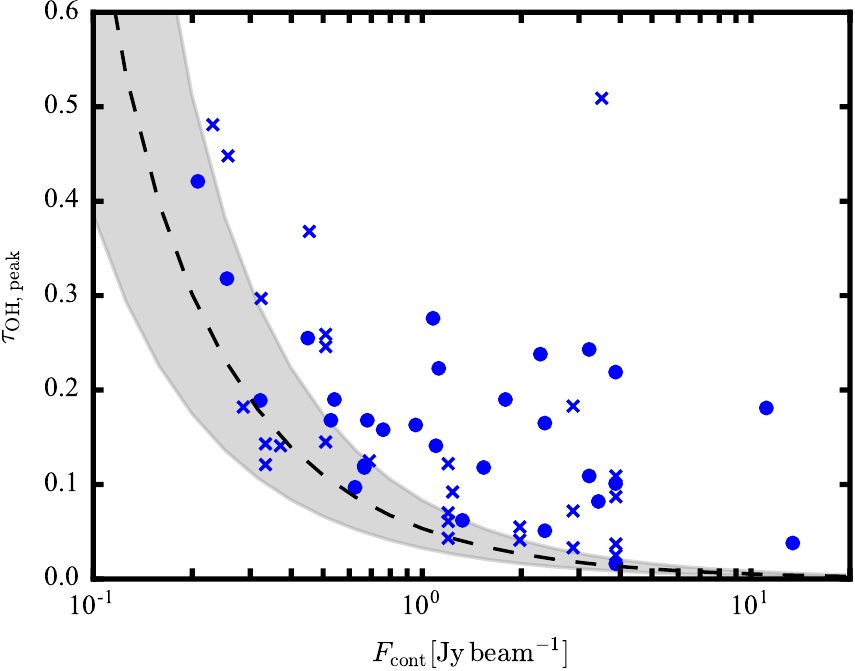} 
 \caption{Peak optical depth of the 1665-MHz~transition (circles) and the 1667-MHz~transition (crosses) versus continuum flux density at 46\arcsec. The sensitivity in \ce{OH}~optical
  depth is indicated by an average $4$-$\sigma$~detection~limit (black dashed curve). Variations in the detection limit among sightlines are indicated by the gray shaded area.}
 \label{fig:plot_oh1665_vs_contemp_with_tau_min}
\end{figure}

\subsubsection{Sensitivity of the survey}
The weakest continuum source with OH~absorption has a continuum flux density of 0.1\us\jyb. As shown in Fig.~\ref{fig:plot_oh1665_vs_contemp_with_tau_min}, stronger continuum sources are more sensitive to lower OH column densities. Therefore, the fraction of continuum sources that exhibit OH absorption is dependent on the continuum source strength (which sets the $4\sigma$ detection threshold, $\tau_{\rm min}$): There are 291 continuum sources above a flux density of 0.1\us\jyb\ at ${20\arcsec\times20\arcsec}$ ($\tau_{\rm min} \sim 0.5$), of which 42 show OH absorption lines (14.4\%). Above 1.0\us\jyb\ ($\tau_{\rm min} \sim 0.04$), 13 of 29 LOS have OH absorption lines (44.8\%), while above 2.0\us\jyb\ ($\tau_{\rm min} \sim 0.02$), 3 out of 4 LOS (75.0\%; this value should be taken with caution due to small number statistics). The cumulative detection fraction therefore is an increasing function of the continuum strength.

\smallskip
\noindent
The reason for this increase is the dependence of the sensitivity of the optical depth ($\tau$) on the strength of the continuum source:
\begin{equation}
\tau = - \ln{\left(\frac{F_{\rm line}}{F_{\rm cont}} +1\right)}, 
\label{eq:opticaldepth}
\end{equation}
where $F_{\rm line}$ is the continuum subtracted \ce{OH} absorption spectrum
and $F_{\rm cont}$ the continuum emission. Contributions from OH emission to the observed signal are neglected, assuming the distribution of OH is smooth enough for emission to be filtered out by the interferometric observations and due to the small OH excitation temperatures in comparison to the continuum emission (see sect.~\ref{sec:line_integrals} for a more detailed discussion). While the OH transitions are mapped at rather uniform noise, the noise in OH optical depth, and therefore its sensitivity, is inversely proportional to continuum flux. 

\smallskip
\noindent
The OH ground state main lines show maser emission against many of the strong continuum sources. Non-detection of absorption lines is therefore not indicative of absence of OH in these lines of sight, but points to OH having different excitation conditions. Such regions typically have high dust temperatures and local densities ($T_{\rm dust} > 80$\,K, $n_{\rm H_2} >10^{5}$\,${\rm cm}^{-3}$; e.g., \citealt{CesaroniWalmsley:1991aa}, \citealt{GuibertRieu:1978aa}, \citealt{CsengeriMenten:2012aa}, \citealt{Elitzur:1992aa}, and references therein).

\smallskip
\noindent
Artifacts also influence some of the spectra. This can be due to increased noise caused by residual radio frequency interference (RFI),
which should be
minor, as the data have been closely inspected for RFI prior to imaging. 
Also, strong line emission of non-thermal origin can leave various
traces. At the position of a maser, adjacent channels are affected by
Gibbs ringing, which is a recurring pattern in velocity of emission and absorption. 
 If the maser emission is strong, channel maps around the peak velocity of the emission can be affected by
increased noise levels and sidelobes. In the case of \mbox{W51 Main} (e.g., \citealt{GinsburgBressert:2012aa}; G49.488$-$0.380 in this work), for example, it is difficult to discern negative sidelobes from true
absorption. Absorption is present in this region as it is different from the shape of the interferometry pattern. 
Both, however, overlap and therefore a quantitative analysis of the affected velocity channels is not possible. 

\subsubsection{\ion{H}{ii}~region associations} \label{sec:hiiregion_ass}
In order to classify the continuum sources, we compare their location and the velocity of the detections to the emission from Radio Recombination Lines (RRLs) as reported in the WISE\footnote{\url{http://astro.phys.wvu.edu/wise/}} catalog of \ion{H}{ii}-regions (\citealt{AndersonBania:2014aa}). The spatial selection criterium is overlap with the \ion{H}{ii}-regions using their angular sizes as reported in the WISE catalog. Since typical velocity differences between \ion{H}{ii} regions and associated molecular gas are lower than 10\us\kms\ \citep{AndersonBania:2009aa,AndersonBania:2014aa}, we use this as criterium for association in velocity. 

\smallskip
\noindent
For completeness, we also search the RRL observations in THOR \citep{BeutherBihr:2016aa} and catalogs of dense molecular gas tracers associated with compact and ultracompact \ion{H}{ii} regions (e.g., \ce{NH3}, \ce{HCO^+}; ATLASGAL survey; \citealt{UrquhartThompson:2013aa,ContrerasSchuller:2013aa,UrquhartCsengeri:2014aa}). While not adding new sources, counterparts of many OH detections could be found also in these datasets.

\smallskip
\noindent
To confirm the presence \ion{H}{ii}-regions, we obtain information on the spatial extent and the spectral index of the continuum sources from the THOR continuum source catalog (\citealt{BihrJohnston:2016aa}; Wang et al. in prep.). As \ion{H}{ii} regions would be located in our Galaxy, they are likely to be spatially resolved within THOR. The spectral index between 1 and 2\,GHz helps to distinguish between thermal (with a spectral index of $\alpha \gtrsim -0.1$) and non-thermal emission sources (with a spectral index of $\alpha \sim -1$). Most of the continuum sources with RRL counterparts show thermal emission and are spatially resolved. 

\smallskip
\noindent
In total, 47 OH absorption components have its origin in molecular gas that is associated with \ion{H}{ii} regions in position and velocity, which represent 80\% of all detections (see discussion in Sect.~\ref{sec:discussion_distribution}).
We find that 38 out of 42 of the cm-continuum sources, against which the detections occur, show evidence of being \ion{H}{ii} regions. 
Three of the four other continuum sources are likely to be of extragalactic origin, as they are spatially unresolved and show non-thermal emission.

\smallskip
\noindent
Twelve velocity components in the 1665{\us}MHz and 1667{\us}MHz transitions, of
which 4 are detected in both, originate from clouds that are not associated with \ion{H}{ii} regions. 
Neither RRL emission in the WISE catalog, nor dense molecular gas tracers are reported at the same $v_{\rm LSR}$. 
The peak optical depth is lower than seen for sources associated with \ion{H}{ii} regions. 
Accordingly, these absorption features are likely to originate from foreground, potentially diffuse clouds.

\subsubsection{Distribution of sources in the Galactic plane}
\begin{figure}
 \centering
 \includegraphics[width=0.99\columnwidth]{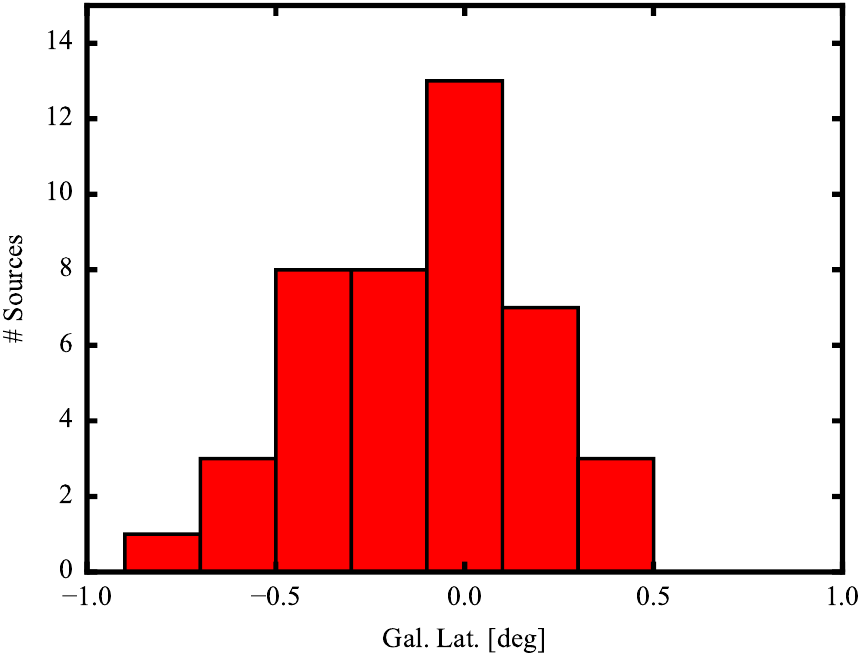} 
 \caption{Number of continuum sources with OH absorption detections versus Galactic latitude.}
 \label{fig:plot_hist_glat}
\end{figure}

\begin{figure}
 \centering
 \includegraphics[width=0.99\columnwidth]{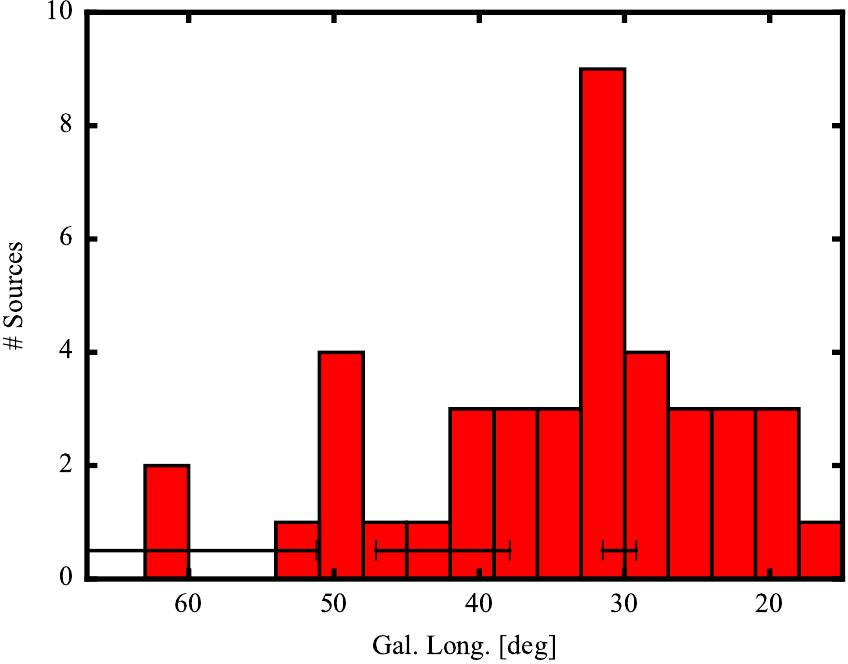} 
 \caption{Number of continuum sources with OH absorption detections versus Galactic longitude. The 1665\, MHz transition was observed over the entire region of the survey. The coverage of the 1667\,MHz transition is indicated by the black bars.
 }
 \label{fig:plot_hist_glon}
\end{figure}

The distribution of OH absorption detections is strongly concentrated
towards the Galactic midplane (Figs.~\ref{fig:plot_hist_glon}~and~\ref{fig:plot_hist_glat}), while relatively few 
detections are made at $|b| > 0.5\degree$. This follows the distribution of 
resolved Galactic continuum sources as a function of Galactic latitude \citep[e.g.,][]{BihrJohnston:2016aa}. 
Figure~\ref{fig:plot_hist_glat} is slightly skewed towards negative Galactic latitudes. This may be due to the sun being located above the true galactic plane, while the sun is located at $b$=0.0\degree\ in the Galactic coordinate system \citep[][]{BlaauwGum:1960aa,RaganHenning:2014aa}. Depending on the distance of the object and the assumptions used to determine the physical location of the galactic plane, sources which lie, e.g., in the Scutum--Centaurus Arm have galactic latitudes of $b=[-0.4\degree, -0.1\degree]$ \citep[see discussion in, e.g.,][]{GoodmanAlves:2014aa}, which agrees well with the observed extent of the source distribution towards lower Galactic latitudes. 

\smallskip
\noindent
The histogram of sources versus Galactic longitude (Fig.~\ref{fig:plot_hist_glon}) 
reflects the Galactic structure, with a peak in the number of OH absorption sources 
at longitudes $l = 30 \degree$ and $50 \degree$, which are the tangential points
of the Scutum and Sagitarius spiral arms, respectively \citep[e.g.,][]{ReidMenten:2014aa}. 
This confirms that most of the continuum background sources, against which OH absorption is seen, are of Galactic origin, 
as already indicated by the large number of \ion{H}{ii} regions in our sample. 

\begin{table*}
\centering
\caption{Lines of sight with detections of \ce{OH} 1665/1667{\us}MHz absorption.}
\small
\setlength\tabcolsep{4pt}
\begin{tabular}{l|rr|rrrr|rrl}
\hline \hline
Name & RA (J2000) & Dec. (J2000) & \multicolumn{1}{c}{${F_{c}}(20\arcsec)$}& \multicolumn{1}{c}{${F_{c}}(46\arcsec)$}& Ext. & $\alpha$ & OH velocities & Ass. \ion{H}{ii} &\ion{H}{ii}-region\\
	 & \multicolumn{1}{c}{${\rm ({}^{h}\,{}^{m}\,{}^{s})}$}&\multicolumn{1}{c|}{ (\degree\ \arcmin\ \arcsec)} &\multicolumn{1}{c}{$\left[\frac{\rm Jy}{\rm beam}\right]$}&\multicolumn{1}{c}{$\left[\frac{\rm Jy}{\rm beam}\right]$}& & &$\left[\frac{\rm km}{\rm s}\right]$ & &\\
\hline
G14.490$+$0.021 & 18 16 46.666 & $-$16 20 33.15 & 0.2 & 0.2 & 1 & +0.0 & (23.1) & (1) & G014.489+00.020 \\
G14.996$-$0.738 & 18 20 33.756 & $-$16 15 22.46 & 1.1 & 3.4 & 1 & +0.7 & 21.8 & 1 & G015.035$-$00.677; M17; S45 \\
G15.033$-$0.679 & 18 20 25.129 & $-$16 11 43.00 & 3.4 & 13.4 & 1 & +0.7 & 12.9 & 1 & G015.035$-$00.677; M17; S45  \\
G18.148$-$0.283 & 18 25 01.009 & $-$13 15 30.78 & 0.4 & 1.1 & 1 & +0.1 & 56.4 & 1 & G018.144$-$00.286 \\
G18.303$-$0.390 & 18 25 42.284 & $-$13 10 18.26 & 0.8 & 1.1 & 1 & +0.4 & (27.2), 32.2, (36.2) & (1), 1, (1) & G018.305$-$00.391 \\
G19.075$-$0.288 & 18 26 48.526 & $-$12 26 28.66 & 0.4 & 0.7 & 1 & +0.4 & 64.6 & 1 & G019.066$-$00.281; W39 \\
G21.347$-$0.629 & 18 32 20.815 & $-$10 35 11.37 & 1.0 & 1.0 & 0 & +0.0 & 56.1 & 0 &  \\
G21.874$+$0.007 & 18 31 02.524 & $-$09 49 30.49 & 0.5 & 0.6 & 1 & +0.3 & 21.8 & 1 & G021.870+00.010 \\
G23.956$+$0.150 & 18 34 25.303 & $-$07 54 46.37 & 0.9 & 1.3 & 1 & +0.6 & 80.0 & 1 & G023.956+00.152 \\
G25.396$+$0.033 & 18 37 30.623 & $-$06 41 16.38 & 0.3 & 0.4 & 1 & +0.4 & $-$12.1 & 1 & G025.396+00.034 \\
G25.397$-$0.141 & 18 38 08.202 & $-$06 45 58.85 & 1.6 & 2.4 & 1 & +0.8 & 67.1, 94.5 & 1, 1 & G025.383$-$00.177 \\
G26.609$-$0.212 & 18 40 37.495 & $-$05 43 19.00 & 0.2 & 0.2 & 1 & +0.3 & $-$33.2 & 1 & G026.610$-$00.212a \\
G27.563$+$0.084 & 18 41 19.379 & $-$04 44 17.57 & 0.1 & 0.1 & 1 & +0.2 & 86.0 & 1 & G027.562+00.084 \\
G28.806$+$0.174 & 18 43 16.959 & $-$03 35 28.97 & 0.3 & 0.7 & 1 & +0.1 & 79.8, 103.5 & 0, 1 & G028.801+00.174 \\
G29.935$-$0.053 & 18 46 09.525 & $-$02 41 27.54 & 0.5 & 1.2 & 1 & +0.1 & 7.0, 50.6, 68.0, 98.9 & 0, 0, 0, 1 & G029.945$-$00.039; G29 \\
G29.957$-$0.018 & 18 46 04.241 & $-$02 39 19.25 & 1.4 & 2.0 & 1 & +0.9 & 8.0, 99.8 & 0, 1 &G029.945$-$00.039; G29 \\
G30.535$+$0.021 & 18 46 59.359 & $-$02 07 26.40 & 0.5 & 0.7 & 1 & +0.6 & (43.7, 49.4), 91.6 & (1, 1), 1 & G030.539+00.024 \\
G30.720$-$0.083$^{*}$ & 18 47 41.713 & $-$02 00 23.48 & 0.4 & -- & 0 & +1.0 & 93.6 & 1 & G030.782$-$00.028; W43 \\
G30.783$-$0.028 & 18 47 36.898 & $-$01 55 30.26 & 1.3 & 3.9 & 1 & +0.3 & 77.9, 81.9, 87.1 & 1, 1, 1& G030.782$-$00.028; W43 \\
& &&  &  &&  &  92.1, 98.6 & 1, 1 &  \\
G30.854$+$0.151 & 18 47 06.546 & $-$01 46 49.14 & 0.1 & 0.3 & 1 & $-$0.1 & 95.4 & 1 & G030.852+00.149a \\
G31.242$-$0.110 & 18 48 44.821 & $-$01 33 14.65 & 0.4 & 0.5 & 1 & +0.5 & 19.6, 79.2, 83.7 & 1, 0, 0 & G031.239$-$00.108 \\
G31.388$-$0.383 & 18 49 59.195 & $-$01 32 56.04 & 1.2 & 1.2 & 0 & $-$0.9 & 18.1 & 0 &  \\
G32.151$+$0.132 & 18 49 32.499 & $-$00 38 05.79 & 0.4 & 0.5 & 1 & +0.4 & 93.8 & 1 & G032.160+00.130 \\
G32.272$-$0.226 & 18 51 02.358 & $-$00 41 24.22 & 0.3 & 0.3 & 1 & +0.2 & 22.6 & 1 & G032.272$-$00.226 \\
G32.798$+$0.190 & 18 50 31.084 & $-$00 01 56.81 & 1.2 & 1.5 & 1 & +1.0 & 12.8 & 1 & G032.960+00.276 \\
G32.928$+$0.607 & 18 49 16.315 &  00 16 23.85 & 0.2 & 0.3 & 1 & +0.4 & ($-$33.9) & (1) & G032.928+00.607 \\
G33.915$+$0.110 & 18 52 50.381 &  00 55 29.56 & 0.6 & 0.8 & 1 & +0.5 & (95.3, 101.5), 106.3 & (1, 1), 1 & G033.910+00.110 \\
G34.132$+$0.471 & 18 51 57.102 &  01 16 58.86 & 0.4 & 0.5 & 1 & +0.2 & 33.8 & 1 & G034.133+00.471 \\
G35.467$+$0.139 & 18 55 34.169 &  02 19 11.16 & 0.2 & 0.3 & 1 & +0.6 & 77.0 & 1 & G035.470+00.140 \\
G37.764$-$0.215 & 19 01 02.118 &  04 12 03.76 & 0.4 & 1.1 & 1 & +0.1 & 63.4 & 1 & G037.760$-$00.200 \\
G37.874$-$0.399 & 19 01 53.641 &  04 12 52.52 & 1.3 & 1.8 & 1 & +0.5 & 61.0 & 1 & G037.870$-$00.400; W47 \\
G38.876$+$0.308 & 19 01 12.538 &  05 25 44.28 & 0.2 & 0.2 & 0 & +0.8 & $-$16.3 & 1 & G038.875+00.308 \\
G39.565$-$0.040 & 19 03 43.340 &  05 52 55.41 & 0.4 & 0.5 & 0 & $-$0.9 & 23.2 & 0 &  \\
G39.883$-$0.346 & 19 05 24.156 &  06 01 28.52 & 0.2 & 0.3 & 0 & +0.8 & 56.9 & 1 & G039.883$-$00.346 \\
G41.741$+$0.097 & 19 07 15.655 &  07 52 42.12 & 0.3 & 0.3 & 1 & +0.3 & 14.2 & 1 & G041.740+00.100 \\
G42.027$-$0.604 & 19 10 18.252 &  07 48 32.39 & 0.3 & 0.4 & 1 & $-$1.0 & 65.1 & 0 &  \\
G45.454$+$0.060 & 19 14 21.188 &  11 09 12.92 & 1.4 & 2.9 & 1 & +0.4 & 56.0, 59.4, 64.9 & 1, 1, 1 & G045.454+00.059; K47 \\
G49.206$-$0.342 & 19 23 00.834 &  14 16 50.73 & 0.8 & 2.3 & 1 & +0.2 & 65.3 & 1 & G049.205$-$00.343; W51 \\
G49.369$-$0.302 & 19 23 11.204 &  14 26 37.06 & 1.5 & 3.2 & 1 & +0.5 & 50.9, 62.9 & 1, 1 & G049.384$-$00.298; W51  \\
G49.459$-$0.353 & 19 23 32.908 &  14 29 55.71 & 1.9 & 3.9 & 1 & +1.3 & 62.0, 68.6 & 1, 1 & G049.490$-$00.381; W51 \\
G49.488$-$0.380 & 19 23 42.119 &  14 30 41.99 & 4.0 & 11.1 & 1 & +1.3 & 65.9 & 1 & G049.490$-$00.381; W51 \\
G52.753$+$0.334 & 19 27 32.385 &  17 43 27.32 & 0.3 & 0.3 & 1 & +0.1 & 12.1, 45.2 & 1, 0 & G052.757+00.334 \\
G60.882$-$0.132 & 19 46 20.621 &  24 35 17.59 & 0.2 & 0.3 & 1 & +0.0 & 22.3 & 1 & G060.883$-$00.133; S87 \\
G61.475$+$0.092 & 19 46 48.189 &  25 12 47.06 & 2.1 & 3.5 & 1 & +0.5 & (5.9), 21.1 & (0), 1 & G061.477+00.094; S88 \\
\hline
\label{tbl:detections}
\end{tabular}
\tablefoot{The source name is constructed from the Galactic coordinates at which the peak of the continuum emission measured at 20\arcsec\ resolution occurs. The columns ${F_{c}}$ denotes the flux density at the given coordinates for an angular resolution of 20\arcsec\ and 46\arcsec. The column ``Ext.'' denotes whether the source is resolved (\textit{1} for resolved, \textit{0} for unresolved sources), and the column $\alpha$ is the L$-$band spectral index, defined as $I(\nu) \propto \nu^\alpha$, which are both taken from \citet{BihrJohnston:2016aa} and Wang et al. (in prep.). 
The central velocities of detected OH main line absorption are summarized in column 8 ``OH velocities'' (tentative, 3-$\sigma$ components are listed in brackets in column 8 and 9). 
Association in position and velocity with an \ion{H}{ii} region is indicated with \textit{1} in column 9.
Criteria for the association are a smaller angular separation from the \ion{H}{ii} region than its radius, and a velocity difference of less than 10\us\kms. 
The name of the \ion{H}{ii} regions are obtained from the WISE catalog of \ion{H}{ii}-regions (\citealt{AndersonBania:2014aa}).
{\bf Footnotes:} $^{(*)}$ Used 20\arcsec\ resolution data only in order not to blend OH absorption with adjacent maser.
}
\end{table*}

\begin{table*}
\caption{Line properties of \ce{OH} 1665/1667{\us}MHz absorption and \ce{^{13}CO}(1-0) emission.}
\tiny
\setlength\tabcolsep{2.pt}
\renewcommand{\arraystretch}{0.8}
\begin{tabular}{l|rrrrl|rrrrl|rrrl|c|c|ll}
\hline \hline
&
\multicolumn{5}{c|}{OH 1665{\us}MHz}&
\multicolumn{5}{c|}{OH 1667{\us}MHz}&
\multicolumn{4}{c|}{${\rm{}^{13}CO(1-0)}$}&
&
&
\multicolumn{2}{c }{Notes}
\\
Name &
v & $\Delta$v & $F_{\rm peak}$ & $\tau_{\rm peak}$ & $\int\tau{\rm dv}$ &
v & $\Delta$v & $F_{\rm peak}$ & $\tau_{\rm peak}$ & $\int\tau{\rm dv}$ &
v & $\Delta$v & $T_{\rm mb, p}$ & $\int T_{\rm mb}{\rm dv}$ &
Ass.&
Int. Range&
FWHM & Int.
\\
&
$\left[\frac{\rm km}{\rm s}\right]$ & $\left[\frac{\rm km}{\rm s}\right]$ & $\left[\frac{\rm Jy}{\rm beam}\right]$ & & $\left[\frac{\rm km}{\rm s}\right]$ &
$\left[\frac{\rm km}{\rm s}\right]$ & $\left[\frac{\rm km}{\rm s}\right]$ & $\left[\frac{\rm Jy}{\rm beam}\right]$ & & $\left[\frac{\rm km}{\rm s}\right]$ &
$\left[\frac{\rm km}{\rm s}\right]$ & $\left[\frac{\rm km}{\rm s}\right]$ &[K] & $\left[{\rm K}\frac{\rm km}{\rm s}\right]$ &
 \ion{H}{ii}&
$\left[\frac{\rm km}{\rm s},\frac{\rm km}{\rm s}\right]$&
&
\\
\hline
G14.490$+$0.021 & 23.1 & 2.9 & $-$0.06 & 0.34\tablefootmark{a} & <1.35 & -- & -- & -- & -- & -- & -- & -- & -- & --      			&(1)&[ +19.4, +27.1]& 2(OH)    & 2(OH)\\
G14.996$-$0.738 & 21.8 & 4.0 & $-$0.28 & 0.08 & 0.37 & -- & -- & -- & -- & -- & -- & -- & -- & --                                        			&1&[ +16.0, +28.0]& 1        & 1   \\
G15.033$-$0.679 & 12.9 & 5.4 & $-$0.49 & 0.04 & 0.21 & -- & -- & -- & -- & -- & -- & -- & -- & --                                        			&1&[  +7.0, +16.6]& 1        & 1   \\
G18.148$-$0.283 & 56.4 & 3.5 & $-$0.22 & 0.22 & 0.81 & -- & -- & -- & -- & -- & 55.5 & 3.7 & 6.81 & 32.84                                	 	&1&[ +51.2, +61.0]& 1665     & 1665\\
G18.303$-$0.390 & 27.2 & 4.3 & $-$0.04 & 0.03\tablefootmark{a} & --\tablefootmark{b} & -- & -- & -- & -- & -- & -- & -- & -- & --        	&(1)&             --& 3(OH)    & 3(OH)\\
 & 32.2 & 2.5 & $-$0.15 & 0.14 & 0.47 & -- & -- & -- & -- & -- & 33.0 & 4.1 & 7.69 & 34.90                                               			&1&[ +27.0, +38.0]& 4(CO)    & 1665\\
 & 36.2 & 2.5 & $-$0.02 & 0.02\tablefootmark{a} & --\tablefootmark{b} & -- & -- & -- & -- & -- & -- & -- & -- & --                       		&(1)&             --& 3(OH)    & 3(OH)\\
G19.075$-$0.288 & 64.6 & 7.4 & $-$0.11 & 0.17 & 1.35 & -- & -- & -- & -- & -- & 65.1 & 6.1 & 7.52 & 55.94                                		&1&[ +56.9, +75.0]& 4(OH)    & 1665\\
G21.347$-$0.629 & 56.1 & 2.5 & $-$0.15 & 0.16 & 0.47 & -- & -- & -- & -- & -- & 56.2 & 1.8 & 3.43 & 21.95              				&0&[ +48.0, +58.6]& 1665     & 4(CO)\\
G21.874$+$0.007 & 21.8 & 8.8 & $-$0.08 & 0.10 & 1.01 & -- & -- & -- & -- & -- & 22.3 & 4.9 & 5.86 & 30.52                                		&1&[ +15.5, +27.0]& 5(OH)    & 1665\\
G23.956$+$0.150 & 80.0 & 5.5 & $-$0.08 & 0.06 & 0.38 & -- & -- & -- & -- & -- & 79.6 & 3.6 & 9.31 & 38.99                                		&1&[ +74.9, +86.1]& 4(CO)    & 1665\\
G25.396$+$0.033 & $-$12.1 & 2.7 & $-$0.10 & 0.25 & 0.80 & -- & -- & -- & -- & -- & -- & -- & -- & --                                       		&1&[ $-$15.1,  $-$8.9]& 1        & 1   \\
G25.397$-$0.141 & 67.1 & 5.3 & $-$0.12 & 0.05 & 0.26 & -- & -- & -- & -- & -- & 66.5 & 5.7 & 1.75 & 10.12                                		&1&[ +61.4, +72.5]& 4(CO)    & 1665\\
 & 94.5 & 5.8 & $-$0.35 & 0.17 & 0.96 & -- & -- & -- & -- & -- & 96.3 & 6.9 & 8.04 & 60.09                                               			&1&[ +88.0,+105.0]& 4(CO,OH) & 1665\\
G26.609$-$0.212 & $-$33.2 & 2.5 & $-$0.07 & 0.42 & 0.77 & -- & -- & -- & -- & -- & -- & -- & -- & --                                       		&1&[ $-$37.6, $-$29.9]& 1        & 1   \\
G27.563$+$0.084 & 86.0 & 2.5 & $-$0.06 & 0.61 & 1.81 & -- & -- & -- & -- & -- & 85.8 & 3.2 & 3.30 & 29.90                                		&1&[ +78.0, +89.0]& 5(OH)    & 1665\\
G28.806$+$0.174 & 79.8 & 2.5 & $-$0.08 & 0.12 & 0.39 & -- & -- & -- & -- & -- & 80.6 & 3.1 & 2.26 & 10.72                                         &0&[ +76.4, +83.0]& 4(CO)    & 1665\\
& 103.5 & 3.8 & $-$0.07 & 0.12 & 0.38 & -- & -- & -- & -- & -- & 104.8 & 5.6 & 6.57 & 41.67                              			                &1&[ +98.0,+115.0]& 1665     & 4(CO)\\
G29.935$-$0.053 & -- & -- & -- & -- & <0.14 & 7.0 & 2.5 & $-$0.05 & 0.04 & 0.11 & -- & -- & -- & <2.34                  				&0&[  +2.0, +12.0]& 2(CO)    & 1667\\ 
& -- & -- & -- & -- & <0.15 & 50.6 & 2.5 & $-$0.08 & 0.06 & 0.14 & 49.9 & 2.2 & 3.39 & 8.32              			                			&0&[ +47.0, +53.0]& 5(CO)    & 1667\\
 & -- & -- & -- & -- & <0.14 & 68.0 & 2.5 & $-$0.09 & 0.07 & 0.19 & 68.1 & 3.1 & 1.15 & 4.20                             					&0&[ +61.0, +74.0]& 5(CO)    & 1667\\
 & 98.8 & 4.3 & $-$0.12 & 0.10 & 0.47 & 99.0 & 4.1 & $-$0.16 & 0.12 & 0.64 & 99.4 & 8.3 & 11.09 & 95.48                                    	&1&[ +90.0,+105.6]& 4(CO)    & 1667\\
G29.957$-$0.018 & 8.2 & 2.7 & $-$0.06 & 0.03 & 0.07 & 7.8 & 2.6 & $-$0.11 & 0.05 & 0.17 & 7.4 & 1.7 & 1.31 & 1.61                        &0&[  +2.0, +13.0]& 1667     & 1667\\
& -- & -- & -- & -- & --\tablefootmark{c} & 99.8 & 5.1 & $-$0.08 & 0.04 & 0.19 & 98.0 & 7.2 & 13.33 & 101.98             			        &1&[ +93.0,+107.0]& 4(CO)    & 4(CO)\\
G30.535$+$0.021 & -- & -- & -- & -- & --\tablefootmark{c} & 43.7 & 4.7 & $-$0.05 & 0.07\tablefootmark{a} & --\tablefootmark{b} & -- & -- & -- & --        &1&             --& 2(OH)    & 2(OH)\\
 & -- & -- & -- & -- & 0.28 & 49.4 & 3.5 & $-$0.04 & 0.06\tablefootmark{a} & 0.57 & 47.7 & 4.4 & 4.96 & 42.44                            		&1&[ +33.0, +54.0]& 2(OH)    & 2(OH)\\
 & 91.0 & 5.6 & $-$0.06 & 0.11 & 0.67 & 92.1 & 5.4 & $-$0.08 & 0.12 & 0.74 & 91.9 & 6.4 & 3.57 & 22.10                                     	&1&[ +86.9, +98.0]& 5(OH)    & 1667\\
G30.720$-$0.083\tablefootmark{*} & 93.1 & 5.8 & $-$0.25 & 0.80 & --\tablefootmark{c} & 94.1 & 5.8 & $-$0.34 & 1.29 & --\tablefootmark{c}&94.0&9.2&8.36&73.81&1&[ +87.0,+101.0]& 7        & 7   \\
G30.783$-$0.028  & -- & -- & -- & -- & -- & 77.9 & 3.4 & $-$0.13 & 0.04 & --\tablefootmark{b} & -- & -- & -- & --                                    &1 &             --& 3(OH)    & 3(OH)\\
& 81.6 & 6.3 & $-$0.14 & 0.02 & 0.20 & 82.2 & 2.5 & $-$0.44 & 0.11 & 0.39 & 81.9 & 5.6 & 2.30 & 12.18                                     	&1&[ +79.0, +85.6]& 4(CO)    & 1667\\
 & 87.1 & 2.7 & $-$0.06 & 0.02 & --\tablefootmark{b} & -- & -- & -- & -- & -- & -- & -- & -- & --                                        				&1&             --& 3(OH)    & 3(OH)\\
 & 92.0 & 3.5 & $-$0.25 & 0.07 & 0.39 & 92.2 & 3.5 & $-$0.32 & 0.09 & 0.55 & 93.3 & 8.7 & 7.41 & 70.82                                     	&1&[ +85.0,+103.0]& 4(CO)    & 1667\\
 & 98.3 & 3.4 & $-$0.07 & 0.01 & --\tablefootmark{b} & 98.8 & 4.2 & $-$0.10 & 0.03 & --\tablefootmark{b} & 104.9 & 5.6 & 1.78 & --\tablefootmark{b}        &1 &             --& 3(OH)    & 3(OH)\\
G30.854$+$0.151 & 95.5 & 2.9 & $-$0.14 & 0.52 & 1.36 & 95.3 & 3.2 & $-$0.16 & 0.63 & 2.01 & 95.5 & 4.1 & 6.72 & 27.98               &1&[ +91.0, +98.5]& 1667     & 1667\\
G31.242$-$0.110 & -- & -- & -- & -- & --\tablefootmark{c} & 19.6 & 2.6 & $-$0.12 & 0.26 & 0.83 & 20.8 & 3.8 & 3.33 & 12.45               	&1&[ +15.0, +23.0]& 1667     & 1667\\
 & 79.4 & 4.1 & $-$0.07 & 0.15 & 0.57 & 78.9 & 2.5 & $-$0.11 & 0.25 & 0.65 & 78.8 & 4.0 & 3.09 & 12.57                                     	&0&[ +75.0, +81.8]& 4        & 1667\\
 & -- & -- & -- & -- & <0.29 & 83.7 & 2.5 & $-$0.07 & 0.14 & 0.30 & 84.0 & 2.4 & 1.96 & 5.34                             					&0&[ +81.8, +87.1]& 4        & 1667\\
G31.388$-$0.383 & 18.4 & 2.5 & $-$0.08 & 0.06 & 0.17 & 17.9 & 2.5 & $-$0.11 & 0.09 & 0.20 & -- & -- & -- & <1.79           			&0&[ +15.0, +22.0]& 2(CO)    & 1667\\
G32.151$+$0.132 & 93.8 & 6.0 & $-$0.10 & 0.19 & 1.20 & -- & -- & -- & -- & -- & 94.4 & 4.5 & 7.24 & 33.72                                		&1&[ +88.0,+101.0]& 1665     & 1665\\
G32.272$-$0.226 & 22.6 & 4.6 & $-$0.06 & 0.19 & 0.91 & -- & -- & -- & -- & -- & 22.4 & 5.7 & 1.20 & 6.76                                 		&1&[ +18.0, +27.0]& 4(CO)    & 1665\\
G32.798$+$0.190 & 12.8 & 11.8 & $-$0.17 & 0.12 & 1.45 & -- & -- & -- & -- & -- & 15.1 & 7.4 & 6.61 & 51.69                               		&1&[  +0.0, +25.0]& 1665     & 1665\\
G32.928$+$0.607 & $-$33.9 & 2.9 & $-$0.05 & 0.21\tablefootmark{a} & 0.64 & -- & -- & -- & -- & -- & -- & -- & -- & --                           &1&[ $-$41.0, $-$28.0]& 1        & 1   \\
G33.915$+$0.110 & 95.3 & 3.4 & $-$0.04 & 0.06\tablefootmark{a} & 0.16 & -- & -- & -- & -- & -- & -- & -- & -- & <2.28                     	&1&[ +92.0, +96.5]& 2(OH)    & 2(OH)\\
 & 101.5 & 2.9 & $-$0.06 & 0.08\tablefootmark{a} & 0.28 & -- & -- & -- & -- & -- & -- & -- & -- & <2.28                                   			&1&[ +96.5,+104.0]& 2(OH)    & 2(OH)\\
 & 106.3 & 2.7 & $-$0.11 & 0.16 & 0.46 & -- & -- & -- & -- & -- & 107.5 & 4.6 & 6.80 & 32.47                                             			&1&[+103.4,+113.0]& 1665     & 1665\\
G34.132$+$0.471 & 33.8 & 3.1 & $-$0.08 & 0.17 & 0.59 & -- & -- & -- & -- & -- & 35.1 & 3.8 & 2.93 & 11.88                                		&1&[ +28.0, +39.5]& 1665     & 1665\\
G35.467$+$0.139 & 77.0 & 2.9 & $-$0.07 & 0.32 & 0.86 & -- & -- & -- & -- & -- & 77.4 & 3.2 & 8.73 & 29.92                                		&1&[ +74.0, +80.1]& 1665     & 1665\\
G37.764$-$0.215 & 63.4 & 4.3 & $-$0.32 & 0.28 & 1.64 & -- & -- & -- & -- & -- & 61.5 & 5.6 & 3.29 & 35.33                                		&1&[ +56.5, +69.5]& 4        & 1665\\
G37.874$-$0.399 & 61.0 & 10.1 & $-$0.31 & 0.19 & 1.91 & -- & -- & -- & -- & -- & 61.4 & 7.2 & 4.29 & 32.56                               		&1&[ +52.5, +67.5]& 5(OH)    & 1665\\
G38.876$+$0.308 & $-$16.4 & 2.9 & $-$0.07 & 0.38 & 1.03 & $-$16.1 & 4.1 & $-$0.08 & 0.48 & 1.79 & -- & -- & -- & --                       &1&[ $-$20.0, $-$13.0]& 1        & 1   \\
G39.565$-$0.040 & 23.4 & 2.8 & $-$0.10 & 0.24 & 0.66 & 23.1 & 2.5 & $-$0.15 & 0.37 & 0.96 & 23.4 & 3.1 & 1.25 & 4.65                  &0&[ +17.0, +28.0]& 1667     & 1667\\
G39.883$-$0.346 & 56.9 & 4.8 & $-$0.05 & 0.23 & 1.23 & 56.9 & 3.5 & $-$0.09 & 0.45 & 1.50 & 58.7 & 4.6 & 4.60 & 23.11                &1&[ +50.5, +63.5]& 1667     & 1667\\
G41.741$+$0.097 & 14.4 & 8.2 & $-$0.04 & 0.17 & 0.47 & 13.9 & 4.3 & $-$0.05 & 0.18 & 0.52 & 13.4 & 2.9 & 2.44 & 5.79                 &1&[ +10.0, +14.2]& 5(OH)    & 1667\\
G42.027$-$0.604 & -- & -- & -- & -- & <0.33 & 65.1 & 3.5 & $-$0.05 & 0.14 & 0.59 & 65.5 & 3.9 & 3.02 & 12.48             			&0&[ +62.0, +71.0]& 4(CO)    & 1667\\
G45.454$+$0.060  & -- & -- & -- & -- & -- & 56.0 & 7.2 & $-$0.17 & 0.07 & --\tablefootmark{b} & -- & -- & -- & --                                    &1&             --& 3(OH)    & 3(OH)\\ 
 & 59.3 & 4.1 & $-$0.19 & 0.07 & --\tablefootmark{c} & 59.5 & 2.5 & $-$0.38 & 0.18 & 0.83 & 58.9 & 5.9 & 8.02 & 45.42                      &1&[ +53.2, +62.3]& 4(CO)    & 1667\\    
 & -- & -- & -- & -- & -- & 64.9 & 8.2 & $-$0.09 & 0.03 & --\tablefootmark{c} & -- & -- & -- & --                                                         		&1&             --& 3(OH)    & 3(OH)\\    
G49.206$-$0.342 & 65.3 & 6.4 & $-$0.52 & 0.24 & 1.79 & -- & -- & -- & -- & -- & 65.7 & 6.0 & 4.31 & 28.54                                		&1&[ +55.0, +73.0]& 1665     & 1665\\
G49.369$-$0.302 & 50.9 & 4.4 & $-$0.75 & 0.24 & 1.26 & -- & -- & -- & -- & -- & 51.1 & 4.5 & 11.71 & 61.22                               		&1&[ +44.5, +57.5]& 1665     & 1665\\
 & 62.9 & 5.2 & $-$0.33 & 0.11 & 0.63 & -- & -- & -- & -- & -- & 61.1 & 2.3 & 4.11 & 24.18                                               			&1&[ +56.5, +69.5]& 4        & 1665\\
G49.459$-$0.353 & 62.0 & 7.4 & $-$0.36 & 0.10 & 0.68 & -- & -- & -- & -- & -- & 61.1 & 8.2 & 4.55 & 34.25                                		&1&[ +55.0, +65.0]& 4        & 1665\\
 & 68.6 & 3.7 & $-$0.81 & 0.22 & 1.01 & -- & -- & -- & -- & -- & 68.6 & 4.7 & 7.47 & 40.51                                               			&1&[ +65.0, +73.0]& 1665     & 1665\\
G49.488$-$0.380 & 65.9 & 2.5 & $-$1.89 & 0.18 & 0.48 & -- & -- & -- & -- & -- & 67.8 & 4.0 & 6.13 & 44.89                                		&1&[ +63.0, +72.5]& 4        & 4   \\
G52.753$+$0.334 & 13.6 & 2.9 & $-$0.02 & 0.07\tablefootmark{a} & <0.35 & 10.6 & 4.8 & $-$0.04 & 0.12 & 0.31 & 15.2 & 2.3 & 2.86 & 7.22      &1&[ +11.5, +18.5]& 7        & 7   \\
 & 45.3 & 2.5 & $-$0.01 & 0.04\tablefootmark{a} & <0.32 & 45.0 & 2.5 & $-$0.05 & 0.14 & 0.36 & 44.5 & 2.8 & 2.00 & 5.18                 &0&[ +43.0, +47.0]& 5(OH)    & 1667\\
G60.882$-$0.132 & 22.3 & 3.7 & $-$0.06 & 0.19 & 0.70 & 22.3 & 4.2 & $-$0.08 & 0.30 & 1.35 & 22.4 & 3.9 &  13.73 &  56.08             &1&[ +17.0, +28.0]& 1667     & 1667\\
G61.475$+$0.092 & -- & -- & -- & -- & <0.03 & 5.9 & 4.2 & $-$0.04 & 0.01\tablefootmark{a} & 0.05 & 6.6 & 0.6 & 1.18 & 0.24              &(0)&[  +2.5, +10.0]& 2(OH)    & 2(OH)\\ 
& 21.2 & 2.5 & $-$1.38 & 0.40 & 1.22 & 21.1 & 2.5 & $-$1.77 & 0.51 & 1.66 & 21.6 & 3.6 & 15.02 & 56.84                   			&1&[ +17.5, +25.0]& 1667     & 1667\\    

\hline
\label{tbl:lineparameters}
\end{tabular}
\tablefoot{Gaussian profiles are fitted to the OH 1665\,MHz and OH 1667\,MHz absorption spectra to determine the central velocity (v), the full width of half maximum ($\Delta$v) and the minimum line depth ($F_{\rm peak}$). Similarly, the line parameters of the \ce{^{13}CO} emission are determined. The peak optical depth ($\tau_{\rm peak}$) is determined by fitting Gaussians to the optical depth spectra. The velocity integrated optical depth ($\int \tau {\rm dv}$) and \ce{^{13}CO} emission is determined by summing all channels over a common range of \ce{OH} absorption and \ce{^{13}CO} emission (Column ``Int. Range''). For lines that were not detected, 3-$\sigma$ upper limits (``<'') are given under the assumption of an average line width of 3.6\us\kms\ for the OH 1665\,MHz transition and 4.0\us\kms\ for the \ce{^{13}CO} transition. Notes on each detection indicate if the OH transition at 1665\,MHz or at 1667\,MHz (``1665''/``1667'') was used for the linewidth comparison (``FWHM''; Fig.~\ref{fig:plot_fwhm_oh_vs_co}) or the OH abundance (``Int.''; Figs.~\ref{fig:plot_oh_vs_h2}, \ref{fig:plot_ohbynh2_vs_nh2}, \ref{fig:plot_ohbynh_vs_nh} and \ref{fig:plot_ohbynh2_vs_nh2_wiesemeyer}). Numbers specify why a transition was not used: (1) No \ce{^{13}CO} data available; (2) Weak or non-detection; (3) Auxiliary fit component; (4) Blended components could not be separated; (5) Bad fit; (6) No velocity overlap between \ce{OH} and \ce{^{13}CO}; (7) OH spectrum affected by emission of a maser at this position or close-by.\\
{\bf Footnotes:} 
$^{(a)}$ 3-$\sigma$ detection;
$^{(b)}$ Included in integration of an adjacent component;
$^{(c)}$ This transition is affected by emission of a maser at this position or close-by;
$^{(*)}$ Data at 20\arcsec\ resolution was used in order not to blend with an adjacent maser.
}
\end{table*}

\subsection{OH abundance} \label{sec:ohabundance}
\subsubsection{Line integrals} \label{sec:line_integrals}
The \ce{OH} column density is derived from the integrated optical depth in the main line 
transitions under the assumption that all molecules are in the four sublevels of the ground 
state arising from the $\Lambda$~doubling and hyperfine structure \citep[e.g.,][]{Elitzur:1992aa}. 
The derived column densities are listed in Table~\ref{tbl:integrated_properties}.

\smallskip
\noindent
Optical depths are computed from the line-to-continuum ratio. Contributions from large-scale emission are assumed to be filtered out by the interferometer. The emission term in the radiative transfer equation which includes the excitation temperature is therefore negligible ($T_{\rm line}=(T_{\rm ex}-T_{\rm cont})(1-e^{-\tau})$; where $T_{\rm line}$ and $T_{\rm cont}$ can be derived in Raleigh-Jeans approximation from the continuum-subtracted line flux and the continuum flux, $F_{\rm line}$ and $F_{\rm cont}$). Even if this assumption did not hold true, at excitation temperatures of about ${T_{\rm ex} = 5-10{\rm{\us}K }}$, the approximation ${T_{\rm cont}\gg T_{\rm ex}}$ underestimates the 
optical depth for a 5-$\sigma$ detection by less than $\ll$10\% at 20\arcsec\ resolution and for sources with $F_{\rm cont}>0.5 $\us\jyb\ at 46\arcsec\ resolution. For weaker sources  which show detectable extended continuum emission, the underestimation would be between 6-16\% after smoothing to 46\arcsec.

\smallskip
\noindent
The integrated optical depth is determined by
summation over all spectral bins of the absorption feature. The ${\rm{}^{13}CO}$(1-0) emission is
integrated over the same velocity range. If a corresponding ${\rm{}^{13}CO}$(1-0) feature exists that is broader than the OH feature,
the velocity range is chosen to enclose the \ce{^{13}CO}(1-0) feature (see Fig.~\ref{fig:plot_positions_single}). 
For lines that have no ${\rm{}^{13}CO}$(1-0) detection counterpart, 3-$\sigma$ upper limits
are given for integrated emission and all derived quantities, under the assumption of an average line width of 4.0\us\kms\ of the detected \ce{^{13}CO} emission (Table~\ref{tbl:integrated_properties}). 

\smallskip
\noindent
Similarly, we derive the optical depth of \ion{H}{i} from the line-to-continuum ratio, and give lower limits
in the case of saturated absorption. A saturated channel is defined here as observed flux that is within 3-$\sigma$ of the zero level. 
This value is then used to calculate the lower limit (see cyan circles in the lowermost panel of Fig.~\ref{fig:plot_positions_single}). 
Analogously to the discussion above on $\tau_{\rm OH}$, emission in principle also affects the \ion{H}{i} optical depth, but is likely to be filtered out here. For this reason, 
we do not attempt to correct for it, but note that the integrated optical depth here always represents a lower limit.

\subsubsection{OH column density}
The \ce{OH} column density is inferred for each main line
separately, under the assumption that all \ce{OH} molecules are in the ground
state, ${}^2\Pi_{3/2}(J=3/2)$. The OH column density is given by \citep[e.g.,][]{StanimirovicWeisberg:2003aa}
\begin{equation}
\frac{N_{\rm OH}}{T_{\rm ex}} = \frac{C_0}{f}\int{\tau \, dv}, 
\end{equation}
where $N_{\rm OH}$ is the total OH column density in ${\rm cm}^{-2}$, $T_{\rm ex}$ the
excitation temperature in Kelvin, $C_0 = 4.0\times10^{14} \us{\rm cm}^{-2} {\rm
 {\us}K}^{-1}{\rm km^{-1}s}$ for the 1665{\us}MHz transition and $C_0 =
2.24\times10^{14} \us{\rm cm}^{-2} {\rm{\us}K}^{-1}{\rm km^{-1}s}$ for the
1667{\us}MHz transition \citep[e.g.,][calculated using Einstein coefficients from
\citealt{Turner:1966aa}]{Goss:1968aa,TurnerHeiles:1971aa,StanimirovicWeisberg:2003aa}. The filling factor $f$ 
describes the solid angle fraction of the continuum source that is covered by the OH cloud. We assume that $f=1$. The 1667{\us}MHz transition is expected to be 
detected at higher signal-to-noise than the 1665{\us}MHz transition because of its larger statistical weight, and therefore 
the 1667{\us}MHz transition is used for further analysis whenever available. 
The calculated $\frac{N_{\rm OH}}{T_{\rm ex}}$ ratios are given in Table~\ref{tbl:integrated_properties}.

\smallskip
\noindent
The excitation temperatures of the OH transitions cannot be derived independently from their optical depth, 
as local thermal equilibrium between the levels of the two main lines is not necessarily given, and thus the
excitation temperatures of both main lines may be different \citep[e.g.,][]{Crutcher:1979aa,DawsonWalsh:2014aa}. 
A determination of both $T_{\rm ex}$ and $\tau$ is in principle possible if additional emission observations 
were obtained at a position slightly offset from the continuum source. 
The \ce{OH} emission is not detectable in the present dataset, as the \ce{OH} emission is expected to be dominated 
by warm gas \citep[e.g.,][]{WannierAndersson:1993aa} that varies on scales larger than the interferometric observations 
presented here. Hence, emission is filtered out. Observations of transitions to higher rotational levels of OH are not available 
for most sources and would require detailed modelling to constrain the excitation conditions of 
the hyperfine transitions in the ground state, which is beyond the scope of this paper. 
We therefore assume an excitation temperature from the literature in order to determine 
\ce{OH} column densities \citep[e.g.,][]{StanimirovicWeisberg:2003aa}. 

\smallskip
\noindent
The excitation temperatures of the OH main lines have been found to differ relative to each other by
$0.5-2.0${\us}K \citep[e.g.,][]{Goss:1968aa,Crutcher:1979aa, DawsonWalsh:2014aa}. This has also been seen in models
\citep[e.g.,][]{GuibertRieu:1978aa}. The level populations may also be affected by 
radiative pumping \citep{CsengeriMenten:2012aa,WiesemeyerGusten:2016aa}.
Previous investigations reported excitation temperatures for the OH main line transitions 
between 3 and 10{\us}K \citep[e.g.,][]{Goss:1968aa,Turner:1973aa,Crutcher:1977aa,ColganSalpeter:1989aa,LiGoldsmith:2003aa,BourkeMyers:2001aa,Yusef-ZadehWardle:2003aa}. We therefore assume a uniform excitation temperature of $T_{\rm ex}(1665) = 5${\us}K for
the 1665\,MHz transition, which in principle can be higher by up to about a factor of two for OH gas associated 
with \ion{H}{ii} regions. We consider this in the systematic uncertainties (see Sect.~\ref{sec:systematics}).
For transitions in which both main lines are detected, the medians of the column density distributions of each main line are offset. 
A better agreement between the samples is reached for a slightly higher excitation temperature for the 1667\,MHz line of $T_{\rm ex}(1667)=6.7\,{\rm K}$ . 
We therefore adopt this value for the OH 1667 MHz transition in the following, while using $T_{\rm ex}(1665) = 5${\us}K in the OH 1665 MHz transition.

\subsubsection{\ce{H2} column density}
\begin{figure}
 \centering
 \includegraphics[width=0.99\columnwidth]{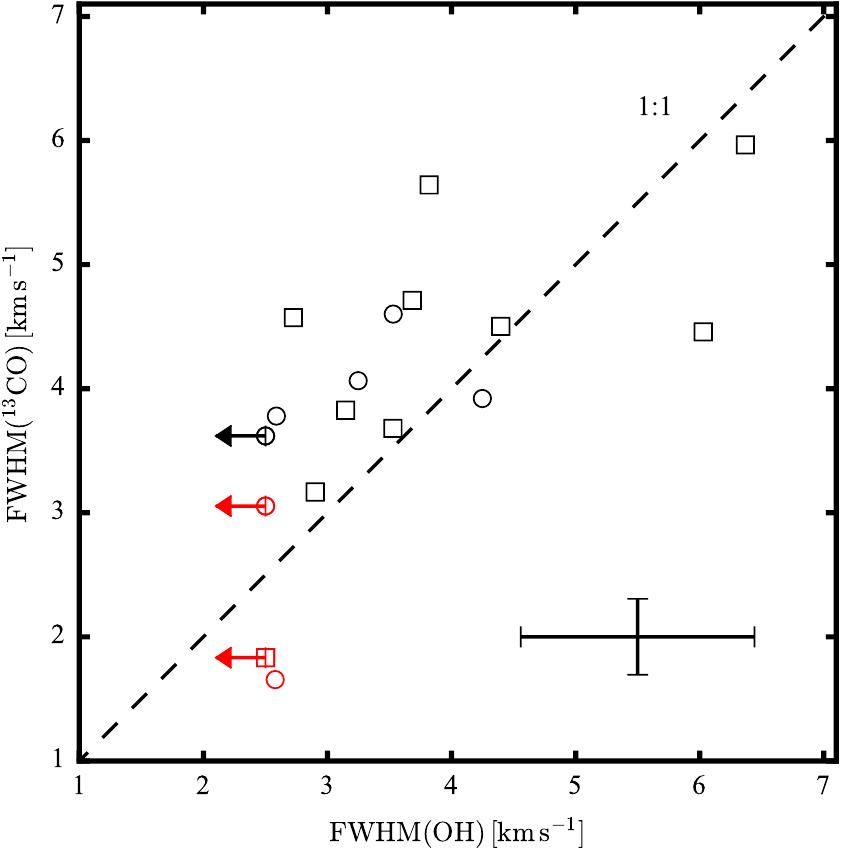} 
 \caption{Comparison of FWHM of \ce{OH} and \ce{^{13}CO}(1-0) lines. OH line widths of the 1665{\us}MHz transition (squares) are used whenever the 1667{\us}MHz transition (circles) was not available. Detections associated with \ion{H}{ii} regions are drawn in black, others in red. Arrows indicate spectrally unresolved lines. The dashed line indicates a 1:1 correlation. The bars in the lower right-hand corner of the figure indicate typical errors in both quantities.}
 \label{fig:plot_fwhm_oh_vs_co}
\end{figure}

As proxy for \ce{H2}, we use \ce{^{13}CO} emission. Kinematically, \ce{^{13}CO} is related to the OH gas. The line widths of the OH main lines and \ce{^{13}CO} are compared in Fig.~\ref{fig:plot_fwhm_oh_vs_co}. The channel spacing of 1.5\us\kms\ in the OH observations poses a limit of 2.5\us\kms\ on the narrowest resolved line width for OH lines, while the full spectral resolution of 0.21\us\kms\ is used for \ce{^{13}CO} to disentangle different velocity components. Excluding the unresolved OH lines, the line widths of both tracers are found to be correlated\footnote{This sample of 15 datapoints is described by a Pearson's correlation coefficient $\rho = 0.8$ at a statistical significance of $\approx3\sigma$.
}. In some cases, the \ce{^{13}CO} emission features larger line widths than OH. One possible explanation is that larger parts of the molecular cloud contribute to the \ce{^{13}CO} antenna temperature than to the OH absorption. For many continuum sources, only OH absorption from scales less than 46'' is recovered. 
As emission from within the entire beam contributes to the \ce{^{13}CO} antenna temperature, the \ce{^{13}CO} emission may average over larger parts of the cloud. Similarly, if the continuum source is located within the molecular cloud, \ce{^{13}CO} emission contains information from all cloud depths (if not optically thick), while only parts of the cloud between the continuum source and the observer affect the OH absorption (see also sect.~\ref{sec:systematics}). Depending on the velocity substructure of the cloud on scales smaller than the beam and along the line of sight, this can lead in both cases to larger line widths in the \ce{^{13}CO} emission than in the OH absorption. 

\smallskip
\noindent
The column density of \ce{{}^{13}CO} is determined from the integrated line profile ($\int{T_{\rm mb} {\rm dv}}$) under 
the assumption that the gas is optically thin \citep[e.g.,][eq. 15.37]{WilsonRohlfs:2009aa} : 
\begin{equation}
N({}^{13}{\rm CO}) \approx 3.0\times10^{14} \frac{\int{\frac{T_{\rm mb}}{\rm K} {\rm dv}}}{1-\exp{(-5.3/T_{\rm ex})}} {\rm cm}^{-2}.
\label{eq:coldens_13co}
\end{equation}
Average excitation temperatures of ${\rm{}^{13}CO}$ in molecular clouds are typically between 10$-$15\,K, with values of up to 25\,K in some cases \citep[e.g.,][]{PinedaGoldsmith:2010aa,NishimuraTokuda:2015aa,FrerkingLanger:1982aa,AndersonBania:2009aa}. We select an excitation temperature towards the upper end of this range of $T_{\rm ex} = 20\,{\rm K}$, as most of the OH detections are associated with \ion{H}{ii} regions. To account for possibly lower excitation temperatures, we assume an uncertainty of a factor of two on $T_{\rm ex}$, which results in an uncertainty of approximately a factor of two for $N_{\rm {}^{13}CO}$. 

\smallskip
\noindent
The column density of
molecular gas is determined by assuming a constant \ce{^{13}CO} abundance relative to molecular hydrogen molecules of 
$N_{\rm H_2}/N_{\rm {}^{13}CO} = 3.8\times10^{5}$ \citep{BolattoWolfire:2013aa,PinedaCaselli:2008aa}. 
Uncertainties in this estimate due to optical thickness of ${\rm ^{13}CO }$ or local variations in the ${\rm^{13}CO/^{12}CO}$ ratio 
\citep{SzucsGlover:2016aa} are discussed in Sect.~\ref{sec:systematics}. If the ${}^{13}{\rm CO}(1-0)$ emission is not detected, we report upper limits for both $N_{\rm H_2}$ and 
$N_{\rm H}$ and lower limits on $X_{\rm OH}$.

\subsubsection{\ion{H}{i} column density} \label{sec:hicoldense}
To derive the \ion{H}{i} column density, we assume a spin~temperature~($=
T_{\rm ex}$) of $T_{\rm spin} \sim 100 {\rm{\us}K}$ \citep{BihrBeuther:2015aa}. The \ion{H}{i} column density is given by \citep[e.g.,][]{WilsonRohlfs:2009aa}
\begin{equation}
N_\ion{H}{i} = 1.8224\times10^{18}\frac{T_{\rm spin}}{\rm K} \int{\tau({\rm v})\left(\frac{\rm dv}{\rm km\,s^{-1}}\right)}\us{\rm cm^{-2}}.
\end{equation}
Since the optical depth is a lower limit here (see sect.~\ref{sec:line_integrals}), the \ion{H}{i} column density is a lower limit as well. As $T_{\rm spin}$ may vary for individual regions by a factor of two, these limits are subject to a systematic uncertainty of the same factor. 

\smallskip
\noindent
In the following, we determine the OH abundance, $X_{\rm OH}$, both in terms of the column density of molecular hydrogen, $N_{\rm H_2}$, and in terms of the total column density of hydrogen nuclei, which includes both atomic and molecular hydrogen, $N_{\rm \ion{H}{i}}$ and $N_{\rm H_2}$. $N_{\rm H}$ is given by $N_{\rm H} = N_{\rm \ion{H}{i}} + 2N_{\rm H_2}$. The lower limits on $N_\ion{H}{i}$ yield upper limits on $X_{\rm OH}$. The derived quantities from this section are listed in Table~\ref{tbl:integrated_properties}. 

\subsubsection{Systematic uncertainties} \label{sec:systematics}
The systematic uncertainties of $X_{\rm OH}$, $N_{\rm H_2}$ and $N_{\rm H}$ are estimated as follows. 
The $N_{\rm H_2}/N_{\rm {}^{13}CO}$ ratio varies within molecular clouds and may be affected by global changes in the \ce{^{13}C} isotope abundance with Galactocentric radius. First, the scatter of $N_{\rm H_2}/N_{\rm {}^{13}CO}$ measurements has been found to be up to a factor of two within individual clouds \citep[e.g., in Perseus; ][]{PinedaCaselli:2008aa}. We assume that this inflicts an uncertainty of a factor of two on $N_{\rm H_2}$. Second, the conversion used here has been determined in nearby Gould Belt clouds, but the \ce{^{12}C}/\ce{^{13}C} isotope ratio increases with Galactocentric radius \citep{MilamSavage:2005aa}. With absorption sources located at Galactocentric radii between $R_{\rm gc} =3-10\us{\rm kpc}$, we assume that global variations in the \ce{^{13}C} isotope abundance introduce an additional uncertainty of a factor of two on $N_{\rm H_2}$. 

\smallskip
\noindent
As described above, the assumptions regarding the excitation
temperatures of \ce{OH}, \ce{^{13}CO}(1-0) and \ion{H}{i} are likely to be valid
within a factor of two as well. These result in a combined uncertainty of approximately 
a factor of 3.5 in $N_{\rm H_2}$ and $N_{\rm H}$, and a factor of four
in $X_{\rm OH}$. As the \ion{H}{i} absorption saturates in most cases, we can
only determine lower limits to $N_{\ion{H}{i}}$. 

\smallskip
\noindent
Also, some of the ${}^{13}{\rm CO}(1-0)$ emission may be optically thick and give lower limits on $N_{\rm H_2}$, and therefore upper
limits on $X_{\rm OH}$. While $N_{\rm H_2}$ as derived from \ce{^{13}CO} 
may also be underestimated due to chemical effects \citep[e.g.,][]{SzucsGlover:2016aa}, 
the derived \ce{H2} column densities have been compared with column densities derived from 870\,$\mu$m dust
emission (ATLASGAL), and we find reasonable agreement also towards high column
densities within a factor of two. 

\smallskip
\noindent
At low \ce{CO} column densities, molecular cloud regions may be traced that contain a significant fraction of ``CO-dark'' \ce{H2}. This means that the column density of \ce{H2} would be underestimated and the OH abundance overestimated. The amount of ``CO-dark'' gas depends on the environment, i.e., on metallicity and the strength of the external radiation field. This makes a quantitative correction difficult, but the effect may influence detections for which \ce{H2} and \ion{H}{i} column densities are comparable.

\smallskip
\noindent
Geometrical uncertainties should also be considered. For the sake of simplicity, they are mentioned here only briefly, because they are difficult to quantify. First, the \ce{^{13}CO} emission may trace molecular gas that is not accessible by the OH absorption observations. OH absorption occurs in material between the observer and a continuum source, while material at any distance along the sightline contributes to the \ce{^{13}CO} emission if optically thin. Most of the OH absorption detections are associated in velocity with the continuum source itself and by our definition separated from the continuum source by less than $10$\us\kms. Assuming that the \ce{OH} and \ce{^{13}CO} gas, as well as the \ion{H}{ii}-region are part of the same molecular cloud, a fraction of the \ce{^{13}CO} emission emerges from behind the continuum source as seen by the observer. This fraction depends on the structure of the molecular cloud and the relative position of the continuum source. For example, 50\% of the \ce{^{13}CO} emission will come from behind the \ion{H}{ii} region, if it is embedded in the middle of a spherically symmetric molecular cloud. No \ce{OH} absorption can be measured for this part of the cloud, and the OH abundance will be underestimated. Velocity shifts between the different tracers in our data (<2\us\kms) indicate that at least some sources are affected by this. As we cannot constrain the structure of the cloud and  quantify this effect, we choose not to correct for it here. 

\smallskip
\noindent
Second, crowded regions may contain multiple, overlapping \ion{H}{ii}-regions, which contribute to the observed continuum flux, if the continuum emission is optically thin. If OH absorption occurs along the line of sight in between such continuum sources, the observed continuum will overestimate the true continuum incident on the absorbing cloud. Therefore, the optical depth of OH would be underestimated. This is likely to affect lines of sight towards galactic continuum sources that are located in crowded regions, such as the tangent points of spiral arms (G29, W43 and W51). 

\subsubsection{OH vs. \ce{H2} column density} \label{sec:oh_vs_h2_coldens}
\begin{figure}
 \centering
 \includegraphics[width=0.99\columnwidth]{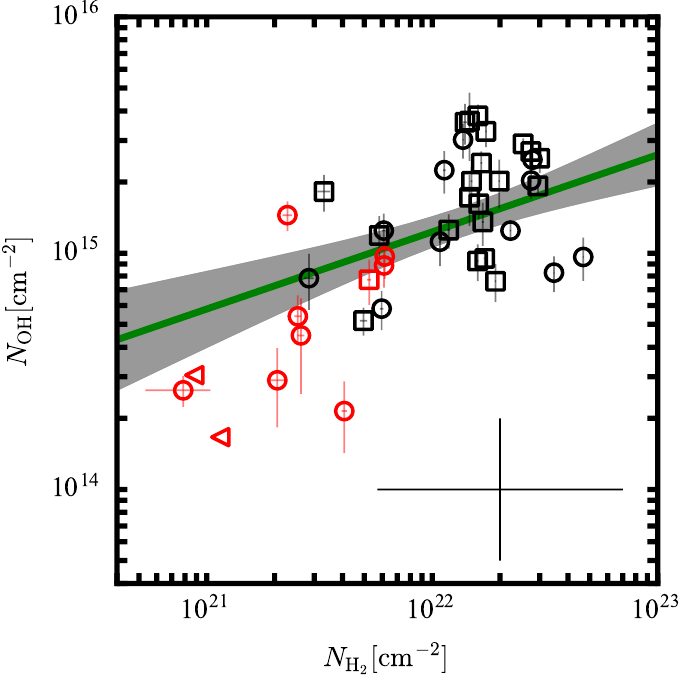} 
 \caption{
  Comparison of the \ce{OH} column density
  from the 1665{\us}MHz (squares) and 1667{\us}MHz transitions (circles)
  to that of \ce{H_2} as inferred from \mbox{\ce{^{13}CO}(1-0)}
  emission. 
  Column densities from absorption features overlapping 
  in velocity with \ion{H}{ii} regions (black) and with no such
  counterpart (red) divide the plot into regions 
  with higher and lower hydrogen column densities. 
  Triangles denote upper limits on $N_{\rm H_2}$ as determined from \ce{^{13}CO}(1-0) non-detections. 
  The green line is the result of parameter estimation of a correlation between $N_{\rm OH}$ and $N_{\rm H_2}$, 
  using errors on x and y axes and not including upper limits. The gray shaded regions shows the 16\%- and 84\%-percentiles.
  The black error bars in the lower right corner show the systematic errors. 
}
 \label{fig:plot_oh_vs_h2}
\end{figure}

OH and \ce{H2} column densities are shown in Fig.~\ref{fig:plot_oh_vs_h2}. $N_{\rm OH}$ is derived from the 1667{\us}MHz transition
(circles) if available, and from the 1665{\us}MHz transitions in the rest of
the cases (squares). The absorption features are separated by color into
regions that are associated (black) or not associated (red) with \ion{H}{ii}~regions. 

\smallskip
\noindent
To investigate the correlation between $N_{\rm OH}$ and $N_{\rm H_2}$, we perform a linear regression in log-space, \mbox{$\log(N_{\rm OH})  = m\times\log(N_{\rm H_2})+t$}, to determine the slope $m$. The uncertainties are dominated by the systematic errors, i.e. possible variations of $N_{\rm OH}$ and $N_{\rm H_2}$ by a factor of 2 and 3.5, respectively (see sect.~\ref{sec:systematics}). To properly take into account their impact on the correlation, we estimate the distribution of slopes $m$ given these uncertainties and the stochastic measurement errors. We do not include upper limits. We sample the {\it posterior} distribution of $m$ by performing linear regressions on multiple, artificial representations of the data, which are inferred from the uncertainty distributions of the measurements. 

\smallskip
\noindent
We create a large number of artificial datasets ($n_{\rm datasets} = 10^{5}$). Each artificial dataset has the same number of points as our measured sample, but instead of containing the measurements itself, each point is replaced by randomly drawing an artificial datapoint from the uncertainty distribution. We also ``bootstrap'' each of these simulated datasets, i.e. randomnly assigning weights to the points to reduce the importance of each individual measurement. From the linear regression on each of these artificial datasets, we obtain a distribution of slopes with median and \mbox{16\%-,} 84\%-percentiles of $m = 0.33^{+0.14}_{-0.13}$ (see also Fig.~\ref{fig:marginalized_distributions}). The green line in Fig.~\ref{fig:plot_oh_vs_h2} shows the median slope.

\smallskip
\noindent
We interpret this result as an indication of a weak, sublinear correlation between $N_{\rm OH}$ and $N_{\rm H_2}$. A direct proportionality between the two parameters is unlikely, given the distribution of slopes. The sublinear relation between $N_{\rm OH}$ and $N_{\rm H_2}$ yields a decreasing OH abundance ($X_{\rm OH} = N_{\rm OH}/N_{\rm H_2}$), as discussed in the following sections. The analysis shows that a correlation is present in the data, but for tighter constraints, follow-up studies are needed to provide more data and/or to decrease the systematic uncertainties.

\subsubsection{OH abundance at different hydrogen column densities}
\begin{figure}
 \centering
 \includegraphics[width=0.99\columnwidth]{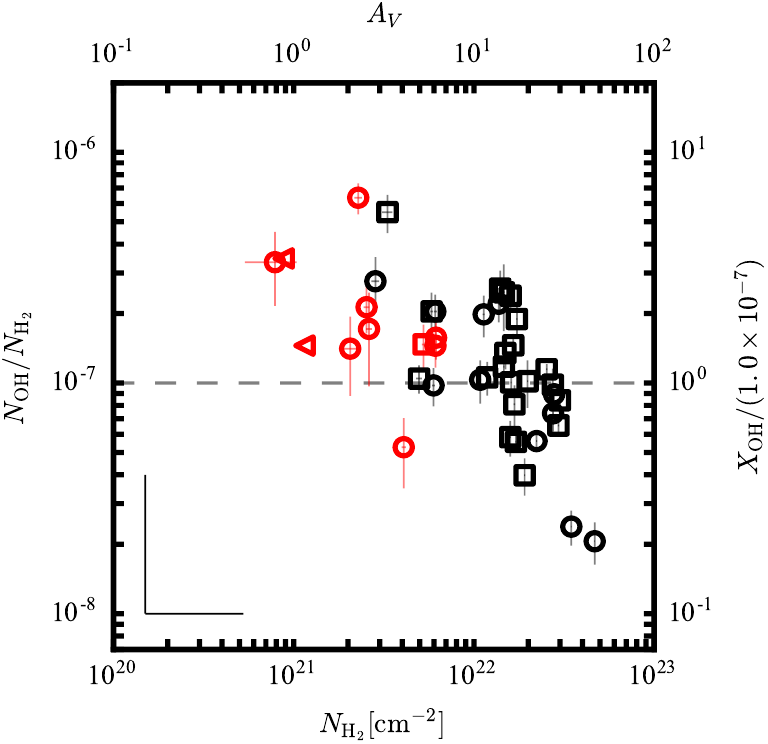} 
 \caption{OH abundance $X_{\rm OH} = N_{\rm OH}/N_{\rm H_2}$ vs. $N_{\rm H_2}$.  $N_{\rm OH}$ is inferred from 1665{\us}MHz (squares) or 1667{\us}MHz absorption (circles) and  $N_{\rm H_2}$ from \ce{^{13}CO}(1-0) emission. Absorption features associated with \ion{H}{ii} regions are shown in black; those not associated are shown in red. Triangles denote upper limits on $N_{\rm H_2}$ (inferred from the non-detection of ${\rm {}^{13}CO}$). 
  The right axis shows the OH abundance in units of literature (molecular) OH
  abundance of $1\times10^{-7}$ \citep[e.g.,][indicated also by the dashed
  gray line]{LisztLucas:1999aa, LisztLucas:2002aa}.
  The black error bars in the lower left corner show the systematic errors
  (only the upper halfs of the error bars are shown here).
}
 \label{fig:plot_ohbynh2_vs_nh2}
\end{figure}

Fig.~\ref{fig:plot_ohbynh2_vs_nh2} shows the OH abundance in terms of the
molecular hydrogen column density ($X_{\rm OH} = N_{\rm OH}/N_{\rm H_2}$)
versus $N_{\rm H_2}$. The literature value for $X_{\rm OH} = N_{\rm OH}/N_{\rm H_2} \approx 1\times10^{-7}$ is
plotted as a dashed gray line, and the right axis shows the data in terms of
this value \citep[e.g.,][]{Guelin:1985aa,LangerGraedel:1989aa,van-Langeveldevan-Dishoeck:1995aa,LisztLucas:1999aa,
LisztLucas:2002aa}. The OH abundance is found to be anti-correlated with $N_{\rm H_2}$ over the
range of probed cloud depths ($8\times10^{20}\us{\rm cm^{-2}}<N_{\rm H_2}
<5.8\times10^{22}\us{\rm cm^{-2}}$). Above $N_{\rm H_2}>1.9\times10^{22}\us{\rm cm^{-2}}$
($A_V \sim 20\,{\rm mag}$)\footnote{Assuming $N_{\rm H}/(A_V/R_V) =
  5.8\times10^{21}\us{\rm cm^{-2}\,mag^{-1}}$ \citep{BohlinSavage:1978aa} and
  the average ISM value for the total-to-selective extinction of $R_V = 3.1$,
  the total hydrogen column density can be related to visual extinction as
  $N_{\rm H} = 1.9\times10^{21}\us{\rm cm^{-2}\,mag^{-1}}\times A_V$. For
  large column densities ($N_{\rm H_2}\gg 5\times10^{21}{\rm cm}^{-2}$) we
  assume the contribution of atomic hydrogen to be negligible, i.e.,  $N_{\rm H}\approx 2\times N_{\rm H_2}$.},
most of the abundances are lower than the literature value, while the
abundances at lower \ce{H2} column densities are slightly higher.

\begin{figure}
 \centering
 \includegraphics[width=0.99\columnwidth]{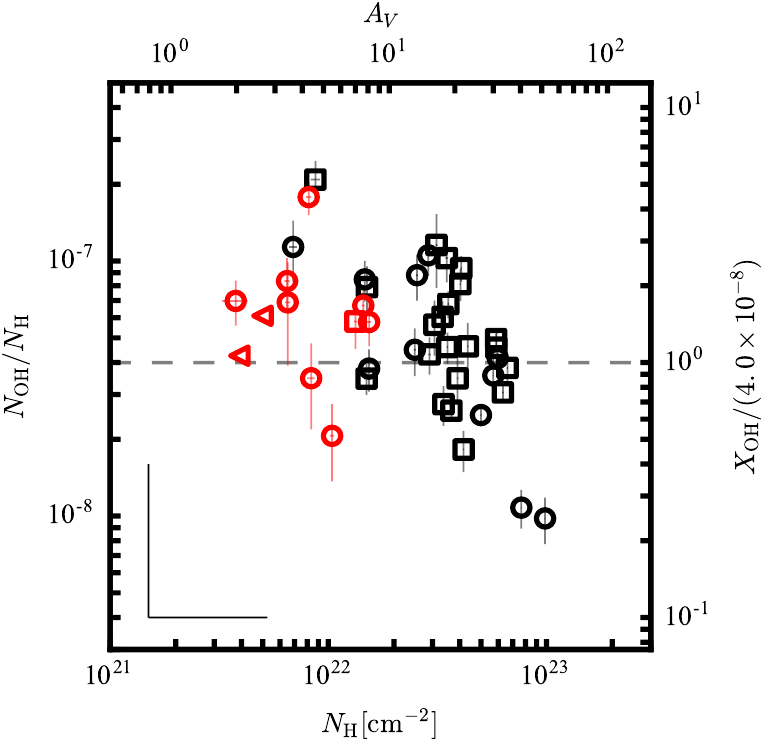} 
 \caption{OH abundance $X_{\rm OH} = N_{\rm OH}/N_{\rm H}$ vs. $N_{\rm H}$. The total column density of hydrogen nuclei, $N_{\rm H}$, is inferred from \ce{^{13}CO}(1-0) emission and \ion{H}{i} absorption. As the \ion{H}{i} column density represents lower limits, the OH abundances represent upper limits. Colors and symbols are as in
  Fig.~\ref{fig:plot_ohbynh2_vs_nh2}. The right axis shows the OH abundance in units of the typical OH abundance in diffuse clouds 
  of $4\times10^{-8}$ \citep[][indicated also by the dashed gray line]{Crutcher:1979aa}.}
 \label{fig:plot_ohbynh_vs_nh}
\end{figure}

\smallskip
\noindent
We use the \ion{H}{i} absorption data as lower limit for the column density of atomic hydrogen,
and show the OH abundance with respect to the total number of hydrogen nuclei 
in Fig.~\ref{fig:plot_ohbynh_vs_nh}. While the overall trend is similar to that seen in
Fig.~\ref{fig:plot_ohbynh2_vs_nh2}, the atomic hydrogen content probed along
the line-of-sight is comparable to molecular hydrogen for a few detected
components with low molecular hydrogen column densities
(e.g., for G29.935$-$0.053 at 7.0\us\kms\ and G29.957$-$0.018 at 8.0\us\kms).

\smallskip
\noindent
We see a clear anti-correlation between $X_{\rm OH}$ and $N_{\rm H_2}$ for OH associated with \ion{H}{ii} regions (black).
Measurements which are not associated (red), follow this trend in Fig.~\ref{fig:plot_ohbynh2_vs_nh2}. In the abundance with respect to all hydrogen nuclei ($N_{\rm OH}/N_{\rm H}$) in Fig.~\ref{fig:plot_ohbynh_vs_nh}, this trend appears not to be present in the red data points, since the lowest data points have significant contribution from atomic hydrogen. 
As the \ion{H}{i} column densities are lower limits, the abundances are upper limits, favoring an even shallower trend of $X_{\rm OH}$ in this $N_{\rm H}$ column density regime.
As mentioned in Sect.~\ref{sec:systematics}, also the fraction of ``CO-dark'' \ce{H2} may be significant 
for most of the detections which are not associated with \ion{H}{ii} regions and the 
lowest of the measurements associated with \ion{H}{ii} regions, which would place them at higher 
$N_{\rm H_2}$ (and $N_{\rm H}$) and lower $X_{\rm OH}$ in Figs.~\ref{fig:plot_ohbynh2_vs_nh2}~and~\ref{fig:plot_ohbynh_vs_nh}. 

\smallskip
\noindent
The red data points fall into similar ranges of visual extinction as probed by many earlier studies. 
The OH abundance for visual extinctions of $A_V<7\us{\rm mag}$ has been
reported in the literature to be constant at $X_{\rm OH} = N_{\rm OH}/N_{\rm
  H} = 4\times10^{-8}$ \citep[e.g.,][]{Goss:1968aa,Crutcher:1979aa}. The median OH abundance for this group of points here is
$X_{\rm OH}(N_{\rm H_2}) =1.6\times10^{-7}$ with a scatter of $\Delta X_{\rm OH} =
0.27\,{\rm dex}$ and $X_{\rm OH}(N_{\rm H}) = 6.0\times10^{-8}$ and a scatter of $\Delta X_{\rm OH} =
0.22\,{\rm dex}$.

\begin{table*}
\caption{Derived quantities from \ce{OH} 1665/1667{\us}MHz and \ion{H}{i} absorption and \ce{^{13}CO}(1-0) emission.}
\tiny
\renewcommand{\arraystretch}{0.9}
\centering
\begin{tabular}{l|HrHrHr|rrr|rrrr}
\hline \hline
Name 
& v
& $\varv_{\rm OH}$
& v
& ${N_{\rm OH\,1665}}/{T_{\rm ex}}$ 
& v
& ${N_{\rm OH\,1667}}/{T_{\rm ex}}$ 
& $\varv_{\rm {}^{13}CO}$
& $N_{\rm {}^{13}CO}$ 
& $N_{\rm H_2}$
& $N_{\rm \ion{H}{i}}$
& Notes
& $X_{\rm OH}(N_{\rm H})$ 
& $X_{\rm OH}(N_{\rm H_2})$ 
\\
& $\left[\frac{\rm km}{\rm s}\right]$
& $\left[\frac{\rm km}{\rm s}\right]$
& $\left[\frac{\rm km}{\rm s}\right]$
& $\left[\times\frac{10^{14}}{\rm cm^{2}K}\right]$
& $\left[\frac{\rm km}{\rm s}\right]$
& $\left[\times\frac{10^{14}}{\rm cm^{2}K}\right]$
& $\left[\frac{\rm km}{\rm s}\right]$
& $\left[\times\frac{10^{16}}{\rm cm^{2}}\right]$
& $\left[\times\frac{10^{21}}{\rm cm^{2}}\right]$
& $\left[\times\frac{10^{21}}{\rm cm^{2}}\right]$
&
& [$\times10^{-7}$]
& [$\times10^{-7}$]
\\
\hline
G14.490$+$0.021 & 23.2   &  23.1 & 22.9 & <5.4 $\pm$  2.4\tablefootmark{a} & -- & -- & -- & -- & -- & >1.3		       & 2(OH)   &   -- &   -- \\
G14.996$-$0.738 & 22.0   &  21.8 & 21.9 & 1.5 $\pm$  0.4 & -- & -- & -- & -- & --      	       	    & >6.3 		       & 1       &   -- &   -- \\
G15.033$-$0.679 & 11.8   &  12.9 & 12.5 & 0.8 $\pm$  0.2 & -- & -- & -- & -- & -- 		    & >4.6 		       & 1       &   -- &   -- \\
G18.148$-$0.283 & 56.1   &  56.4 & 56.4 & 3.2 $\pm$  0.4 & -- & -- & 55.5 & 4.2 & 16.1		    & >3.0 		       & 1665    &  >0.5 &  1.0 \\
G18.303$-$0.390 & nan    &  27.2 & 27.3 & --\tablefootmark{b} & -- & -- & -- & -- & -- 		    & --   		       & 3(OH)   &   -- &   -- \\
		& 32.5   &  32.2 & 32.2 & 1.9 $\pm$  0.2 & -- & -- & 33.0 & 4.5 & 17.1 		    & >2.5 		       & 1665    &  >0.3 &  0.6 \\
		& nan    &  36.2 & 36.2 & --\tablefootmark{b} & -- & -- & -- & -- & -- 		    & --   		       & 3(OH)   &   -- &   -- \\
G19.075$-$0.288 & 66.0   &  64.6 & 64.6 & 5.4 $\pm$  0.9 & -- & -- & 65.1 & 7.2 & 27.4 		    & >4.1 		       & 1665    &  >0.5 &  1.0 \\
G21.347$-$0.629 & 53.3   &  56.1 & 56.1 & 1.9 $\pm$  0.3 & -- & -- & 56.2 & 2.8 & 10.7 		    & >3.3 		       & 4(CO)   &   -- &   -- \\
G21.874$+$0.007 & 21.2   &  21.8 & 20.7 & 4.0 $\pm$  0.6 & -- & -- & 22.3 & 3.9 & 14.9 		    & >3.5 		       & 1665    &  >0.6 &  1.3 \\
G23.956$+$0.150 & 80.5   &  80.0 & 80.1 & 1.5 $\pm$  0.3 & -- & -- & 79.6 & 5.0 & 19.1 		    & >3.5 		       & 1665    &  >0.2 &  0.4 \\
G25.396$+$0.033 & $-$12.0  & $-$12.1 & $-$12.1 & 3.2 $\pm$  0.6 & -- & -- & -- & -- & -- 		    & >1.3 		       & 1       &   -- &   -- \\
G25.397$-$0.141 & 67.0   &  67.1 & 67.2 & 1.0 $\pm$  0.1 & -- & -- & 66.5 & 1.3 & 5.0 		    & >5.1 		       & 1665    &  >0.3 &  1.0 \\
		& 96.5   &  94.5 & 94.4 & 3.9 $\pm$  0.3 & -- & -- & 96.3 & 7.7 & 29.4 		    & >4.2 		       & 1665    &  >0.3 &  0.7 \\
G26.609$-$0.212 & $-$33.8  & $-$33.2 & $-$33.2 & 3.1 $\pm$  1.3 & -- & -- & -- & -- & -- 		    & >1.5 		       & 1       &   -- &   -- \\
G27.563$+$0.084 & 83.5   &  86.0 & 85.9 & 7.2 $\pm$  2.3 & -- & -- & 85.8 & 3.9 & 14.6 		    & >2.0 		       & 1665    &  >1.2 &  2.5 \\
G28.806$+$0.174 & 79.7   &  79.8 & 79.8 & 1.5 $\pm$  0.3 & -- & -- & 80.6 & 1.4 & 5.3 		    & >2.8 		       & 1665    &  >0.6 &  1.5 \\
                & 106.5  & 103.5 & 103.5 & 1.5 $\pm$  0.6 & -- & -- & 104.8 & 5.4 & 20.4 	    & >2.5  		       & 4(CO)   &   -- &   -- \\
G29.935$-$0.053 & 7.0    &   7.0 & -- & <0.6 $\pm$  0.2 & 7.0 & 0.2 $\pm$  0.1 & -- & <0.3 & <1.1 & >1.6       		       & 1667    &  >0.4 &  1.5 \\
                & 50.0   &  50.6 & -- & <0.6 $\pm$  0.2 & 50.5 & 0.3 $\pm$  0.1 & 49.9 & 1.1 & 4.1 & >2.3 	     	       & 1667    &  >0.2 &  0.5 \\
		& 67.5   &  68.0 & -- & <0.6 $\pm$  0.2 & 68.1 & 0.4 $\pm$  0.2 & 68.1 & 0.5 & 2.1 & >4.2  		       & 1667    &  >0.3 &  1.4 \\
		& 97.8   &  98.9 & 98.9 & 1.9 $\pm$  0.5 & 99.0 & 1.4 $\pm$  0.3 & 99.4 & 12.3 & 46.8 & >4.9 		       & 1667    &  >0.1 &  0.2 \\
G29.957$-$0.018 & 7.5    &   8.0 & 8.2 & 0.3 $\pm$ 0.1 & 7.8 & 0.4 $\pm$  0.1 & 7.4 & 0.2 & 0.8 & >2.2 			       & 1667    &  >0.7 &  3.3 \\
                & 100.0  &  99.8 & --  & --\tablefootmark{c} & 100.0 & 0.4 $\pm$  0.1 & 98.0 & 13.1 & 49.9 & >4.5 	       & 4(CO)   &   -- &   -- \\
G30.535$+$0.021 & nan    &  43.7 & -- & --\tablefootmark{c} & 43.8 & --\tablefootmark{b} & -- & -- & -- & -- 		       & 2(OH)   &   -- &   -- \\
		& 43.5   &  49.4 & -- & 1.1 $\pm$  0.5 & 49.3 & 1.3 $\pm$  0.4 & 47.7 & 5.5 & 20.8 & >4.0      	    	       & 2(OH)   &   -- &   -- \\
		& 92.5   &  91.6 & 91.3 & 2.7 $\pm$  0.5 & 92.0 & 1.7 $\pm$  0.4 & 91.9 & 2.8 & 10.8 & >3.3    		       & 1667    &  >0.4 &  1.0 \\
G30.720$-$0.083 & 94.0   &  93.6 & 92.7 & --\tablefootmark{c} & 93.5 & --\tablefootmark{c} & 94.0 & 9.5 & 36.1 & >4.4 	       & 7       &   -- &   -- \\
G30.783$-$0.028 & nan    &  77.9 & -- & -- & 77.4 & --\tablefootmark{b} & -- & -- & -- & --  	    	       	 	       & 3(OH)   &   -- &   -- \\
		& 82.3   &  81.9 & 80.8 & 0.8 $\pm$  0.1 & 81.9 & 0.9 $\pm$  0.2 & 81.9 & 1.6 & 6.0 & >3.4    		       & 1667    &  >0.4 &  1.0 \\
                & nan    &  87.1 & 86.6 & --\tablefootmark{b} & -- & -- & -- & -- & -- & --                 		       & 3(OH)   &   -- &   -- \\
		& 94.0   &  92.1 & 91.9 & 1.5 $\pm$  0.3 & 92.1 & 1.2 $\pm$  0.2 & 93.3 & 9.1 & 34.7 & >7.2 		       & 1667    &  >0.1 &  0.2 \\
		& nan    &  98.6 & 98.1 & --\tablefootmark{b} & 98.7 & --\tablefootmark{b} & 104.9 & -- & -- 		       & 3(OH)   &   -- &   -- \\
G30.854$+$0.151 & 94.8   &  95.4 & 95.5 & 5.4 $\pm$  0.9 & 95.3 & 4.5 $\pm$  0.8 & 95.5 & 3.6 & 13.7 & >1.3    		       & 1667    &  >1.1 &  2.2 \\
G31.242$-$0.110 & 19.0   &  19.6 & -- & --\tablefootmark{c} & 19.6 & 1.9 $\pm$  0.3 & 20.8 & 1.6 & 6.1 & >2.5 		       & 1667    &  >0.8 &  2.0 \\
		& 78.4   &  79.2 & 79.4 & 2.3 $\pm$  0.4 & 78.9 & 1.4 $\pm$  0.3 & 78.8 & 1.6 & 6.2 & >2.1 		       & 1667    &  >0.7 &  1.6 \\
		& 84.4   &  83.7 & -- & <1.1 $\pm$  0.3 & 83.8 & 0.7 $\pm$  0.3 & 84.0 & 0.7 & 2.6 & >1.3 		       & 1667    &  >0.7 &  1.7 \\
G31.388$-$0.383 & 18.5   &  18.1 & 18.5 & 0.7 $\pm$  0.2 & 18.0 & 0.5 $\pm$  0.1 & -- & <0.2 & <0.9 & >3.2    		       & 1667    &  >0.6 &  3.5 \\
G32.151$+$0.132 & 94.5   &  93.8 & 93.8 & 4.8 $\pm$  0.6 & -- & -- & 94.4 & 4.3 & 16.5 & >2.4  			       	       & 1665    &  >0.7 &  1.5 \\
G32.272$-$0.226 & 22.5   &  22.6 & 22.6 & 3.6 $\pm$  0.6 & -- & -- & 22.4 & 0.9 & 3.3 & >2.1 				       & 1665    &  >2.1 &  5.5 \\
G32.798$+$0.190 & 12.5   &  12.8 & 12.7 & 5.8 $\pm$  0.3 & -- & -- & 15.1 & 6.7 & 25.3 & >8.1 				       & 1665    &  >0.5 &  1.1 \\
G32.928$+$0.607 & $-$34.5  & $-$33.9 & $-$34.0 & 2.5 $\pm$  1.1\tablefootmark{a} & -- & -- & -- & -- & -- & >2.8 		       & 1       &   -- &   -- \\
G33.915$+$0.110 & 94.2   &  95.3 & 95.3 & 0.6 $\pm$  0.2\tablefootmark{a} & -- & -- & -- & <0.3 & <1.1  & >0.6 		       & 2(OH)   &   -- &   -- \\
		& 100.2  & 101.5 & 101.5 & 1.1 $\pm$  0.2\tablefootmark{a} & -- & -- & -- & <0.3 & <1.1 & >1.5 		       & 2(OH)   &   -- &   -- \\
                & 108.2  & 106.3 & 106.3 & 1.9 $\pm$  0.3 & -- & -- & 107.5 & 4.2 & 15.9 & >1.9    			       & 1665    &  >0.3 &  0.6 \\
G34.132$+$0.471 & 33.8   &  33.8 & 33.9 & 2.4 $\pm$  0.5 & -- & -- & 35.1 & 1.5 & 5.8 & >3.4 				       & 1665    &  >0.8 &  2.0 \\
G35.467$+$0.139 & 77.0   &  77.0 & 76.9 & 3.4 $\pm$  0.8 & -- & -- & 77.4 & 3.9 & 14.6 & >1.3 				       & 1665    &  >0.6 &  1.2 \\
G37.764$-$0.215 & 63.0   &  63.4 & 63.4 & 6.6 $\pm$  0.9 & -- & -- & 61.5 & 4.6 & 17.3 & >5.7 				       & 1665    &  >0.8 &  1.9 \\
G37.874$-$0.399 & 60.0   &  61.0 & 60.8 & 7.6 $\pm$  0.9 & -- & -- & 61.4 & 4.2 & 15.9 & >8.8 				       & 1665    &  >0.9 &  2.4 \\
G38.876$+$0.308 & $-$16.5  & $-$16.3 & $-$16.4 & 4.1 $\pm$  0.7 & $-$16.2 & 4.0 $\pm$  0.6 & -- & -- & -- & >1.7 		       & 1       &   -- &   -- \\
G39.565$-$0.040 & 22.5   &  23.2 & 23.3 & 2.6 $\pm$  0.4 & 23.2 & 2.2 $\pm$  0.3 & 23.4 & 0.6 & 2.3 & >3.6 		       & 1667    &  >1.8 &  6.4 \\
G39.883$-$0.346 & 57.0   &  56.9 & 56.9 & 4.9 $\pm$  1.0 & 57.0 & 3.4 $\pm$  0.7 & 58.7 & 3.0 & 11.3 & >2.9 		       & 1667    &  >0.9 &  2.0 \\
G41.741$+$0.097 & 12.1   &  14.2 & 14.7 & 1.9 $\pm$  0.7 & 13.9 & 1.2 $\pm$  0.3 & 13.4 & 0.7 & 2.8 & >1.2 		       & 1667    &  >1.1 &  2.8 \\
G42.027$-$0.604 & 66.5   &  65.1 & -- & <1.3 $\pm$  0.4 & 65.1 & 1.3 $\pm$  0.3 & 65.5 & 1.6 & 6.1 & >3.1 		       & 1667    &  >0.6 &  1.5 \\
G45.454$+$0.060 & nan    &  56.0 & -- & -- & 55.2 & --\tablefootmark{b} & -- & -- & -- & --    	     			       & 3(OH)   &   -- &   -- \\
		& 60.0   &  59.4 & 59.2 & --\tablefootmark{c} & 59.4 & 1.9 $\pm$  0.2 & 58.9 & 5.9 & 22.2 & >5.5 	       & 1667    &  >0.2 &  0.6 \\
		& nan    &  64.9 & -- & -- & 66.6 & --\tablefootmark{c}  & -- & -- & -- & --    	     	    	       & 3(OH)   &   -- &   -- \\
G49.206$-$0.342 & 64.0   &  65.3 & 65.3 & 7.2 $\pm$  1.4 & -- & -- & 65.7 & 3.7 & 14.0 & >6.9 				       & 1665    &  >1.0 &  2.6 \\
G49.369$-$0.302 & 51.0   &  50.9 &a 50.9 & 5.0 $\pm$  0.7 & -- & -- & 51.1 & 7.9 & 30.0 & >6.2 				       & 1665    &  >0.4 &  0.8 \\
		& 63.0   &  62.9 & 62.8 & 2.5 $\pm$  0.4 & -- & -- & 61.1 & 3.1 & 11.8 & >5.4 				       & 1665    &  >0.4 &  1.1 \\
G49.459$-$0.353 & 60.0   &  62.0 & 61.5 & 2.7 $\pm$  0.6 & -- & -- & 61.1 & 4.4 & 16.8 & >5.5 				       & 1665    &  >0.3 &  0.8 \\
		& 69.0   &  68.6 & 68.3 & 4.0 $\pm$  0.9 & -- & -- & 68.6 & 5.2 & 19.8 & >3.9 				       & 1665    &  >0.5 &  1.0 \\
G49.488$-$0.380 & 67.8   &  65.9 & 65.9 & 1.9 $\pm$  0.4 & -- & -- & 67.8 & 5.8 & 22.0 & >4.7 				       & 4       &   -- &   -- \\
G52.753$+$0.334 & 15.0   &  12.1 & 13.6 & <1.4$\pm$  0.4\tablefootmark{a} & 10.5 & 0.7 $\pm$  0.3 & -- & 0.9 & 3.5 & >2.1      & 7       &   -- &   -- \\
		& 45.0   &  45.2 & 45.2 & <1.3$\pm$  0.3\tablefootmark{a} & 45.0 & 0.8 $\pm$  0.2 & 44.5 & 0.7 & 2.5 & >1.4    & 1667    &  >0.8 &  2.1 \\
G60.882$-$0.132 & 22.5   &  22.3 & 22.2 & 2.8 $\pm$  0.7 & 22.3 & 3.0 $\pm$  0.5                   & 22.4 & 7.2 & 27.5 & >2.1    & 1667    &  >0.4 &  0.7 \\
G61.475$+$0.092 & 6.2    &   5.9 & --   & <0.12 $\pm$0.03 & 5.9 & 0.1 $\pm$  0.03\tablefootmark{a} & 6.6 & 0.03 & 0.12 & >1.6      & 2(OH)   &   -- &   -- \\
                & 21.2   &  21.1 & 21.2 & 4.9 $\pm$  0.2 & 21.1 & 3.7 $\pm$  0.2 & 21.6 & 7.3 & 27.8 & >4.0  	      	       & 1667    &  >0.4 &  0.9 \\
\hline
\label{tbl:integrated_properties}
\end{tabular}
\tablefoot{${N_{\rm OH\,1665\,MHz}}/{T_{\rm ex}}$, ${N_{\rm OH\,1667\,MHz}}/{T_{\rm ex}}$ and $N_{\rm \ion{H}{i}}$ are determined from the integrated optical depth (Table~\ref{tbl:lineparameters}; see text for the conversions used). The velocity in Col.~2 is the mean of the center velocities of the OH\,1665\,MHz and OH\,1667\,MHz absorption. The column density of molecular hydrogen, $N_{\rm H_2}$, is derived from \ce{^{13}CO} emission (for assumptions and conversions see text). $X_{\rm OH}(N_{\rm H})$ is defined as $N_{\rm OH}$/$N_{\rm H}$, with $N_{\rm H} = N_{\rm \ion{H}{i}} + 2N_{\rm H_2}$, while $X_{\rm OH}(N_{\rm H_2})$ is the ratio $N_{\rm OH}$/$N_{\rm H_2}$. Notes and footnotes are as in Table~\ref{tbl:lineparameters}.}
\end{table*}

\subsection{Satellite line transitions} \label{sec:satellite}
Satellite lines of the \ce{OH} ground state transitions are rarely found in 
local thermodynamic equilibrium with the main line transitions. While the main lines
are seen in absorption, the satellite lines can be anomalously excited and may show 
conjugate emission and absorption, often at equal strength (e.g., towards G18.303$-$0.390 and G25.396+0.033; see Fig.~\ref{fig:plot_alltrans_00}). 
Table~\ref{tbl:satellites} catalogs conjugate satellite transitions found in this survey. 
The qualitative satellite line behavior reflects the physical conditions of the gas, such that we can use this to estimate OH column densities for a subsample of sources. 

\smallskip
\noindent
Emission in the 1612{\us}MHz transition with absorption in the 1720{\us}MHz transition is found in 14 instances. Emission in the 1720{\us}MHz transition with absorption in the 1612{\us}MHz transition occurs in three cases (e.g., towards G32.798+0.190 at 90\us\kms; see Table~\ref{tbl:satellites}).

\smallskip
\noindent
Satellite line reversal, i.e., the transition from absorption to emission (or vice versa) across the line, is seen along 3 lines of sight. Both satellites mirror each other: At lower velocities, the 1720{\us}MHz line is in emission, while the 1612{\us}MHz line is in absorption. At a certain velocity, the behavior reverses. This is found in the following cases in this work: G19.075$-$0.288 at 67.5\us\kms, G32.798+0.190 at 14.0\us\kms, G49.369$-$0.302 at 63.5\us\kms. Lines of sight, for which the full reversal profile is not detected in both satellite lines, but which are indicative of this behavior, are: G29.935$-$0.053 at 98.5\us\kms, for which reversal in the 1612{\us}MHz transition is detected, but the 1720{\us}MHz line is seen only in absorption at higher velocities, and not detected in emission at lower velocities. Towards G37.764$-$0.215 at 65.0\us\kms, G49.206$-$0.342 at 63.5\us\kms\ and G49.369$-$0.302 at 49.5\us\kms, 1612{\us}MHz absorption is seen at low velocities without conjugate emission in the 1720{\us}MHz transition, while at higher velocities, the 1720{\us}MHz transition is in absorption without conjugate emission in the 1612{\us}MHz transition. 

\smallskip
\noindent
In the other cases, the satellite lines are not detected or only one transition of the two is detected. Indication of absorption in both lines can be seen in G61.475+0.092 at +21.2\us\kms, with the satellite line strengths differing from each other, as is expected since the lines are typically not in local thermal equilibrium. 

\smallskip
\noindent
The conjugate emission and absorption of the satellite lines has been noted in previous studies \citep[e.g.,][]{Goss:1968aa,Crutcher:1977aa,van-Langeveldevan-Dishoeck:1995aa,BrooksWhiteoak:2001aa,DawsonWalsh:2014aa}, and is the result of overpopulation of either the F=1 or the F=2 hyperfine energy levels of the ground state and mutual depletion of the others \citep[e.g.,][\S 9.1]{Elitzur:1992aa}. As the satellite lines are transitions with $|\Delta F| =1$, they are affected by the relative population changes, while the main line transitions with $|\Delta F| =0$ may not be affected by this particular inversion mechanism. 

\smallskip
\noindent
There are different pumping mechanisms that may be responsible for the population inversion \citep[see discussion in, e.g.,][]{FrayerSeaquist:1998aa}. In all cases, transitions from higher rotational levels to the ground state need to become optically thick \citep[e.g.,][]{Elitzur:1992aa, van-Langeveldevan-Dishoeck:1995aa}: If the infrared transitions from either the ${}^2\Pi_{3/2}(J=5/2)$ or the ${}^2\Pi_{1/2}(J=1/2)$ states into the ground state become optically thick, the 1720{\us}MHz or the 1612{\us}MHz transition, respectively, is seen in inversion. If both transitions are optically thick, inversion of the 1612{\us}MHz transition is seen. As the transitions from the ${}^2\Pi_{3/2}(J=5/2)$ excited state become optically thick at lower $N_{\rm OH}$ than from the ${}^2\Pi_{1/2}(J=1/2)$ state, there exists a typical OH column density at which the transition from 1720{\us}MHz to 1612{\us}MHz inversion takes place. This has been used and modeled by \citet{van-Langeveldevan-Dishoeck:1995aa} for a molecular cloud heated by a background \ion{H}{ii} region and satellite line reversal was found to take place at $\approx1\times 10^{15} \us{\rm cm^{-2}\,km^{-1}\,s}$. 

\smallskip
\noindent
Assuming this geometry also for the sources in this sample, 
this model provides a possibility to estimate $N_{\rm OH}$.
The column density depends on the velocity dispersion of the gas ($N_{\rm OH} \approx \Delta v \times 1\times 10^{15} \us{\rm cm^{-2}\,km^{-1}\,s}$). 
As we have no direct measure of the line width at the velocity of the reversal of the inversion we use as approximation the
full width of half maximum of the 1665{\us}MHz main. The transition occurs at 
$N_{\rm OH} \approx 7.4\times10^{15}\us{\rm cm^{-2}}$ for G19.075$-$0.288, 
at $N_{\rm OH} \approx 1.2\times10^{16}\us{\rm cm^{-2}}$ for G32.798+0.190 and 
at $N_{\rm OH} \approx 4.4\times10^{15}\us{\rm cm^{-2}}$ for G49.369$-$0.302.

\smallskip
\noindent
We compare the estimates for G32.798+0.190 and G49.369$-$0.302 to $N_{\rm OH}$ derived from the main lines in Sect.~\ref{sec:ohabundance}.
G19.075$-$0.288 is omitted here, as the reversal velocity does not match the velocity of the maximum optical depth of the 1665{\us}MHz transition (Fig.~\ref{fig:plot_alltrans_00}). For G32.798+0.190 and G49.369$-$0.302, $N_{\rm OH}$ determined from the main lines is a factor of 3--4 lower than the estimate from the satellite lines. The line width used for the $N_{\rm OH}$ estimate from the satellite lines could be an overestimate if multiple components are blended into the feature. Alternatively, the discrepancy could be an indication of higher main line excitation temperatures than the assumed $T_{\rm ex}(1665)=5$\,K. To match the estimates from the satellite lines, excitation temperatures of $T_{\rm ex}\approx15 - 20${\us}K would be required.

\smallskip
\noindent
A similar discrepancy has been noted in \citet{XuLi:2016aa}, who find lower OH column densities than needed to reproduce the observed emission in the 1612{\us}MHz transition. They attribute this to other excitation mechanisms, such as collisional excitation in shocks \citep[e.g.,][]{PihlstromFish:2008aa}, which are also not taken into account in the model by \citet{van-Langeveldevan-Dishoeck:1995aa}.

\smallskip
\noindent
Recently, employing non-LTE modeling of all four 18{\us}cm OH emission lines, \citet{EbisawaInokuma:2015aa} have used the relative intensities of main-line and 1720{\us}MHz emission and 1612{\us}MHz absorption in the Heiles~Cloud~2 and $\rho$~Oph to derive kinetic temperatures that are significantly higher in translucent than in dark molecular regions. This indicates that \ce{OH} appears to be able to probe the interface between molecular and warmer atomic material. Such an analysis is beyond the scope of the present paper, but can be included in a future study.

\begin{table}
\setlength\tabcolsep{6pt}
\renewcommand{\arraystretch}{1.0}
\caption{Conjugate inversion and anti-inversion of satellite lines}
\begin{tabular}{l|l|l|r}
\hline
\hline
Name             & 1612 & 1720 & v\\
                 & & & $\left[{\rm km}/{\rm s}\right]$\\
\hline
G18.303$-$0.390	&E	&A	&31.5	\\
G19.075$-$0.288	&A	&E	&58.5	\\ 
				&R	&R	&67.5	\\ 
				&E	&A	&70.0	\\ 
G25.396+0.033		&E	&A	&$-$12.0	\\
G25.397$-$0.141	&E	&A	&96.0	\\
G29.935$-$0.053	&A	&--	&97.0	\\ 
				&R	&--	&98.5	\\ 
				&E	&A	&100.0	\\ 
G29.957$-$0.018	&E	&A	&100.5	\\
G30.535+0.021		&E	&A	&45.0	\\
G30.535+0.021		&A	&E	&92.0	\\
G30.720$-$0.083	&E	&A	&93.0	\\
G30.783$-$0.028	&E	&A	&82.0	\\
G32.151+0.132		&E	&A	&94.5	\\
G32.798+0.190		&A	&E	&12.0	\\
				&R	&R	&14.0	\\
				&E	&A	&18.0	\\
G32.798+0.190		&A	&E	&90.0	\\
G33.915+0.110		&E	&A	&105.0	\\
G37.764$-$0.215	&A	&--	&62.5	\\
				&R	&R	&65.0	\\
				&--	&A	&67.5	\\
G37.874$-$0.399	&E	&A	&58.5	\\
G38.876+0.308		&E	&A	&$-$16.5	\\
G39.883$-$0.346	&E	&A	&57.0	\\
G41.741$+$0.097	&E	&A	&14.0	\\
G49.206$-$0.342	&A	&--	&65.0	\\
				&R	&--	&66.0	\\
				&E	&A	&68.0	\\
G49.369$-$0.302	&A	&--	&49.0	\\
				&R	&--	&49.5	\\
				&E	&A	&53.0	\\
G49.369$-$0.302	&A	&E	&62.5	\\
				&R	&R	&63.5	\\
				&E	&A	&65.0	\\
G49.459$-$0.353	&A	&E	&69.0	\\
G60.882$-$0.132	&E	&A	&22.5	\\
\hline
\label{tbl:satellites}
\end{tabular}
\tablefoot{Columns 2 and 3 indicate conjugate absorption ({\it A}) and emission ({\it E}) of the 
OH 1612\,MHz and OH 1720\,MHz transitions. The corresponding velocity is given in column 4. 
In some cases, 1612\,MHz absorption and 1720\,MHz emission transform 
into 1612\,MHz emission and 1720\,MHz absorption at higher velocities. 
For this kind of profile we give three entries: the central velocities of absorption and emission 
as well as the velocity, at which the reversal (R) of the line profile occurs. This is defined here
as the velocity at which the satellite lines are equal. Components of the reversal profile that are not 
detected significantly are indicated by a horizontal dash.}
\end{table}

\subsection{Extended OH absorption: the example of W43} \label{sec:analysisw43}
Spatially resolved OH absorption is seen against a subsample of the continuum
sources (examples of these are the Galactic \ion{H}{ii}~regions M17, 
G18.148$-$0.283, G37.764$-$0.215, G45.454+0.060 and G61.475+0.092). 
This allows for a comparison of column density and kinematic
structure in different physical regimes: The ionized gas phase is traced by continuum emission and
RRLs, the cold neutral medium is traced by \ion{H}{i}~absorption and the molecular gas regime
is traced by different \ce{CO} isotopologues and far-IR continuum emission. We
present the star-forming region W43, as an example of what can be
learned from this data. 

\smallskip
\noindent
W43 is one of the largest molecular cloud complexes in our Galaxy. It is located at the
intersection of the Scutum-Centaurus spiral arm with the Galactic bar, is actively forming stars at a high rate and
is dynamically complex \citep[e.g.,][]{Nguyen-LuongMotte:2011aa,MotteNguyen-Luong:2014aa,BihrBeuther:2015aa}. 
It is composed of multiple sub-regions, most prominently the W43-main and W43-south
regions, which themselves break down into smaller regions of molecular gas
\citep[e.g.,][]{CarlhoffNguyen-Luong:2013aa}. In W43-main, complex structure is
also indicated by the high \ion{H}{i} column densities
\citep[e.g.,][]{Liszt:1995aa,MotteNguyen-Luong:2014aa,BihrBeuther:2015aa},
which suggests the presence of several molecular clouds along the
line-of-sight \citep{BialyBihr:2017aa}. 

\smallskip
\noindent
Different evolutionary stages of stars and clouds coexist and appear to be influencing each other: An OB cluster at the center of W43-main, 
which contains Wolf-Rayet stars, includes a strong source of
ultraviolet photons \citep[e.g.,][]{SmithBiermann:1978aa,BlumDamineli:1999aa}; these provide the ionization and heating of the central
\ion{H}{ii} region \citep[e.g.,][]{ReifensteinWilson:1970aa,LesterDinerstein:1985aa}. 
There is evidence for a second generation of star formation, indicated by
clumps of dense gas and ultra-compact \ion{H}{ii} regions
\citep[e.g.,][]{MotteSchilke:2003aa,BallyAnderson:2010aa,BeutherTackenberg:2012aa},
which conveys the picture of gas compression driven by the central \ion{H}{ii}
regions \citep[e.g.,][]{BlumDamineli:1999aa,BalserGoss:2001aa}. In the
environment of W43-main, pre-stellar cores manifest higher gas temperatures than
in quiescent regions due to the heating by the central cluster, possibly
affecting the number of stars formed in the future
\citep{BeutherTackenberg:2012aa}. Adding to the complexity of the region,
observations of molecular and ionized gas tracers revealed several velocity
gradients and substructures of different morphologies
\citep[e.g.,][]{Liszt:1995aa,BalserGoss:2001aa,CarlhoffNguyen-Luong:2013aa}. 
The gas streams on global scales in molecular and atomic gas indicate that the H to \ce{H2} conversion is ongoing
\citep{MotteNguyen-Luong:2014aa}. Dynamical interaction between clouds has been investigated on
smaller scales with SiO emission, which possibly emerges from low velocity shocks in mm-emission peaks
\citep[e.g.,][]{Nguyen-LuongMotte:2013aa,LouvetMotte:2016aa}. 

\begin{figure*}
 \centering
 \includegraphics[width=0.99\textwidth]{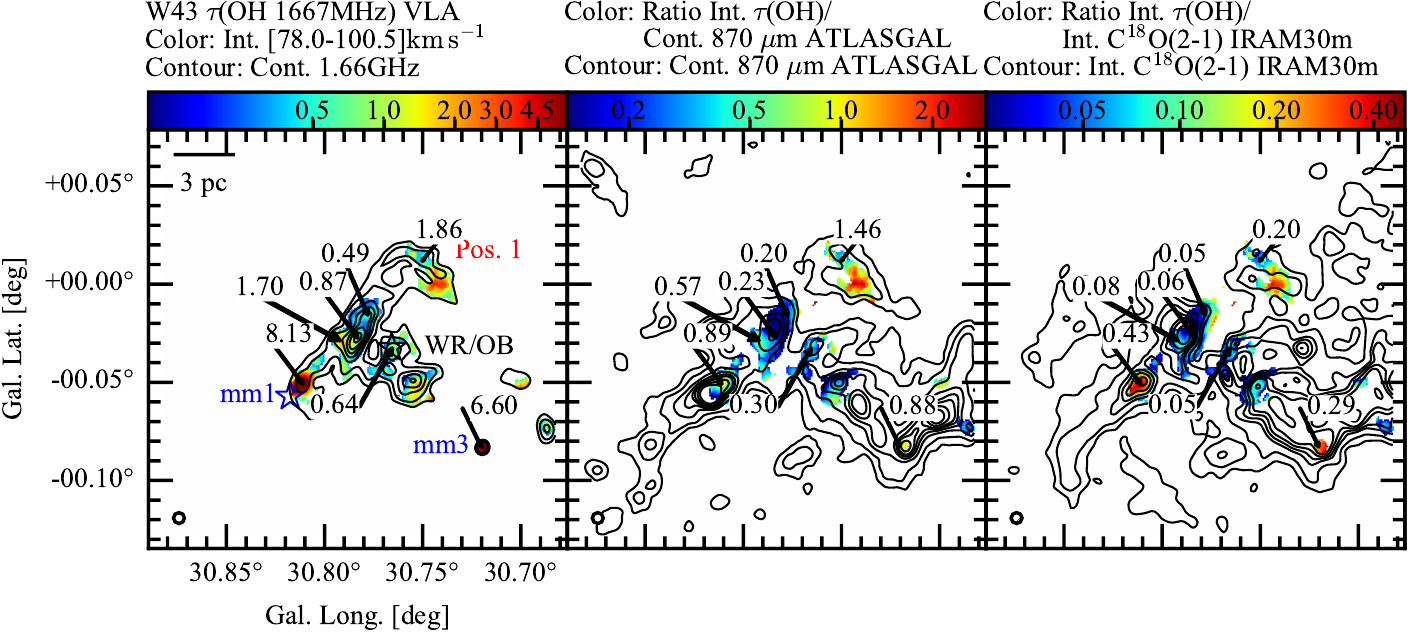}
 \caption{Comparison of $\tau(\ce{OH}\,1667\,{\rm MHz})$, dust continuum emission at $\lambda = 870\,\mu{\rm m}$, and \ce{C^{18}O}(2-1) emission in the W43
  star-forming region. In the left panel, the integrated optical depth of the OH
  1667{\us}MHz transition between 78.0 and 100.5\us\kms\ is displayed in
  colors. For each pixel, only channels that are detected at a 3-$\sigma$
  level contribute to the integrated $\tau$-map. It is overlayed with contours of the 
  18~cm continuum emission (black, at levels of 0.1, 0.2,
  0.4, 0.6, 0.8, 1.0, 1.25, 1.5 and 1.75\us\jyb). The middle panel shows the
  ratio of the integrated $\tau({\rm OH})$ map to ${\rm 870\,\mu m}$ ATLASGAL
  emission \citep{SchullerMenten:2009aa}, which traces dense gas (the
  dust emission is overlayed in black contours, at levels of 0.5, 1.0, 2.0,
  3.0, 4.0, 5.0, 7.0 and 10.0\us\jyb). The right panel shows the ratio of $\tau({\rm OH})$ to
  \ce{C^{18}O}(2-1) emission \citep{CarlhoffNguyen-Luong:2013aa},
  integrated over the same velocity range (the velocity integrated \ce{C^{18}O}(2-1) emission 
  is overplotted in black contours at levels of 6, 9, 13, 15, 17, 19,
  23{\us}K\us\kms). All data have been smoothed
  to a common resolution of 20\arcsec, corresponding to spatial scales of 0.5{\us}pc. In the left panel, the central Wolf-Rayet/OB
  cluster is marked by a black star, while the dense clumps MM1 and MM3
  \citep{MotteSchilke:2003aa} are marked by blue stars. The upper end of the \mbox{T-bar-shaped} continuum emission is marked as Pos. 1 for easier reference in the text. For 
  readability, the values of the ratio are displayed for selected positions in the figure.}
 \label{fig:plot_moment_map_w43_ratio_moments}
\end{figure*}

\begin{figure*}
 \centering
 \includegraphics[width=0.99\textwidth]{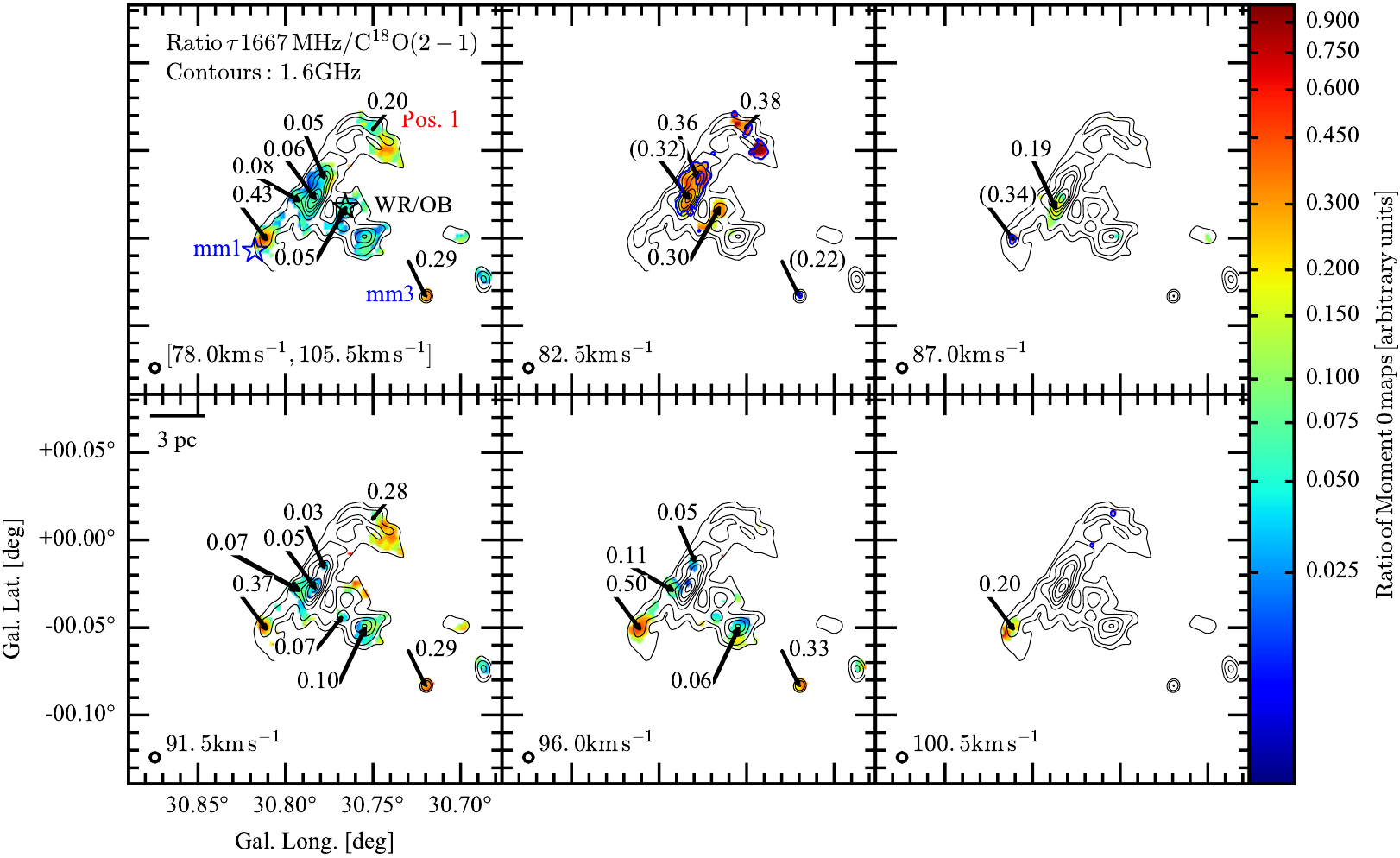}
 \caption{Ratio of integrated $\tau(\ce{OH}\,1667\,{\rm MHz})$ absorption and
  \ce{C^18O}(2-1) emission in W43. The top-left panel is the same as the rightmost panel in 
  Fig.~\ref{fig:plot_moment_map_w43_ratio_moments} and is shown for orientation.
  The other panels show the ratio of $\tau(\ce{OH}\,1667\,{\rm MHz})$ and \ce{C^18O}(2-1) at the 
  indicated velocities after binning three channels of 1.5\us\kms\ width. 
  Overlayed on all panels are contours of 18~cm continuum emission 
  (black, in levels of 0.1, 0.2, 0.4, 0.6, 0.8, 1.0, 1.25, 1.5 and 1.75\us\jyb). 
  The 1667{\us}MHz optical depth has been masked at 3-$\sigma$ detection levels in the
  original \ce{OH} absorption data. For pixels with no \ce{C^{18}O} emission
  counterpart, 3-$\sigma$ detection limits have been used, and are indicated
  by blue contours. The ratio is quoted in brackets for these locations.}
 \label{fig:plot_channel_map_w43_tau_1667_by_c18o21_00}
\end{figure*}

\begin{figure}
 \centering
 \includegraphics[width=0.99\columnwidth]{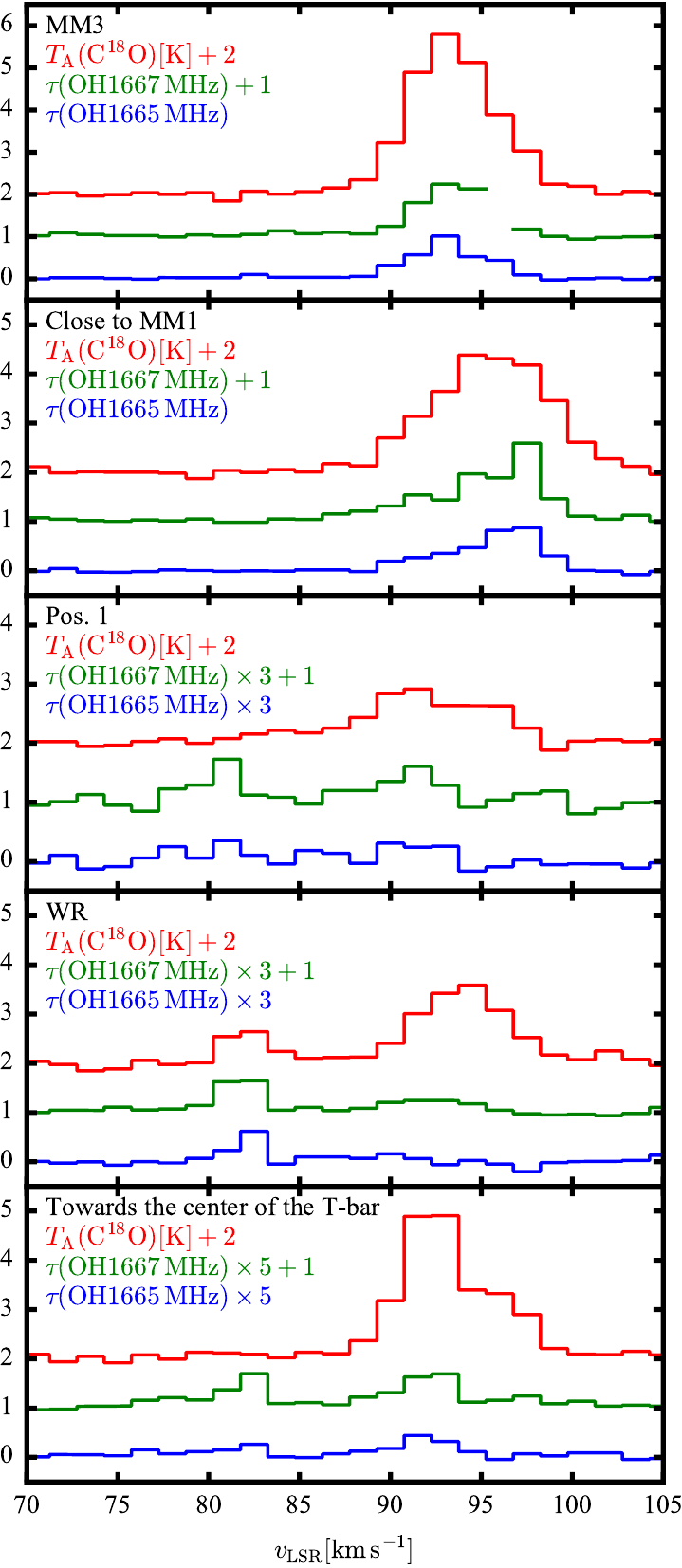}
 \caption{Spectra of $\tau_{\rm OH}$ in the 1665 and 1667\,MHz lines, as well as of emission in \ce{C^{18}O}(2--1) line. The spectra are extracted towards positions MM1, MM3$^{6}$, WR and ``Pos.~1'' as indicated in Figs.~\ref{fig:plot_moment_map_w43_ratio_moments}~and~\ref{fig:plot_channel_map_w43_tau_1667_by_c18o21_00}, as well as towards the central part of the T-bar. Towards MM3, the channel at 96\,\kms\ of the OH 1667\,MHz absorption is masked because of sidelobes of close-by maser emission.}
 \label{fig:plot_spectra_w43}
\end{figure}

\smallskip
\noindent
The extended OH absorption in W43 is displayed in the left panel of
Fig.~\ref{fig:plot_moment_map_w43_ratio_moments}, in which the optical depth 
has been integrated between 78.0--100.5\us\kms\ and all maps are shown at an angular resolution of ${20\arcsec\times20\arcsec}$, corresponding to a spatial scale of 0.5{\us}pc at a distance of 5.5{\us}kpc \citep{ZhangMoscadelli:2014aa}. 
Intrinsically, absorption is seen towards strong continuum
emission peaks, as the sensitivity to find absorption increases with continuum
emission strength. However, the strongest integrated optical depth peaks do
not coincide with continuum emission peaks, but are seen
towards the mm dust emission sources MM1 and MM3\footnote{At MM3, the OH 1667\,MHz absorption is affected by side lobes of close-by maser emission at 96\,\kms. To determine the moment maps shown in Figs.~\ref{fig:plot_moment_map_w43_ratio_moments} and \ref{fig:plot_channel_map_w43_tau_1667_by_c18o21_00}, we interpolated over the channel at 96\,\kms\ in lines of sight close to MM3. For clarity, this channel is masked in Fig.~\ref{fig:plot_spectra_w43}.} \citep[nomenclature taken from][]{MotteSchilke:2003aa}. The integrated OH absorption varies by a factor of 3 around the central
\ion{H}{ii} region, and is higher by an order of magnitude towards the outer
parts of the T-shaped continuum emission: At MM3 and 
towards MM1, the total line-of-sight column density is approximately $7\times10^{15}\us{\rm cm^{-2}}$ and $9\times10^{15}\us{\rm cm^{-2}}$, respectively
while the column density is between $0.5-2\times10^{15}\us{\rm cm^{-2}}$ around the
central \ion{H}{ii} region, assuming an excitation temperature of $T_{\rm ex} = 5{\rm{\us}K}$ in both cases. 

\smallskip
\noindent
Fig.~\ref{fig:plot_moment_map_w43_ratio_moments} compares the optical depth of the 
OH 1667{\us}MHz transition to ATLASGAL 870\,$\mu$m dust emission 
\citep{SchullerMenten:2009aa} and to \ce{C^{18}O}(2--1) emission 
\citep{CarlhoffNguyen-Luong:2013aa}. The middle panel displays 
the ratio of integrated \ce{OH} optical depth to 870\,$\mu$m dust emission, while 
the right panel displays the ratio to integrated \ce{C^{18}O}(2--1) emission. 
Both the \ce{OH} and \ce{C^{18}O}(2--1) data have been integrated between 
78.0\us\kms\ and 100.5\us\kms. For the \ce{OH} optical depth only such pixels contribute to the 
integral that are significantly detected in the absorption data. 
The value of the ratio maps are shown for relative comparison of different parts of the region, 
motivated by the hypothesis that all the tracers are optically thin and hence contribute 
linearly to the column density of the species (the optical thickness of the tracers is discussed further in sect.~\ref{sec:discussion_w43}). 
Thus, the ratios quoted here represent no physical quantity directly but their
variation across the map can be indicative either of OH abundance or excitation variations. 
 
\smallskip
\noindent
The integrated \ce{C^{18}O} and 870\,$\mu$m emission are shown as contours in 
Fig.~\ref{fig:plot_moment_map_w43_ratio_moments}. There are enhancements towards the central continuum source, towards MM1 and MM3 in both tracers. The strongest peak in the \ce{C^{18}O} emission is towards the central part of the T-bar, and is slightly offset from the peak of the continuum emission. The ATLASGAL emission peaks more strongly towards MM1 and MM3. At MM1, continuum emission and \ce{C^{18}O} emission are slightly shifted away from the 870\,$\mu$m emission.

\smallskip
\noindent
Within the central part of the T-bar, the ratio of OH optical depth to \ce{C^{18}O} emission is around 0.05, and slightly higher on the left side of the central continuum peak. This is similar in the 870\,$\mu$m emission, where the ratio is between 0.2-0.3, and by a factor of 2 higher on the side facing MM1. The ratio is around a factor of 4-9 higher against the MM1 and MM3 sources in both tracers. This increase is slightly higher for the ratio to \ce{C^{18}O} emission than to 870\,$\mu$m emission, which is consistent with the stronger increase of the 870\,$\mu$m emission towards these sources. Against ``Pos. 1'', however, the ratio to ATLASGAL 870\,$\mu$m emission is by a factor of $\sim$5 higher than in the central continuum emission region. The increase in this ratio is also seen in the \ce{C^{18}O} emission.

\smallskip
\noindent
In order to understand this comparison better, Fig.~\ref{fig:plot_channel_map_w43_tau_1667_by_c18o21_00} 
shows the ratio of OH optical depth and \ce{C^{18}O} in velocity ranges of 4.5\us\kms\ (after binning three channels of 1.5\us\kms). Between 81.0 and 84.0\us\kms,
\ce{OH} absorption is seen against the central \ion{H}{ii} region and ``Pos. 1''. \ce{C^{18}O} is significantly detected only in few locations, 
and we include upper limits in the plot (encircled by blue contours). 
Ratios are found between 0.3 and
0.4. A similar ratio is seen between 90\us\kms\ and 93.0\us\kms\ towards MM3, MM1
and ``Pos. 1''. In this velocity range, however, the OH ratio at the center of the T-bar is rather low, between 0.03 and 0.1. The ratio increases
for MM1 towards 0.5 between 94.5 and 97.5\us\kms.

\smallskip
\noindent
To conclude, variations in \ce{OH} to \ce{C^{18}O} ratio are seen also when refining the integration interval. There
seem to be two regimes -- the central part of the \ion{H}{ii} region exhibits a ratio of $\sim0.05$, while for other locations and 
other velocities, we find a ratio of $\sim0.3$. At lower velocities, also the central part of the \ion{H}{ii} region is seen at ratios of $\sim0.3$. 
This is further discussed in Section~\ref{sec:discussion_w43}.

\section{Discussion}    \label{sec:discussion}
\subsection{Distribution of OH in the Galactic plane}\label{sec:discussion_distribution}
Recent single-dish observations of the OH ground state transitions find OH to be extended over wide 
areas in the Galactic plane \citep{DawsonWalsh:2014aa}. The number of absorption 
detections found in this work is at first glance small relative to the number of cm-continuum 
sources available in the Galactic plane \citep[e.g.,][]{BihrJohnston:2016aa} 
and needs to be discussed in terms of the varying 
sensitivity limit with continuum source strength. 

\smallskip
\noindent
We detect \ce{OH} absorption mostly against extended Galactic cm-continuum
background sources that show a spectral index in agreement with that of 
free-free emission from \ion{H}{ii} regions. The lower number of detections of \ce{OH}
in diffuse clouds not associated with \ion{H}{ii} regions is likely due to the
sensitivity limits indicated in Fig.~\ref{fig:plot_oh1665_vs_contemp_with_tau_min}.
While we do detect \ce{OH} absorption at a variety of optical depths below
$\tau \leq 0.2$ at continuum flux density >1\us\jyb, at lower continuum
surface brightness the sensitivity is not high enough to detect sources with
$\tau \le 0.05-0.1$. As the majority of the continuum sources have a 
flux density <1\us\jyb, we pick up largely absorption at higher optical depths. The
increase of the relative number of detections with strength of 
the continuum source is a further indication that some of the diffuse OH gas 
\citep[e.g.,][]{DawsonWalsh:2014aa} remains undetected for this group of sources.

\smallskip
\noindent
As diffuse clouds are typically found to have low optical depths \citep[e.g.,][]{LisztLucas:1996aa}, 
we are therefore biased towards higher column densities. As comparison, 
according to \citet{DickeyCrovisier:1981aa} using the
Nancay telescope at $3\farcm5$ resolution, OH optical depths in diffuse clouds
have been found to be approximately 0.05 in a 1\us\kms\ channel in the 1667{\us}MHz
transition. Higher optical depths were found by, e.g., \citet{Goss:1968aa}, \citet{Yusef-ZadehWardle:2003aa} or \citet{StanimirovicWeisberg:2003aa}. OH gas was
associated with the Galactic continuum sources, \ion{H}{ii} regions or
supernova remnants (SNR). The detections presented here match more with the latter categories. 

\subsection{OH as tracer of hydrogen gas}
In section~\ref{sec:ohabundance}, we compared the OH abundance to the
column densities of molecular hydrogen and hydrogen nuclei (hydrogen atoms and
molecules). These comparisons are shown in Figs.~\ref{fig:plot_ohbynh2_vs_nh2} and \ref{fig:plot_ohbynh_vs_nh}. 
OH abundance is found to be decreasing with increasing hydrogen column
density. The OH column density is not directly proportional to molecular column density. 
Therefore, the \ce{OH} columns densities span a smaller dynamic range than molecular hydrogen. 
This also indicates that OH traces only specific ranges of molecular cloud column densities. 

\smallskip
\noindent
At any given hydrogen column density, the OH abundance shows variations of a factor of two, which is within the systematic uncertainties. 
The median value of the OH abundance with respect to $N_{\rm H_2}$ is
$1.3\times10^{-7}$. Within
the systematic uncertainties of a factor of 4, this is in agreement with the
values reported in the literature, $N_{\rm OH}$/$N_{\rm H_2} = 1\times10^{-7}$
\citep[e.g.,][]{LisztLucas:2002aa}.

\smallskip
\noindent
A constant OH abundance with respect to $N_{\rm H}$, as reported by, e.g., \citet[][]{Crutcher:1979aa}, can be
reproduced - albeit with large scatter - for OH absorption with visual extinctions below $A_V\approx 10-20$. 
A median abundance of $X_{\rm OH} (N_{\rm H}) \approx 6.1\times10^{-8}$ is found for $A_V<20$. 
We include atomic hydrogen, as OH may be present in transition
regions that contain significant amounts of atomic hydrogen \citep[e.g.,][]{XuLi:2016aa,TangLi:2017ab}.   
The \ion{H}{i} column density affects the OH abundances at the lowest
molecular hydrogen column densities probed, when both are of similar strength. $X_{\rm OH}$ may even be lower 
in this regime, since the $N_{\rm \ion{H}{i}}$ measurements are lower limits and the $N_{\rm H_2}$ column densities may be underestimated, if the ``CO-dark'' gas fraction is significant (see Sect.~\ref{sec:systematics}).

\smallskip
\noindent
\citet{Crutcher:1979aa} finds an abundance of $X_{\rm OH} = 4.0\times10^{-8}$,
in a range of visual extinction of $A_V=0.4-7$ \citep[see also review
by][]{HeilesGoodman:1993aa}, which is within the errors of our results. 
These results are based on studies of nearby molecular clouds (Perseus, Ophiuchus and Taurus),
the SNR W44 and line-of-sight observations against extragalactic continuum
sources, therefore mainly including observations towards diffuse
molecular/translucent clouds \citep[e.g.,][]{SnowMcCall:2006aa} that probably have environments similar to those of
clouds that are not associated with \ion{H}{ii} regions in the sample
presented here. 

\smallskip
\noindent
Above visual extinctions of $A_V\approx 10-20$, some OH abundances are found to be lower than the literature abundance, which is in agreement with theoretical predictions \citep[e.g.,][]{HeilesGoodman:1993aa}. Oxygen to form OH at these extinctions is likely to be removed from the gas phase by the formation of CO and through the formation of water and its subsequent freeze-out onto grains. Far ultraviolet (FUV) radiation may counteract this removal: Models of photon dominated regions (PDRs) indicate that the local abundance of \ce{OH} peaks between cloud depths of $A_V\approx 3-7$ \citep[e.g.,][]{HollenbachKaufman:2009aa,HollenbachKaufman:2012aa}. According to \citet{HollenbachKaufman:2009aa}, the abundance of water in these regions depends on the photodesorption of water ice from dust grains, and OH forms by photodissociation of water in the gas phase. Both water and hydroxyl gas phase abundances thus depend on the flux of the far ultraviolet (FUV) radiation, and decrease once the FUV radiation is efficiently attenuated deeper inside the cloud. As $N_{\rm OH}$ in this work represents a line-of-sight averaged density, OH from more embedded regions in the molecular cloud may contribute less to $N_{\rm OH}$ than \ce{^{13} CO} does to $N_{\rm H_2}$, which may yield a decrease in the line-of-sight averaged OH abundance.

\smallskip
\noindent
Another possibility for the low OH abundances at high visual extinctions is that the \ce{OH} excitation temperatures
could be higher, approaching the kinetic temperatures in denser and warmer regions of the star forming molecular clouds in our sample. 
An excitation temperature of, e.g., 20\,K would place most
of the lowest measured OH abundances at $X_{\rm OH} \sim 1\times10^{-7}$ in Fig.~\ref{fig:plot_ohbynh2_vs_nh2}.
As many OH abundances at lower $N_{\rm H_2}$ lie above this value, 
the trend in Fig.~\ref{fig:plot_ohbynh2_vs_nh2} is likely to persist but to be less steep 
if higher excitation temperatures at higher $N_{\rm H_2}$ were assumed. 
This effect is difficult to assess from our data alone, as the
excitation temperature cannot be determined independently of the optical
depth. Hence, more detailed modelling or targeted observations would be necessary to
resolve this ambiguity.

\subsection{Comparison with OH column density measurements from other transitions}
In this section, we briefly discuss results on the OH column density that had been inferred from observations of other \ce{OH} transitions. Section~\ref{sec:satellite} described the morphologies of the satellite line transitions inside the OH ground state. In three regions, reversal of the 1612{\us}MHz transition from absorption to emission and of the 1720{\us}MHz transition from emission to absorption has been seen. As discussed in Section~\ref{sec:satellite}, the column density at the transition velocity was inferred using modeling results from \citet{van-Langeveldevan-Dishoeck:1995aa}. The column densities appear to be by a factor of 3-4 higher than the value inferred from the main lines. As an excitation temperature of 5{\us}K was assumed for the OH ground state transitions, this discrepancy could be remedied by assuming an excitation temperature of $15-20${\us}K, where values up to $\sim$15\,K have been found also in previous works \citep{ColganSalpeter:1989aa}.

\smallskip
\noindent
Additionally, rotational transitions in the far infrared wavelength regime \citep[e.g.,][]{WiesemeyerGusten:2012aa,CsengeriMenten:2012aa} or electronic transitions in the optical regime \citep[e.g.,][]{WeselakGalazutdinov:2010aa} can be used to study the \ce{OH} column density. The crossmatch with our sample yields a match only for G49.488$-$0.380 with the source W51e2 in \citet{WiesemeyerGusten:2016aa}. However, the column density at velocities at which OH is detected in this work, is not reported, as the ${}^{2}\Pi_{3/2}\,J=5/2\leftarrow3/2$ transitions at 2.5\,THz saturate between $50-80$\,\kms. 

\begin{figure}
 \centering
 \includegraphics[width=0.99\columnwidth]{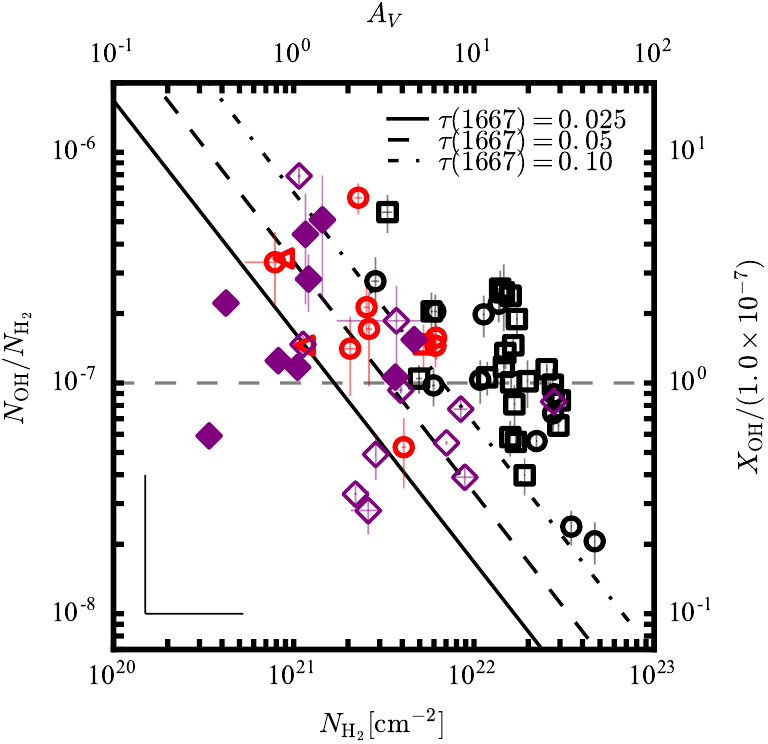}
 \caption{Comparison of the \ce{OH} molecular abundances from this work to abundances measured in \citet{WiesemeyerGusten:2016aa}. Data, symbols and systematic uncertainties as in Fig.~\ref{fig:plot_ohbynh2_vs_nh2}. Measurements from table 4 of \citet{WiesemeyerGusten:2016aa} are overplotted as purple diamonds. Filled diamonds use HF as proxy for the \ce{H2} column density. Empty diamonds use CH. If both HF and CH measurements are available, OH measurements are drawn twice. Black diagonal lines indicate sensitivity limits of $X_{\rm OH}$ for typical 4-$\sigma$ limits on optical depth in the OH main lines and assuming a line width of 2.5\us\kms\ (The detection limits of $\tau=0.025,0.05$ and $0.1$ are acchieved for continuum sources stronger than $F_{\rm cont} = 2.10, 1.05$ and $0.53\us$\jyb, respectively).}
 \label{fig:plot_ohbynh2_vs_nh2_wiesemeyer}
\end{figure}

\smallskip
\noindent
The \ce{OH} abundances determined here agree within our systematic uncertainty with the abundances inferred from optical and infrared transitions \citep{WeselakGalazutdinov:2010aa,WiesemeyerGusten:2016aa}. Figure~\ref{fig:plot_ohbynh2_vs_nh2_wiesemeyer} compares the OH abundances determined here with OH abundances derived from the THz transitions at different lines of sight by \citet{WiesemeyerGusten:2016aa}. Within the sensitivity limits of our survey, for abundances using HF as tracer of \ce{H2}, good agreement is seen between both datasets. This is also true for abundances using CH as tracer of \ce{H2}, although some points are present between $N_{\rm H_2}\approx 6-10\times10^{21}\,{\rm cm}^{-2}$, which show lower abundance, but are still within uncertainties. Also, for low \ce{H2} column densities, our measurements appear to be sensitivity limited (some measurements from \citet{WiesemeyerGusten:2016aa} fall below the sensitivity limits indicated in Fig.~\ref{fig:plot_ohbynh2_vs_nh2_wiesemeyer}). This comparison affirms the conclusion from Sect.~\ref{sec:discussion_distribution} that the sample of OH absorption presented here indeed is biased towards high $N_{\rm OH}$. Conversely, some continuum sources are strong enough to reveal OH absorption in more diffuse molecular cloud regions. Albeit the large systematic uncertainties of $X_{\rm OH}$ here, Fig.~\ref{fig:plot_ohbynh2_vs_nh2_wiesemeyer} shows that the variations in OH abundance at any given $N_{\rm H_2}$ persist when using alternative methods to measure $N_{\rm OH}$ and $N_{\rm H_2}$, which are possibly less prone to systematics. 
 
\subsection{Extended OH absorption towards W43}\label{sec:discussion_w43}
As W43 is structured in a complicated way, and \ce{OH} chemistry and
excitation may vary strongly in different environments, there may be multiple
explanations for variations in the ratios of OH optical depth to \ce{C^{18}O} and 
870\,$\mu$m emission. For example, there are temperature gradients present 
in the entire region, which may affect the dust emission. 

\smallskip
\noindent
The peak optical depth of the 1667 MHz transition is typically at $\tau<1$ in W43. Exceptions are MM1 and MM3, for which optical depth peaks of $\tau_{1667}\sim1.2$ indicate that the line becomes optically thick (Fig.~\ref{fig:plot_spectra_w43}). For MM1, the 1665 MHz transition peaks at $\tau_{1665}\sim0.6$. Therefore, the ratio of the main lines ($\tau_{1667}/\tau_{1665} \approx 2.0$) is within errors of the expected ratio for LTE excitation of 1.8. For MM3, the ratio of the main lines is closer to unity with $\tau_{1665}\sim1.0$, indicating that the transitions are not in LTE. As we have no probe of the excitation temperatures, this cannot be further assessed here. We note, however, that deviations from LTE in the OH hyperfine ground state transitions appear to be a common phenomenon \citep[e.g.,][]{LiTang:2018aa}. 

\smallskip
\noindent
In order to minimize the chance of the CO tracers to become optically thick, we choose data from the \ce{C^{18}O}(2-1) transition. 
Optical thickness even of this transition cannot be ruled out in the entire region, and it will trace regions at higher densities than the \ce{^{13}CO}(1-0) used in the rest of the analysis. However, in order to investigate variations on a 20\arcsec\ scale (0.5{\us}pc) in W43, these data provide information about the molecular gas at matching spatial resolution. 
As the datasets presented here cannot constrain whether these variations are due to 
abundance variations or differences in the excitation behavior, we limit the discussion 
to a qualitative description of the results. 

\smallskip
\noindent
In Figure~\ref{fig:plot_channel_map_w43_tau_1667_by_c18o21_00}, the enhancement in the \ce{OH} to \ce{C^18O} ratio 
around 82.5\us\kms\ coincides with a peak in radio recombination lines 
(see velocity distribution of H92$\alpha$ in fig.~6 of \citealt{BalserGoss:2001aa} or of cm-RRLs from THOR in fig.~2 of \citealt{Nguyen-LuongAnderson:2017aa}). 
At this velocity, possibly a photon-dominated, partly ionized region of the cloud is
seen, in which models of photon-dominated regions predict the peak of the
\ce{OH} abundance \citep[e.g.,][]{HollenbachKaufman:2012aa}. 

\smallskip
\noindent
The \ce{OH} 1667\,MHz to \ce{C^18O} ratios in the center of W43-main at 90$-$93\,\kms\
are found to be lower ($\le0.1$; Fig.~\ref{fig:plot_channel_map_w43_tau_1667_by_c18o21_00}, bottom left panel). 
This may be due to different physical or chemical conditions 
of the OH gas, which are difficult to disentangle in a crowded region like W43. Alternatively, 
this can be a consequence of systematically underestimating the OH optical depth. 
As described in Sect.~\ref{sec:systematics}, the measured continuum is higher than the 
true continuum incident on the OH gas, if \ion{H}{ii}-regions contribute to the continuum 
emission, which lie between the absorbing cloud and the observer. 

\smallskip
\noindent
This geometry is likely to be present here. The continuum emission at the Wolf-Rayet/OB cluster originates from 
the \ion{H}{ii}-region, which emits RRLs between 80$-$90\,\kms. At the T-bar, the 
RRLs peak between 90$-$100\,\kms, or even at higher velocities close to MM1.
Since \ion{H}{ii}-regions are expanding, gas at lower velocities may be 
located closer to the observer than higher velocities gas. 
Absorbing OH gas at 90$-$100\,\kms\ would therefore lie between 
two \ion{H}{ii}-regions and its measured optical depth can be lower than the true value. 

\smallskip
\noindent
The distribution of absorption in OH and emission in \ce{C^18O} is in agreement with this scenario.
In Figure~\ref{fig:plot_spectra_w43}, we show spectra of both molecules at different positions. 
Towards the central Wolf-Rayet cluster, only absorption at 82\,\kms\ is present, while not detected 
between 90$-$95\,\kms\ in spite of the presence of \ce{C^18O} emission. 
Towards the center of the T-bar, absorption becomes visible in both velocity ranges. 
How much and at which lines of sight the optical depth is influenced by this effect, 
depends on the fractional contribution of each \ion{H}{ii}-region to the continuum emission. 
A quantitative assessment of this is beyond the scope of this work. 

\smallskip
\noindent
Close to MM1, the peak of the \ce{OH} absorption occurs between 96 and
97.5\us\kms. The average velocity of the MM1 complex has been found to be at
98\us\kms\ \citep{Nguyen-LuongMotte:2013aa} and more resolved observation in
HCN(1-0) and SiO(2-1) emission show peaks between 97 and 94\us\kms, respectively,
when going from MM1 towards the center of W43-main
\citep{LouvetMotte:2016aa}. Louvet et al. indicate low velocity
shocks in this region. While this needs to be confirmed, 
such shocks may produce temperatures that can enhance the \ce{OH}
abundance by activating neutral-neutral chemistry
\citep{NeufeldKaufman:2002aa}. However, the presence of additional, enhanced UV radiation 
may be required to produce OH, as neutral-neutral chemistry at high temperatures typically leads to water production
\citep[e.g.,][]{van-DishoeckHerbst:2013aa}. Near MM3 we see emission over a
large range of velocities, with a clear peak in optical depth at 93\us\kms\ in
the 1665{\us}MHz line, in agreement with \ce{N2H+} and \ce{SiO} peaks at the
same velocity \citep{Nguyen-LuongMotte:2013aa}. 

\section{Conclusions}    \label{sec:conclusion}
This work gives an overview on the \ce{OH} absorption against
strong continuum background sources as inferred from the THOR survey. This is the first survey-style analysis
with the VLA over a significant fraction of the inner Milky
Way in the range between $l=15\degree$ and $l=67\degree$. We detect 59 distinct absorption features
against 42 continuum background sources. Most of the absorption is found
against Galactic \ion{H}{ii}~regions. We discuss the detection limit in terms
of the continuum source strengths.

\smallskip
\noindent
\begin{itemize}
\item
Using \ce{{}^{13}CO}(1-0) as tracer for $N_{\rm H_2}$, we compare the OH
abundance ($N_{\rm OH}/N_{\rm H_2}$) at different $N_{\rm H_2}$. The OH abundance decreases
with increasing hydrogen column density, especially for OH detections in
molecular clouds that are associated with \ion{H}{ii} regions. This can be due
to probing cloud regions where the OH in the gas phase is significantly
depleted, although varying excitation conditions may provide an alternative
explanation. The median abundance is found at $N_{\rm OH}/N_{\rm H_2} \sim 1.3\times10^{-7}$, in
agreement within errors with previous studies.
\item
At low column densities, the atomic hydrogen fraction of the gas along the line-of-sight 
becomes comparable to molecular hydrogen. The OH abundance ($N_{\rm OH}/N_{\rm H}$) is found to decrease with increasing total hydrogen nucleus column density for $A_v$ > 20, but for lower extinction lines of sight, the data are consistent with a constant abundance having median value $N_{\rm OH}/N_{\rm H}\sim 6.1\times10^{-8}$.
\item
Extended \ce{OH} absorption is seen against W43. The OH absorption is compared
to ancillary data of 870\,$\mu$m and \ce{C^{18}O} emission. At an angular 
resolution of ${20\arcsec\times20\arcsec}$, we find variation in
the ratios of OH optical depth to emission in 870\,$\mu$m and \ce{C^{18}O},
especially towards mm emission sources in the region. 
\end{itemize}

\smallskip
\noindent
Studies of OH provide a unique insight into the physical conditions of the ISM, particularly the transition between diffuse gas and molecular clouds. This first unbiased interferometric survey is a contribution to the characterization of the variation of OH absorption properties throughout the Galaxy. 
This work may provide a starting point for theoretical and observational
follow-up studies with deeper observations at higher velocity
resolution, to expand the sample towards fainter sources with narrower
line widths, and in combination with other observational data to resolve the
physical conditions of the OH gas, and the molecular content of the diffuse gas surrounding molecular clouds. 

\begin{acknowledgements}
We thank Joanne R. Dawson for her very useful and insightful comments that helped to improve the clarity of this paper.

M.R.R. and H.B. are very grateful for the helpful discussions with C. M. Walmsley at the beginning of the project and for detailed comments on early versions of the manuscript. We thank Morgan Fouesneau for his support on the statistical investigation of the correlation between $N_{\rm OH}$ and $N_{\rm H_2}$.

M.R.R. is a fellow of the International Max Planck Research School
for Astronomy and Cosmic Physics (IMPRS) at the University of Heidelberg.

M.R.R., H.B., Y.W. and J.S. acknowledge support from the European Research Council under the Horizon
2020 Framework Program via the ERC Consolidator Grant CSF-648505. 

S.E.R. acknowledges support from the European Union's Horizon 2020 research and innovation programme under the Marie Sk{\l}odowska-Curie grant agreement \# 706390.

F.B. acknowledges funding from the European Union's Horizon 2020 research and innovation programme (grant agreement No 726384 - EMPIRE).

J.K. acknowledges funding from the European Union's Horizon 2020 research and innovation programme under grant agreement No~639459 (PROMISE).

R.S.K., S.C.O.G., M.R.R. and H.B. acknowledge support from the Deutsche Forschungsgemeinschaft in the Collaborative Research Center (SFB 881) ``The Milky Way System'' (subprojects B1, B2, and B8). R.S.K. and S.C.O.G. acknowledge support from the Deutsche Forschungsgemeinschaft in the Priority Program SPP 1573 ``Physics of the Interstellar Medium'' (grant numbers KL 1358/18.1, KL 1358/19.2, GL 668/2-1). R.S.K. furthermore thanks the European Research Council for funding in the ERC Advanced Grant STARLIGHT (project number 339177). This work was carried out in part at the Jet Propulsion Laboratory, which is operated for NASA by the California Institute of Technology. 

N.R. acknowledges support from the Infosys Foundation through the Infosys Young Investigator grant.

N.S. acknowledges support by the French ANR and the German DFG through 
the project ``GENESIS'' (ANR-16-CE92-0035-01/DFG1591/2-1).

RJS gratefully acknowledges support through an STFC Ernest Rutherford Fellowship.

This research made use of Astropy and affiliated packages, a community-developed core Python package for Astronomy \citep{Astropy-CollaborationRobitaille:2013aa} and the VizieR catalogue access tool, CDS, Strasbourg, France.
\end{acknowledgements}

\bibliographystyle{aa}
\bibliography{bibliography}

\clearpage
\begin{appendix}
\section{Detection of \ce{OH} main line absorption -- notes on individual sources} \label{app:detections}
\begin{itemize}
\item G23.956+0.150, +81.0\us\kms: One velocity channel is above $4$-$\sigma$ and two above
$3$-$\sigma$. There is a corresponding \ce{^{13}CO}(1-0) counterpart. 
\item G26.609$-$0.212, $-$33.0\us\kms: One velocity channel is detected at $4$-$\sigma$, another at
$3$-$\sigma$. The \ce{^{13}CO}(1-0) data do not cover these velocities. 
\item G28.806+0.174, +79.5\us\kms: One velocity channel is detected at $4$-$\sigma$, a \ce{^{13}CO}(1-0) counterpart exists
that matches well in velocity.
\item G28.806+0.174, +103.0\us\kms: One velocity channel is detected at $4$-$\sigma$, a \ce{^{13}CO}(1-0) counterpart exists that contains blended components. 
\item G29.935$-$0.053, +51.0\us\kms: This object is detected at $4$-$\sigma$. Also, a
 \ce{^{13}CO}(1-0) counterpart exists. 
\item G29.935$-$0.053, +7.5\us\kms: This object is detected in one velocity channel at 3-$\sigma$ and in one at $5$-$\sigma$ after smoothing to 46\arcsec\ resolution. The detection
is not picked up in the 1665{\us}MHz transition. There is no \ce{^{13}CO}(1-0) emission at this velocity. 
\item G30.535+0.021, +45.0\us\kms: This feature shows one velocity channel at 4-$\sigma$ and
 three at 3-$\sigma$ in the 1667{\us}MHz transition. We compare this feature
 to \ce{^{13}CO}(1-0), after smoothing to 46\arcsec. The feature peaks at 43.5\us\kms. 
 In comparison, the \ce{^{13}CO}(1-0) shows two distinct peaks at 40.3
 and 47.3\us\kms. Interestingly, the trough between both
 features occurs at the position of the OH peak. The trough could be a sign of \ce{CO} self-absorption, i.e. the absorption by cold CO gas of the line emission from a warmer background source \citep[e.g.,][]{PhillipsKnapp:1981aa}. A possible heating source of the background CO gas could be the \ion{H}{ii}-region G30.539$-$00.024, which emits RRLs at 46\us\kms\ (Table~\ref{tbl:detections}; \citealt{AndersonBania:2014aa}).
\item G30.535+0.021, +92.0\us\kms: 
 This feature shows one velocity channel at 4-$\sigma$ and
 three at 3-$\sigma$ in the 1667{\us}MHz transition. The feature peaks at 92\us\kms\
 and we compare it to \ce{^{13}CO}(1-0) emission, after smoothing to 46\arcsec. The \ce{^{13}CO}(1-0) 
 seems to have blended components. 
 In the 1665{\us}MHz transition the feature is weakly detected (at $3$-$\sigma$). 
 However, it has a different shape to the 1667{\us}MHz transition -- most likely 
 due to its weak detection -- and so we do not include it for fitting the line width, 
 but instead merely use it for the column density comparison.
\item G32.272$-$0.226, +22.5\us\kms: In this feature, three velocity channels are detected at
 4-$\sigma$. There is a corresponding feature in \ce{^{13}CO}(1-0) that spreads
 over a similar velocity range but the line profile is not centrally peaked. This is confirmed by investigating the \ce{^{13}CO}(1-0) emission around
 this position, with the strongest individual peak being located at 22.7\us\kms. 
 The noise was checked a few arc minutes away from the
 emission and did not show any anomalies in this velocity range. 
\item G32.928+0.607, $-$34.5\us\kms: This feature contains one pixel at 3-$\sigma$ in the 1665{\us}MHz transition. No \ce{^{13}CO}(1-0) data
 are available for comparison at this velocity. 
\item G35.467+0.139, +78.0\us\kms: This feature consists of two velocity channels close to a signal-to-noise ratio of
 4-$\sigma$. There exists \ce{^{13}CO}(1-0) emission that peaks at a similar velocity. 
\item G60.882$-$0.132, +22.5\us\kms: \ce{OH} absorption in the 1667{\us}MHz transition is detected, with one
 velocity channel at 3-$\sigma$, 4-$\sigma$ and 5-$\sigma$, respectively. In the 1665{\us}MHz transition, one velocity bin is detected at
 5-$\sigma$ . Also, we find a \ce{^{13}CO}(1-0) counterpart at similar
 velocity. 
\end{itemize}

\section{Estimation of the correlation between $N_{\rm OH}$ and $N_{\rm H_2}$}
\begin{figure*}
 \centering
 \includegraphics[width=0.99\textwidth]{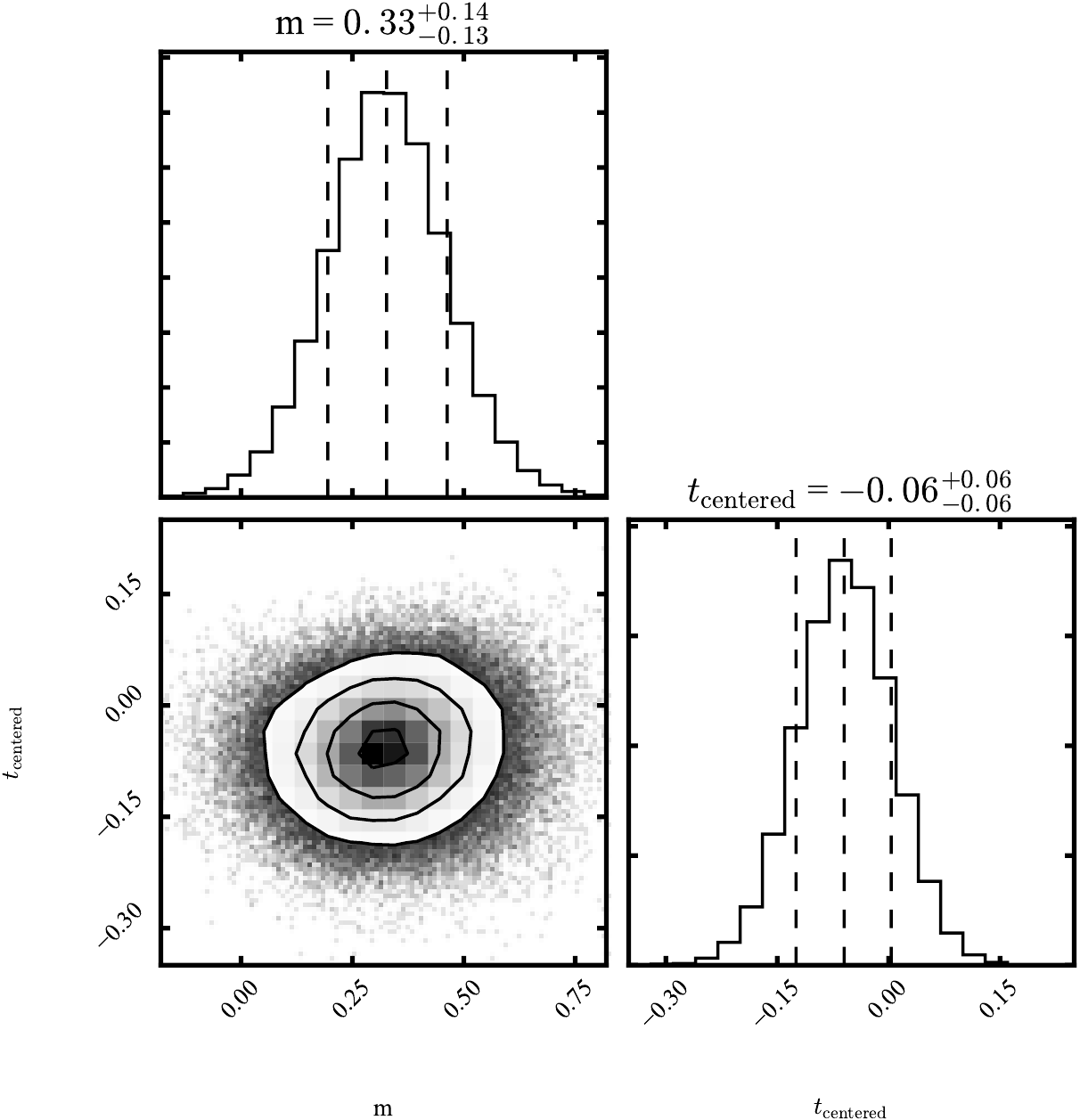}
 \caption{Marginalized distributions of $m$ and $t_{\rm centered}$. These are obtained after centering all datasets by normalizing with the mean $N_{\rm OH}$ and $N_{\rm H_2}$. The uncentered $t$ is given by \mbox{$t = t_{\rm centered}+\log(\overline{N_{\rm OH}}) - m\times\log(\overline{N_{\rm H_2}})$}, as \mbox{$ t= 7.91^{+2.86}_{-2.95}$}.}
 \label{fig:marginalized_distributions}
\end{figure*}

We provide additional information on the estimation of the correlation between $N_{\rm OH}$ and $N_{\rm H_2}$ from Sect.~\ref{sec:oh_vs_h2_coldens}. 
For numerical stability, we center all data on the mean of the measured column densities, $\overline{N_{\rm OH}}$ and $\overline{N_{\rm H_2}}$. We perform the linear regression on $\log(N_{\rm OH}/\overline{N_{\rm OH}})  = m\times\log(N_{\rm H_2}/\overline{N_{\rm H_2}})+t_{\rm centered}$. The parameter $t$ from Sect.~\ref{sec:oh_vs_h2_coldens} relates to the sampled $t_{\rm centered}$ as \mbox{$t = t_{\rm centered}+\log(\overline{N_{\rm OH}}) - m\times\log(\overline{N_{\rm H_2}})$}. The median, 16\%-, and 84\%-percentiles are \mbox{$m = 0.33^{+0.14}_{-0.13}$}, \mbox{$t_{\rm centered} = -0.06^{+0.06}_{-0.06}$} and \mbox{$t= 7.91^{+2.86}_{-2.95}$}. Fig.~\ref{fig:marginalized_distributions} shows the histogram of the marginalized distributions of $m$ and $t_{\rm centered}$. 

\section{Individual sources -- transitions of the OH ground state}
This appendix shows the spectra of the OH ground state transitions. 
\begin{figure*}
 \centering
 \includegraphics[width=.99\textwidth]{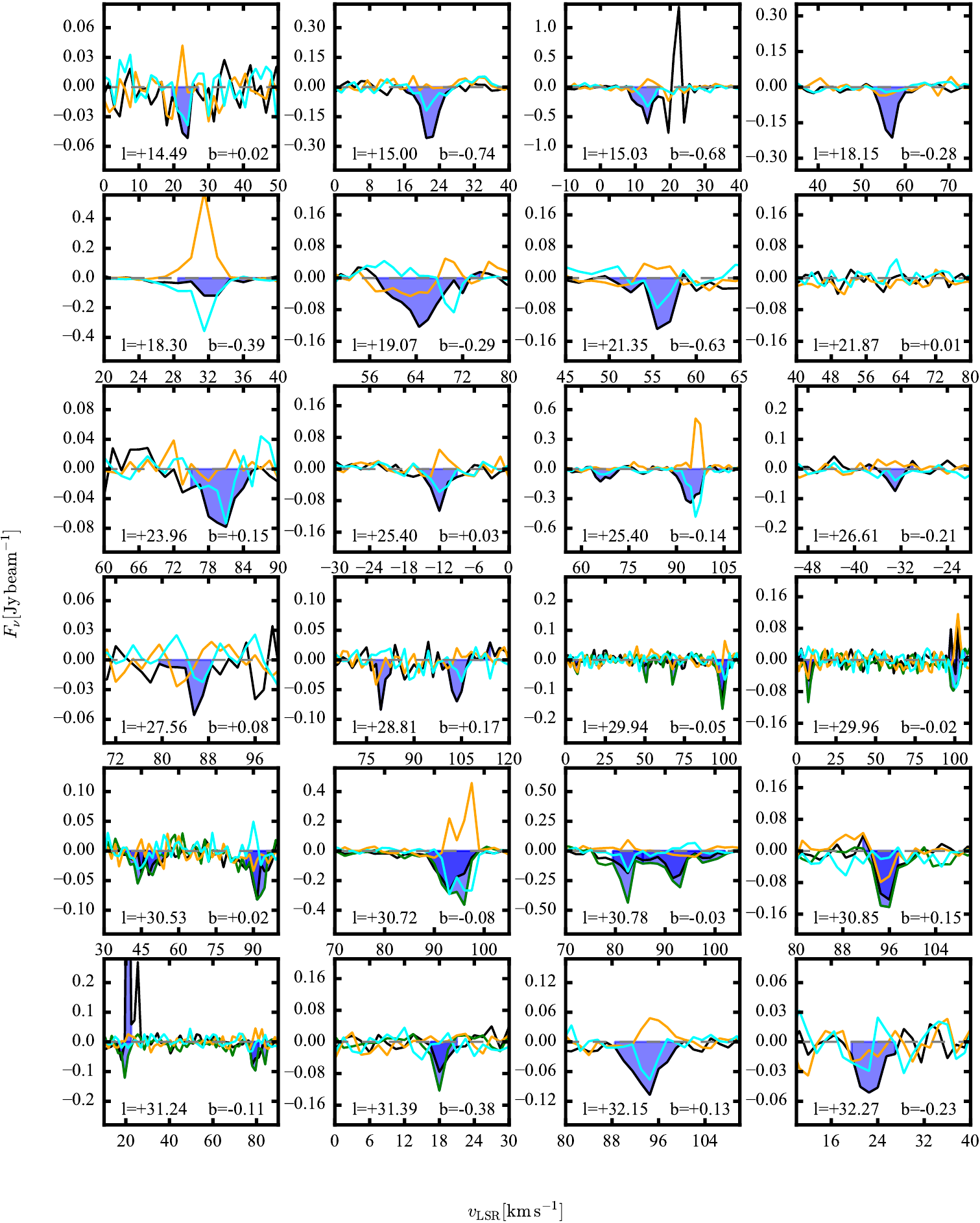}
 \caption{Spectra of OH ground state transitions, especially also of the satellite lines, along lines of
  sight that show a detection in 1665{\us}MHz or 1667{\us}MHz absorption. 
  The spectra are extracted from data cubes
  that have been smoothed to 46\arcsec\ resolution. The transitions at 
  1665{\us}MHz (black), 1612{\us}MHz (orange) and 1720{\us}MHz (cyan) are
  shown for all sources. The 1667{\us}MHz (green) transition is displayed if available. 
  OH absorption detections in the main line transitions are shaded in blue. 
 }
 \label{fig:plot_alltrans_00}
\end{figure*}
\begin{figure*} \centering \includegraphics[width=.99\textwidth]{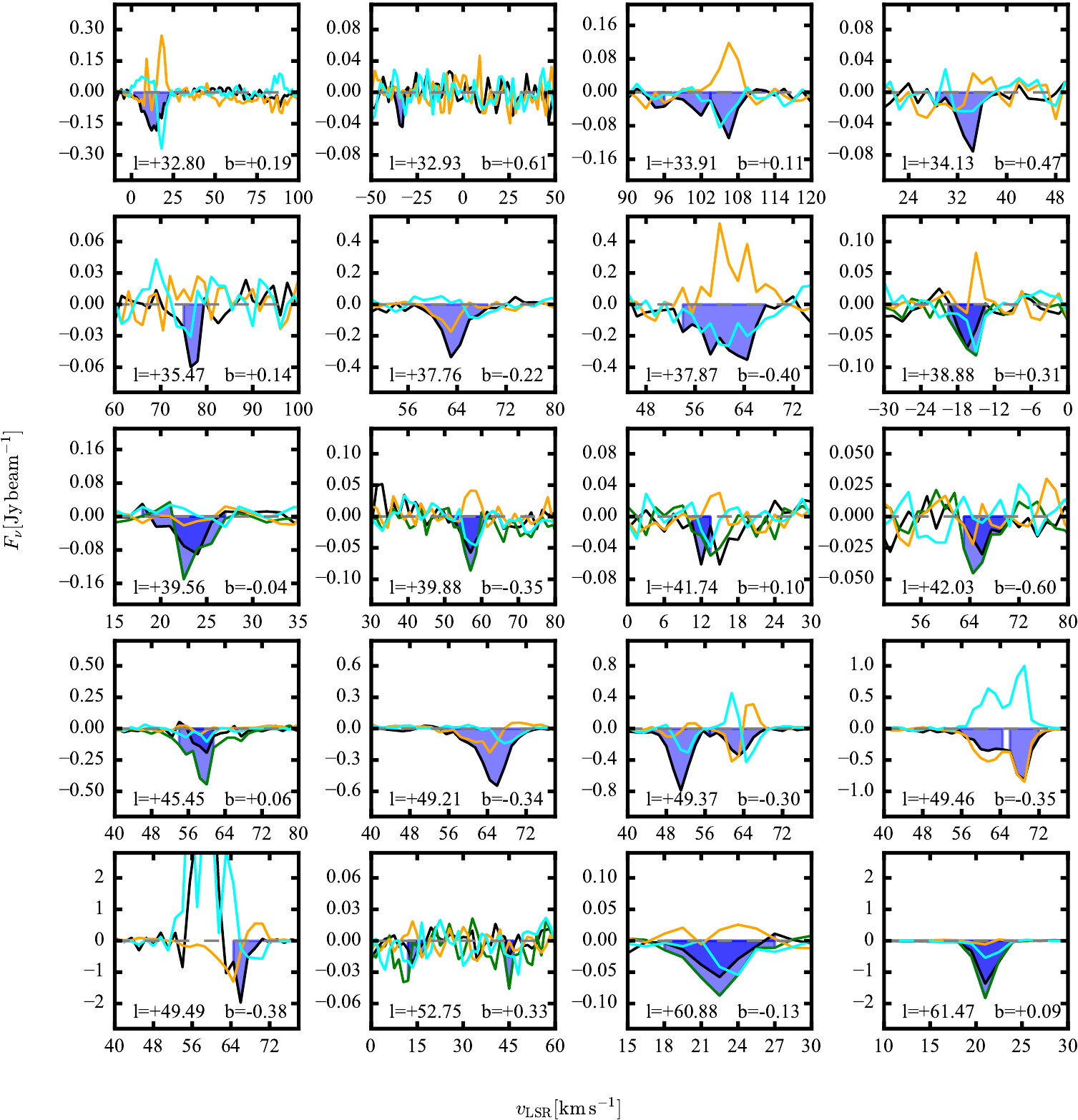}\caption{As Fig. \ref{fig:plot_alltrans_00}.}\label{fig:plot_alltrans_02}\end{figure*}

\section{Individual sources - OH and \ion{H}{I} optical depth and ${\rm {}^{13}CO(1-0)}$ emission}
The appendix shows the OH and \ion{H}{i} optical depth profiles for each source, as well as the \ce{^13CO}(1-0) emission.

\begin{figure*}
 \centering
 \includegraphics[width=.95\textwidth]{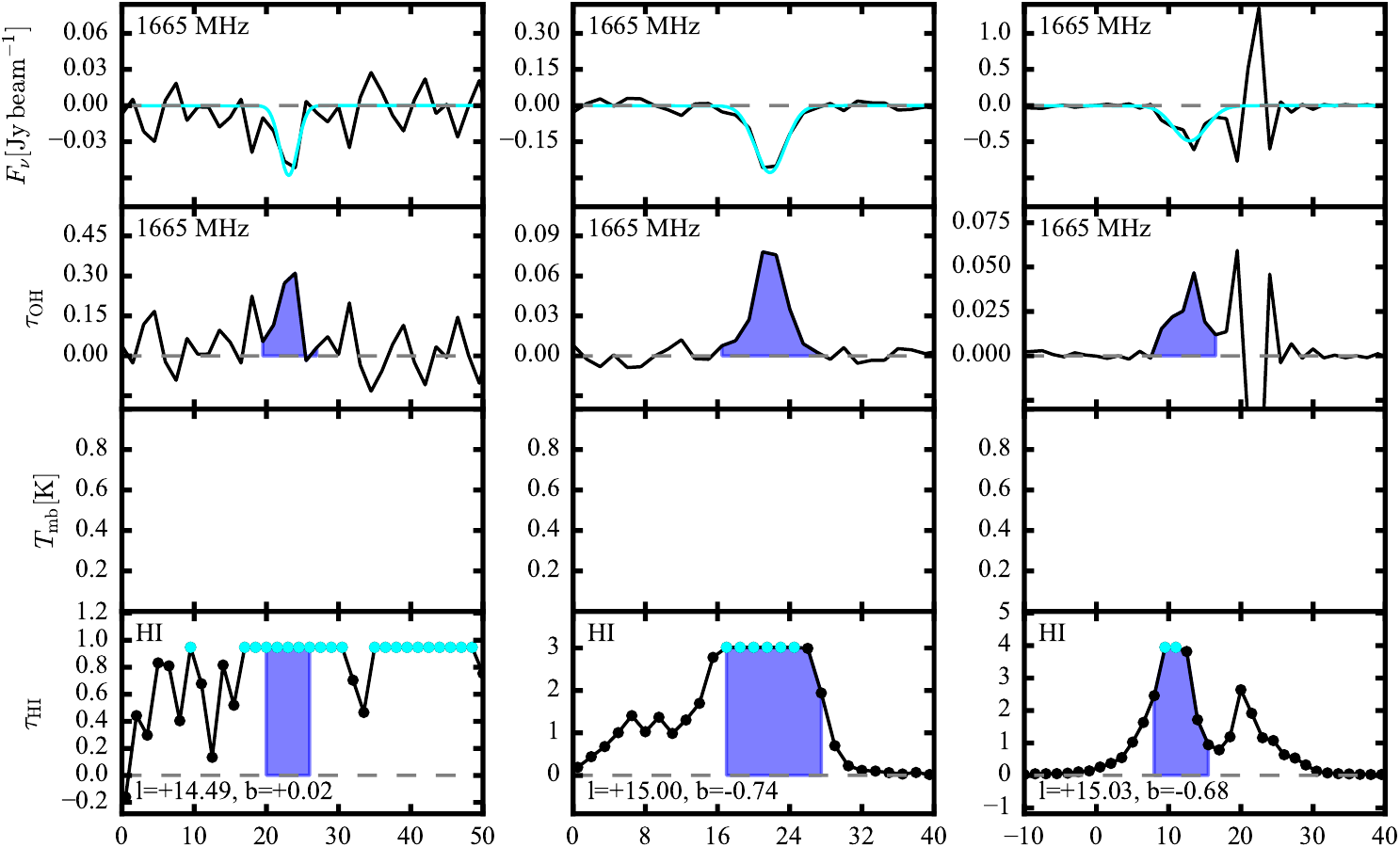}
 \includegraphics[width=.95\textwidth]{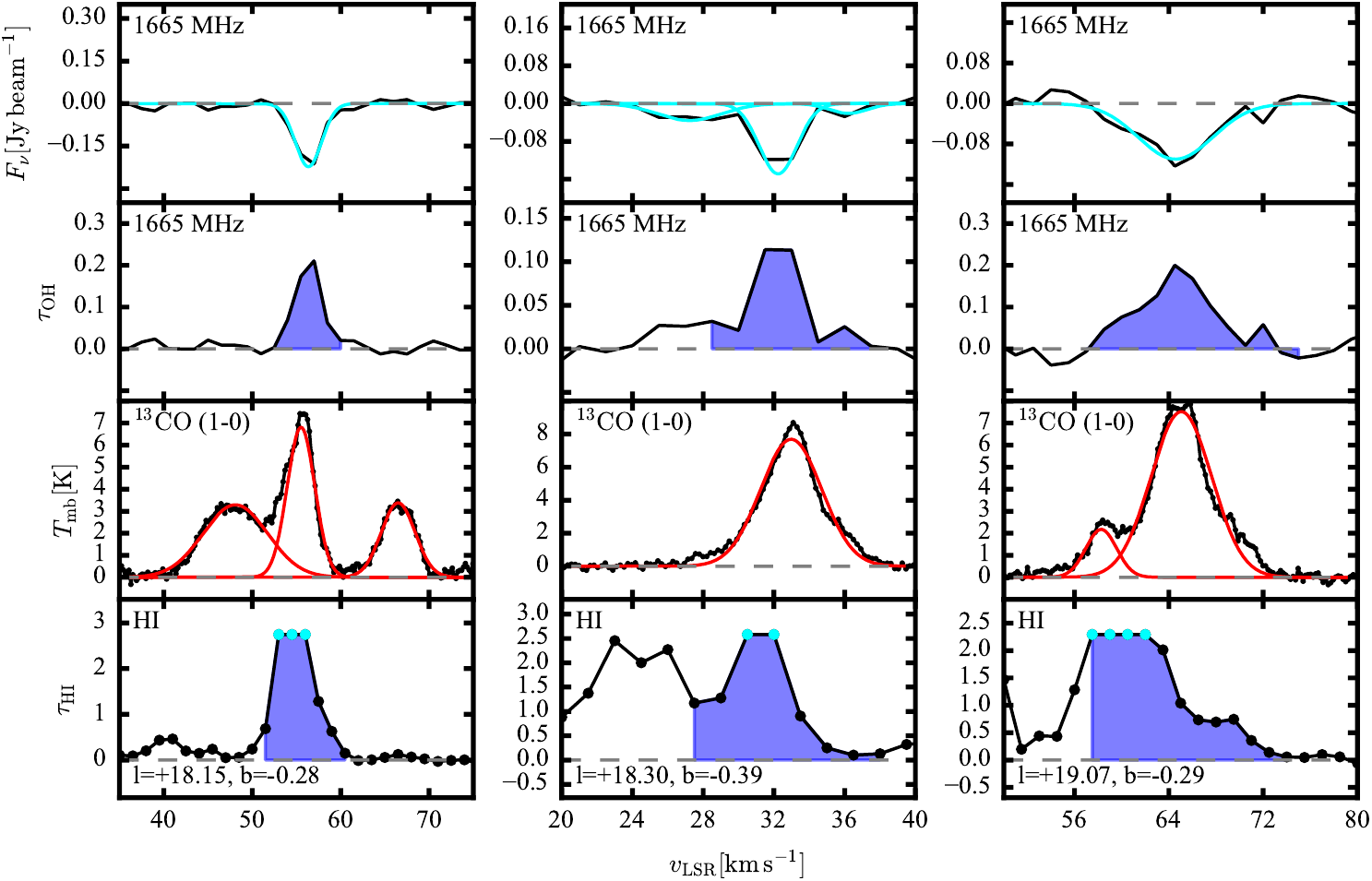}
 \caption{Analyzed 1665 and 1667\,MHz OH absorption features. The topmost panel shows 1665{\us}MHz~(black) and 1667{\us}MHz~(green) absorption features. The second panel from the top shows the spectrum converted to optical depth for the 1665{\us}MHz~(black) and 1667{\us}MHz~(green) transition. The third panel from the top shows ${}^{13}{\rm CO}$(1$-$0) emission in $T_{\rm mb}$ (data from the GRS survey ($l<60\degree$) and Exeter FCRAO CO survey ($l>60\degree$); \citealt{JacksonRathborne:2006aa}, \citealt{MottramBrunt:2010aa}). The line widths were determined by fitting Gaussian profiles to the 1665{\us}MHz~(cyan) and 1667{\us}MHz~(orange) absorption, and the ${}^{13}{\rm CO}$(1$-$0) emission (red). The lowermost panel shows the \ion{H}{i} absorption converted to \ion{H}{i} optical depth. Measured bins are denoted by black dots, while we quote lower limits (cyan) for saturated bins. The 1667\,MHz transition was observed only towards selected regions and is therefore shown only for a subset of lines-of-sight (see Sect.~\ref{sec:observations}). The line-of-sight coordinates are given in degrees of Galactic coordinates. The blue shaded area in the lower three panels shows the line integrals. All data were smoothed to a spatial resolution of 46\arcsec. 
 }
 \label{fig:plot_positions_00}
\end{figure*}

\begin{figure*}
  \centering \includegraphics[width=.95\textwidth]{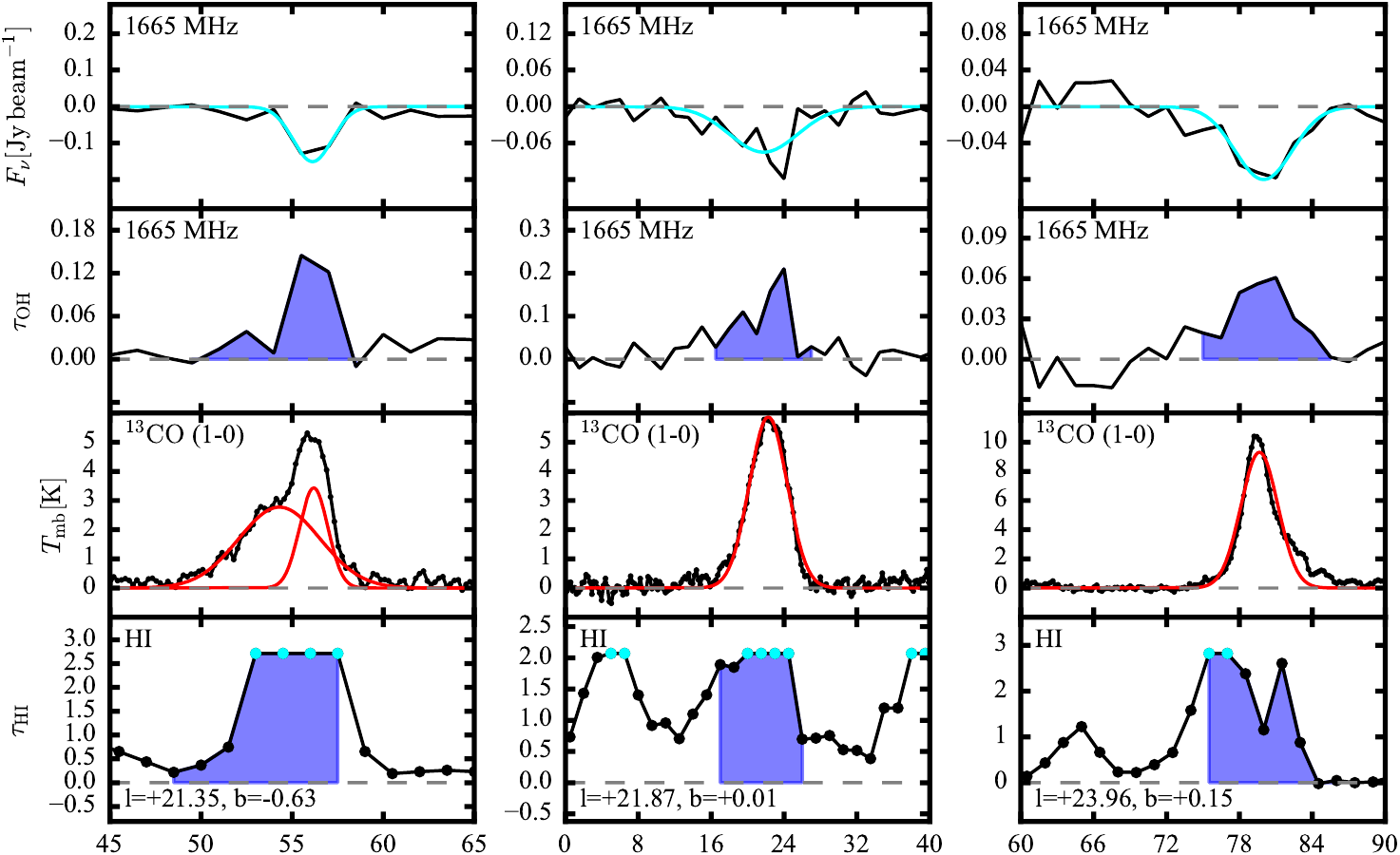}
\includegraphics[width=.95\textwidth]{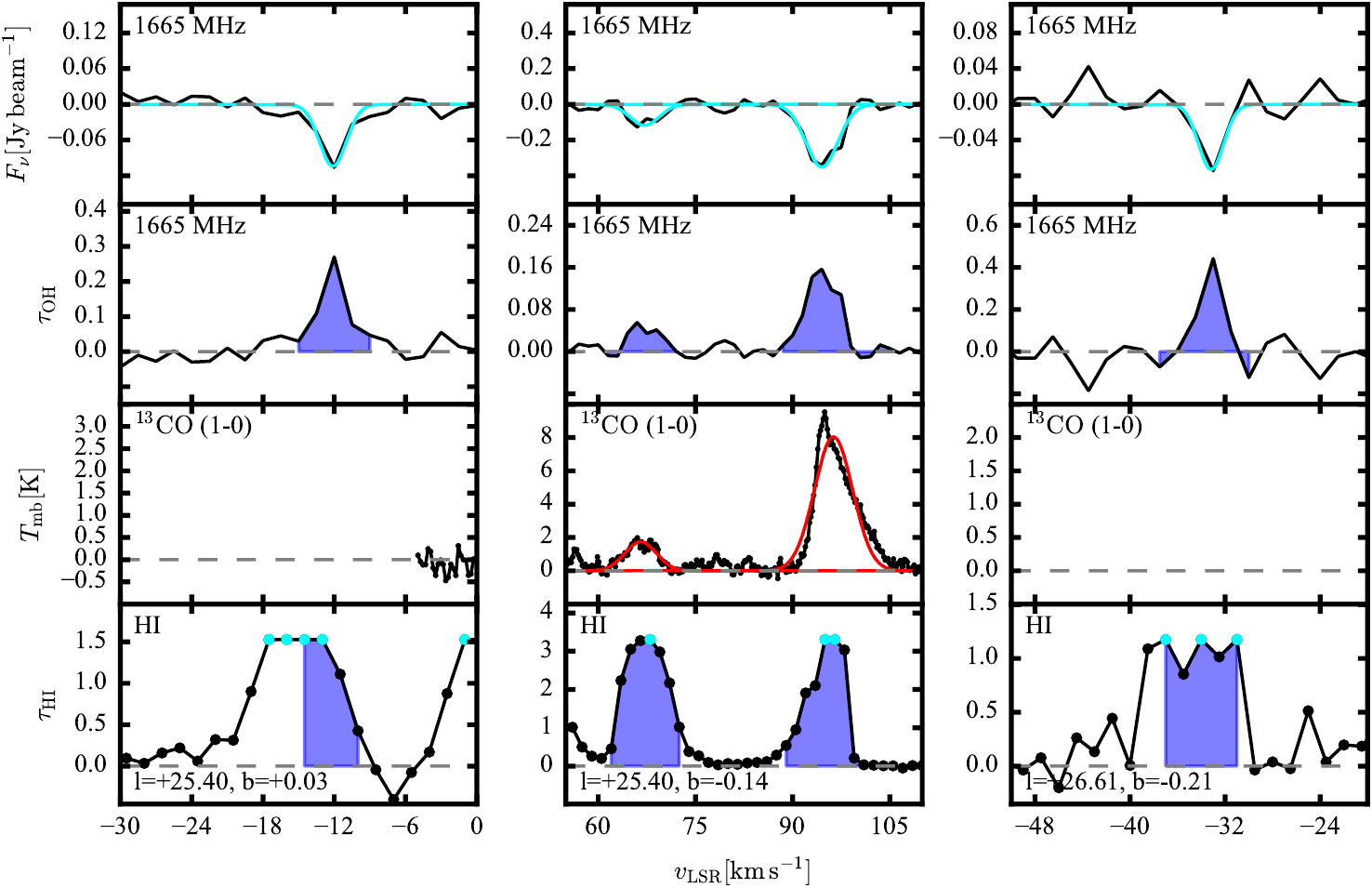}
\caption{As Fig. \ref{fig:plot_positions_00}.}\label{fig:plot_positions_03}\end{figure*}

\begin{figure*} \centering 
\includegraphics[width=.95\textwidth]{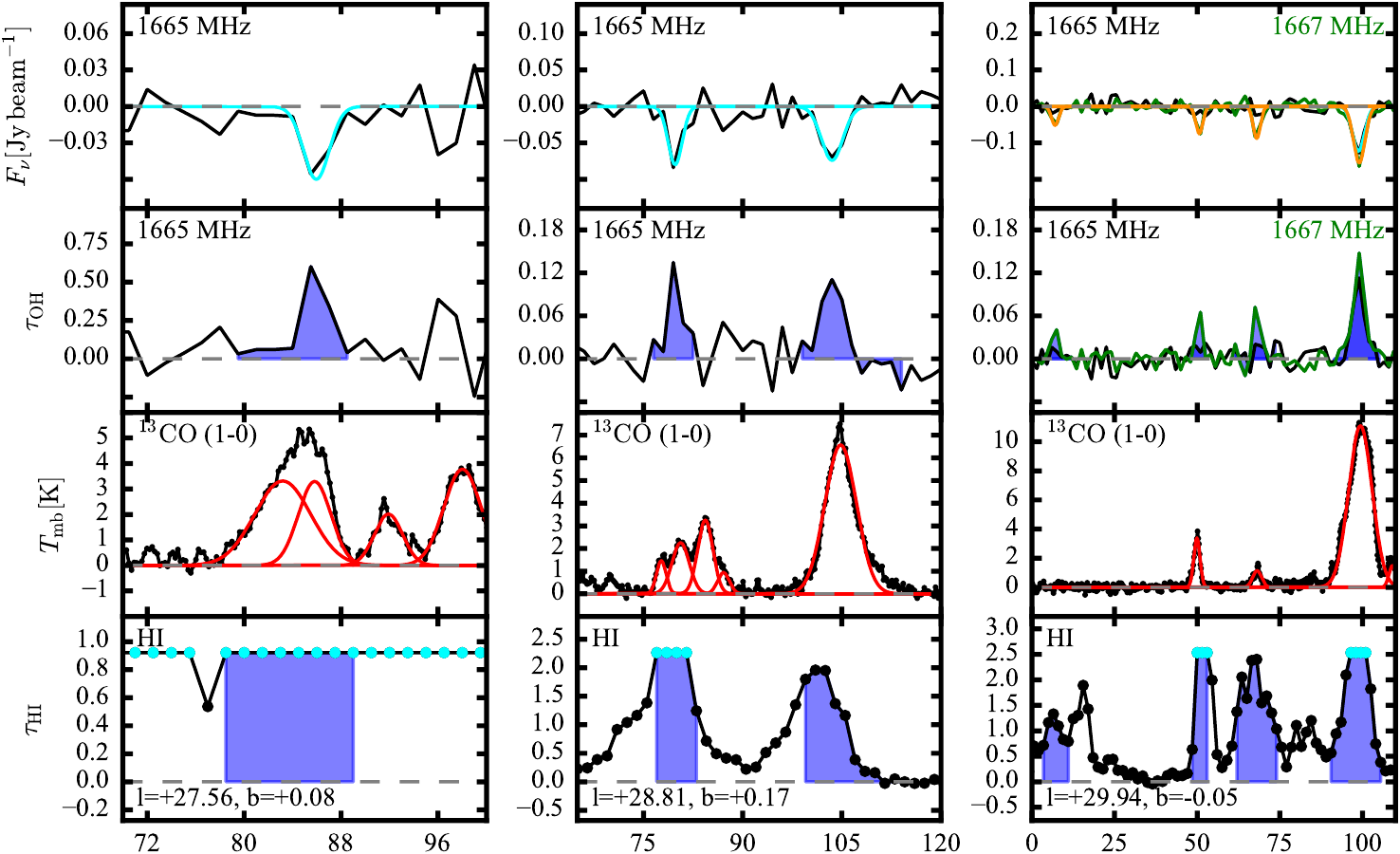}
\includegraphics[width=.95\textwidth]{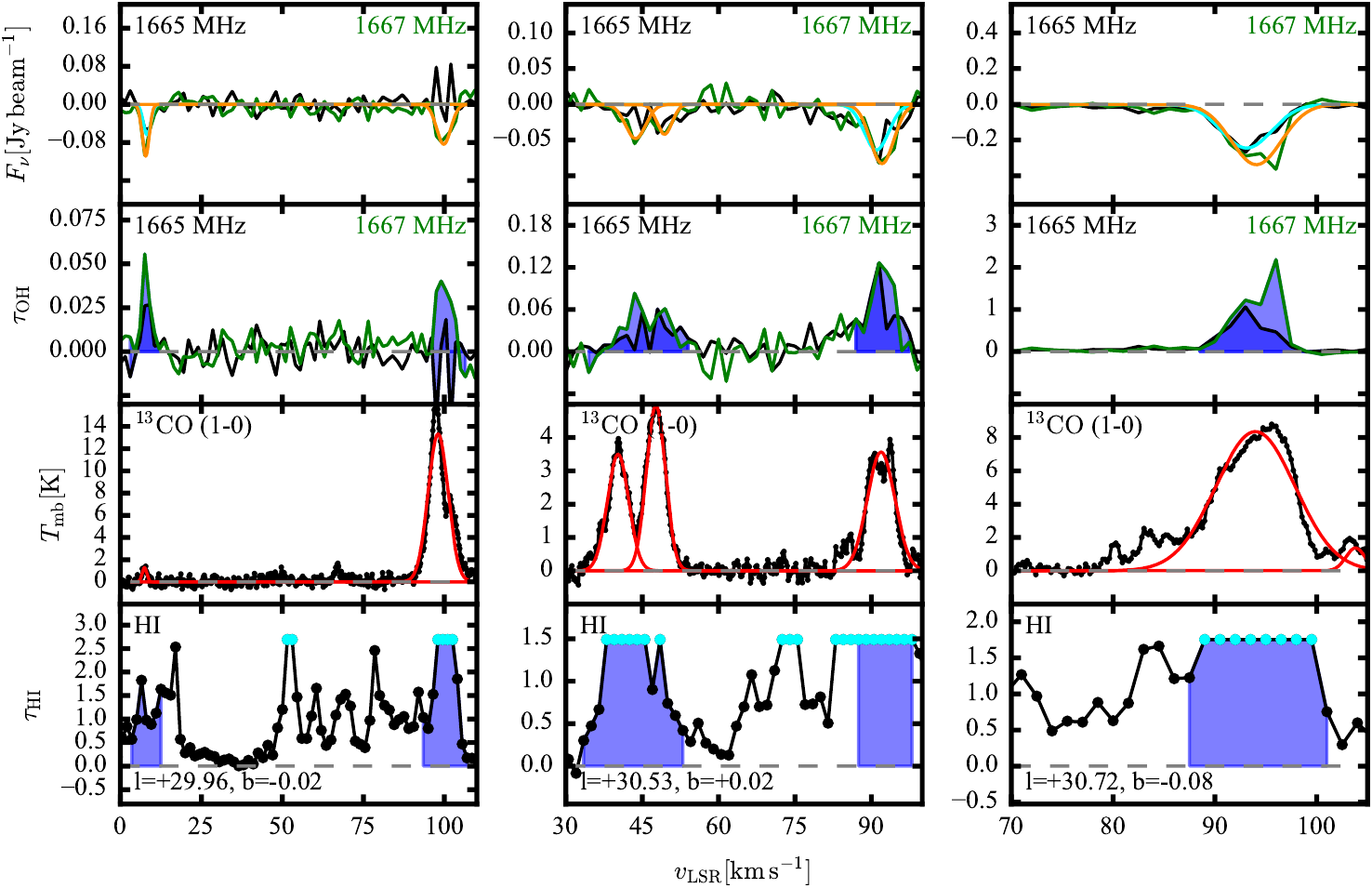}
\caption{As Fig. \ref{fig:plot_positions_00}.}\label{fig:plot_positions_04}\end{figure*}

\begin{figure*} \centering 
\includegraphics[width=.95\textwidth]{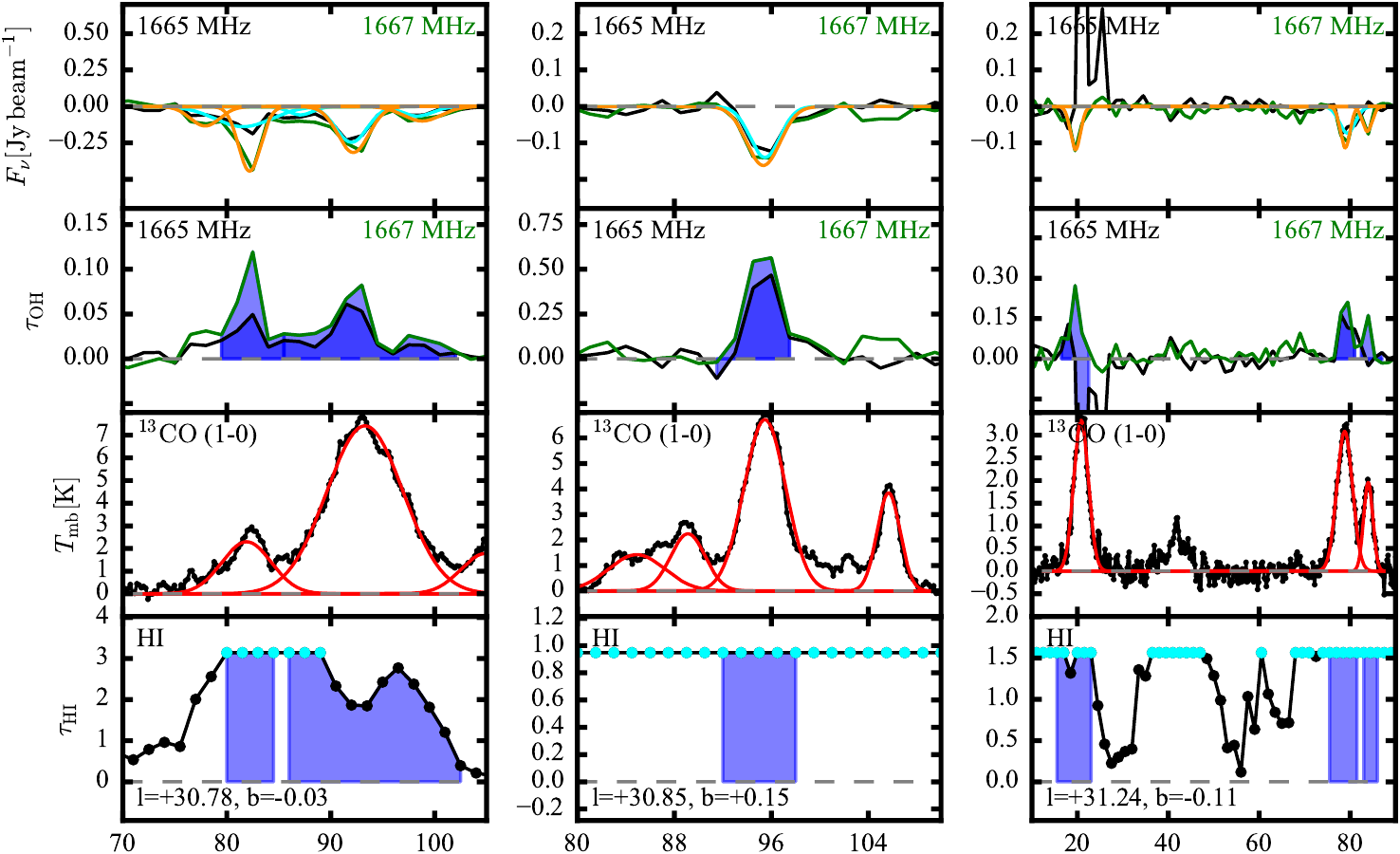}
\includegraphics[width=.95\textwidth]{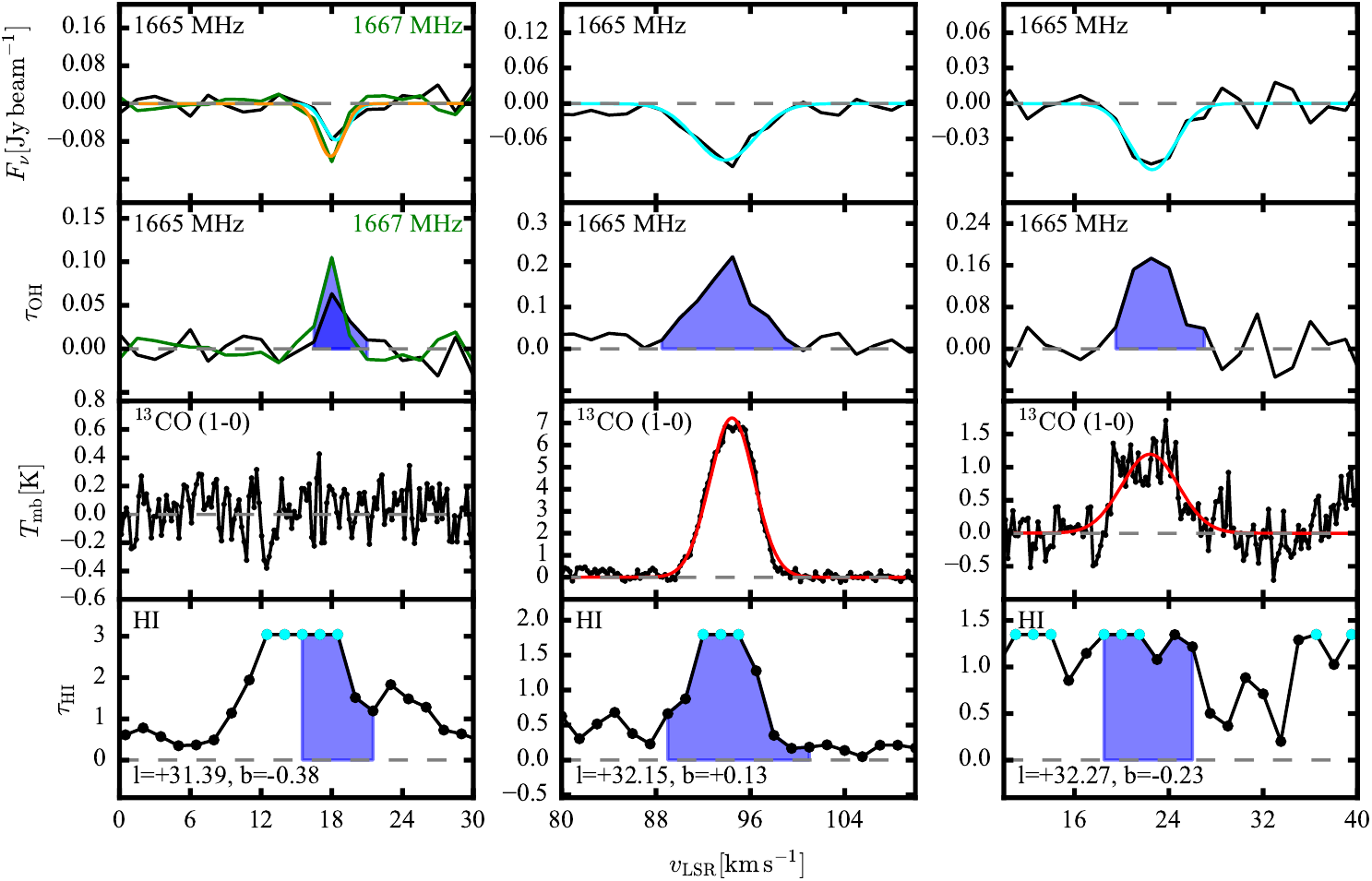}
\caption{As Fig. \ref{fig:plot_positions_00}.}\label{fig:plot_positions_06}\end{figure*}

\begin{figure*} \centering 
\includegraphics[width=.95\textwidth]{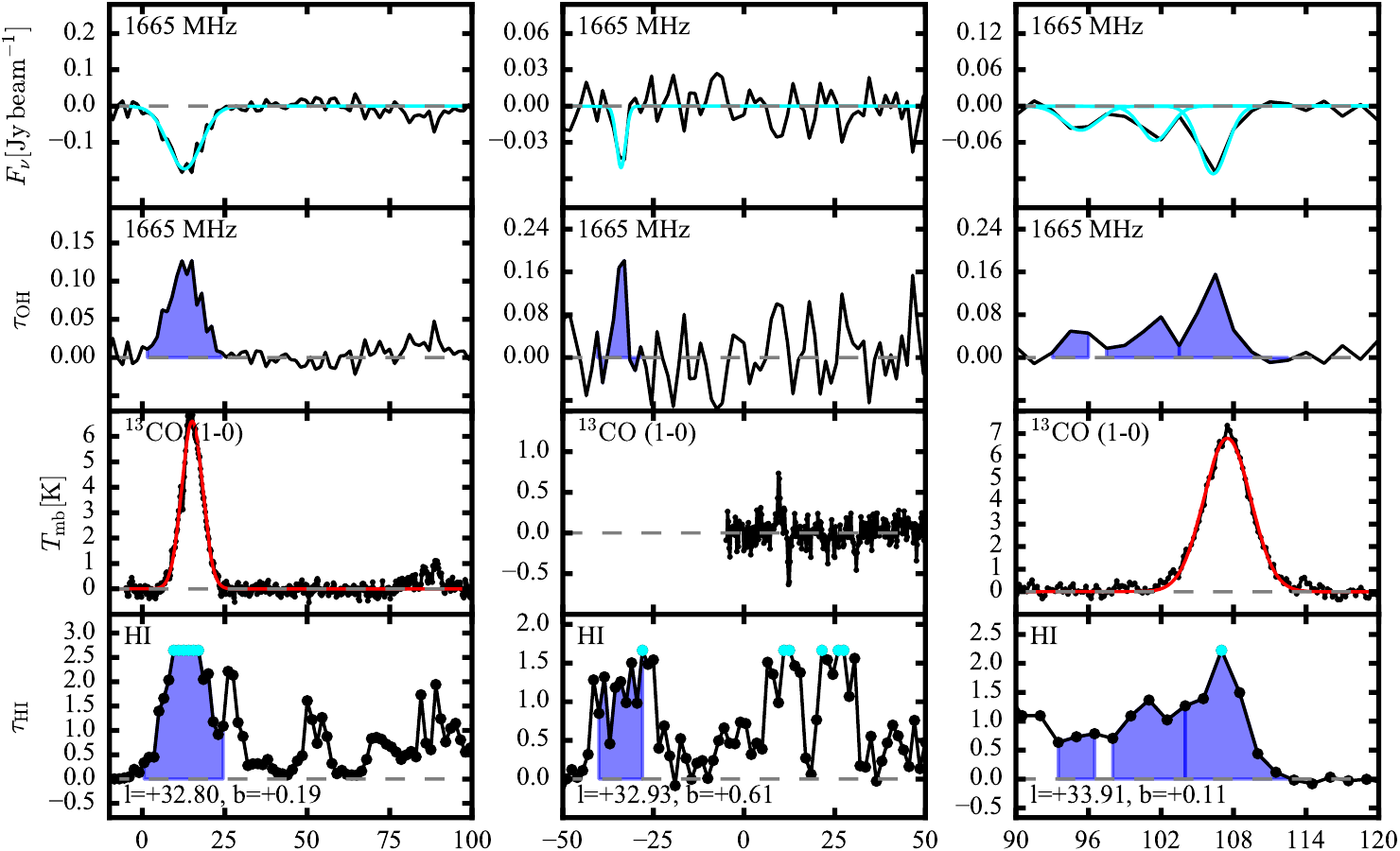}
\includegraphics[width=.95\textwidth]{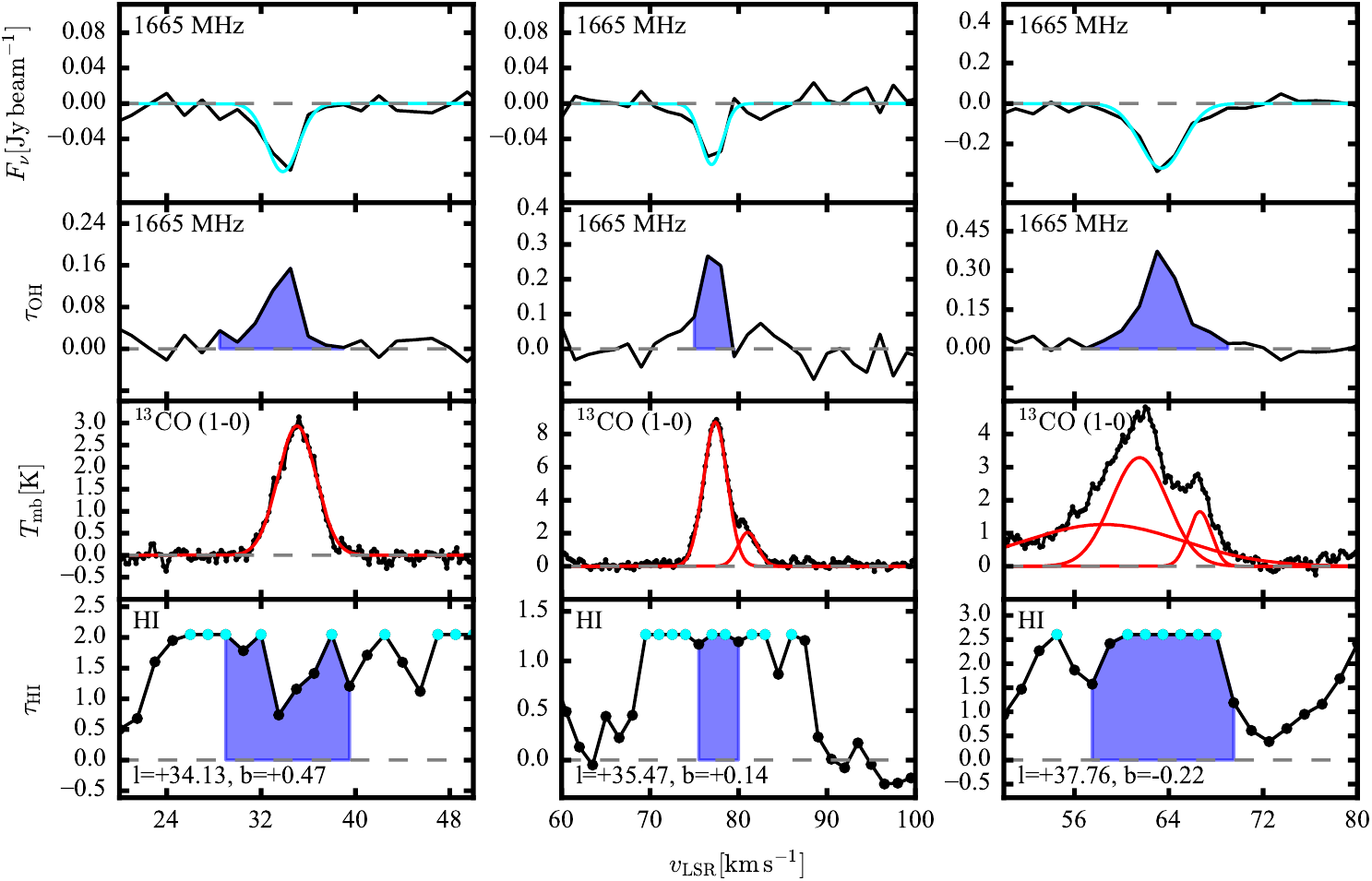}
\caption{As Fig. \ref{fig:plot_positions_00}.}\label{fig:plot_positions_08}\end{figure*}

\begin{figure*} \centering 
\includegraphics[width=.95\textwidth]{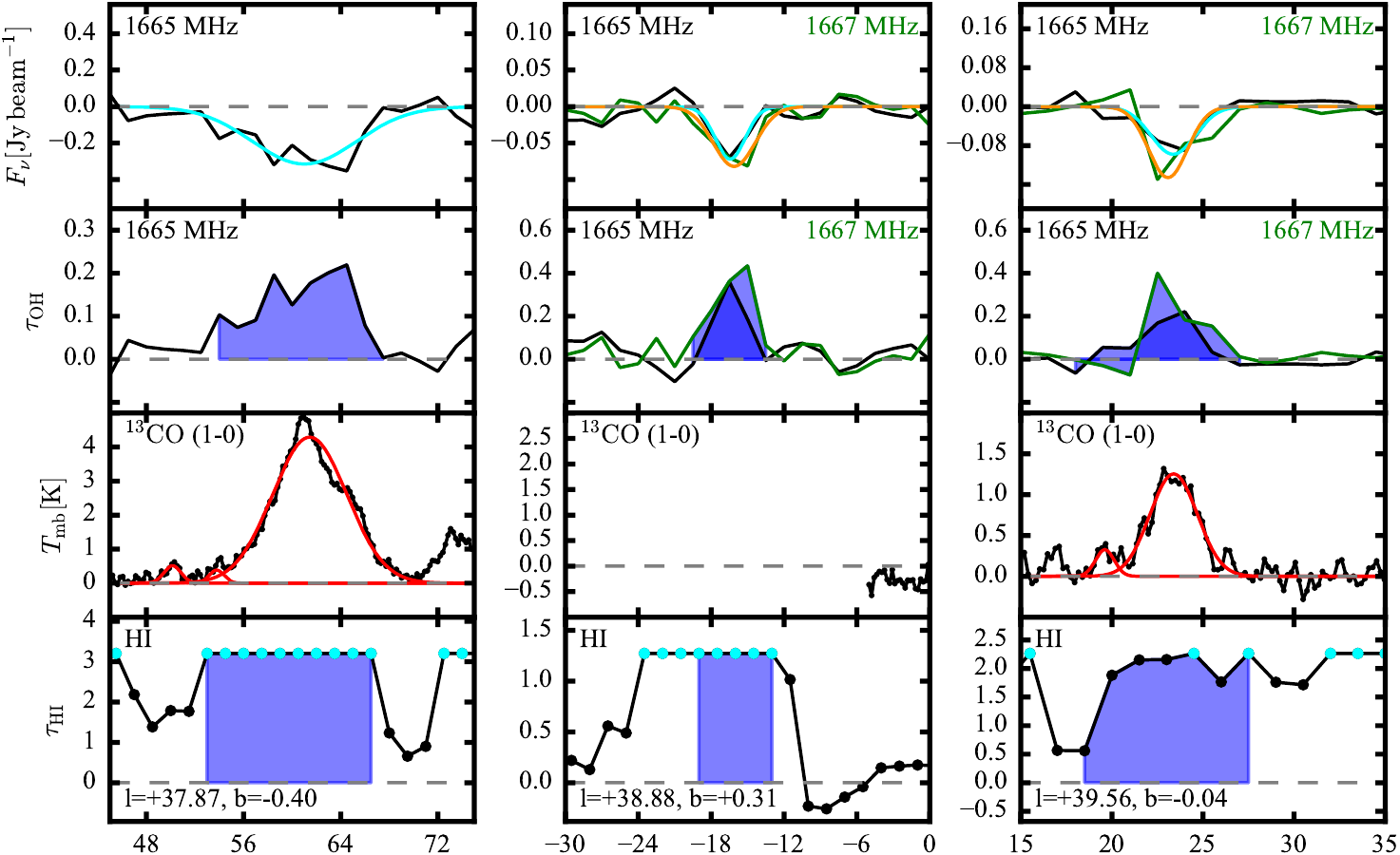}
\includegraphics[width=.95\textwidth]{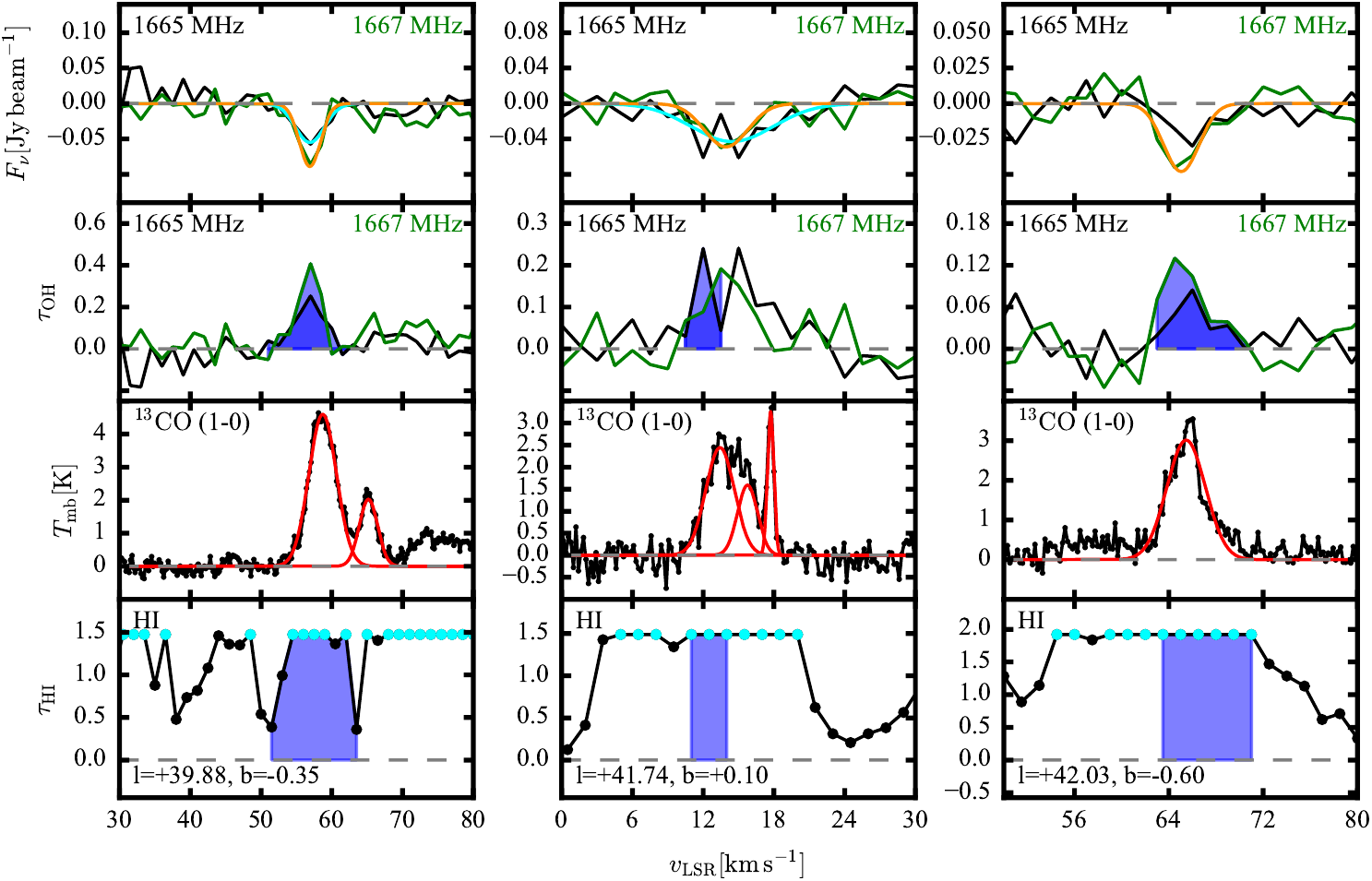}
\caption{As Fig. \ref{fig:plot_positions_00}.}\label{fig:plot_positions_10}\end{figure*}

\begin{figure*} \centering 
\includegraphics[width=.95\textwidth]{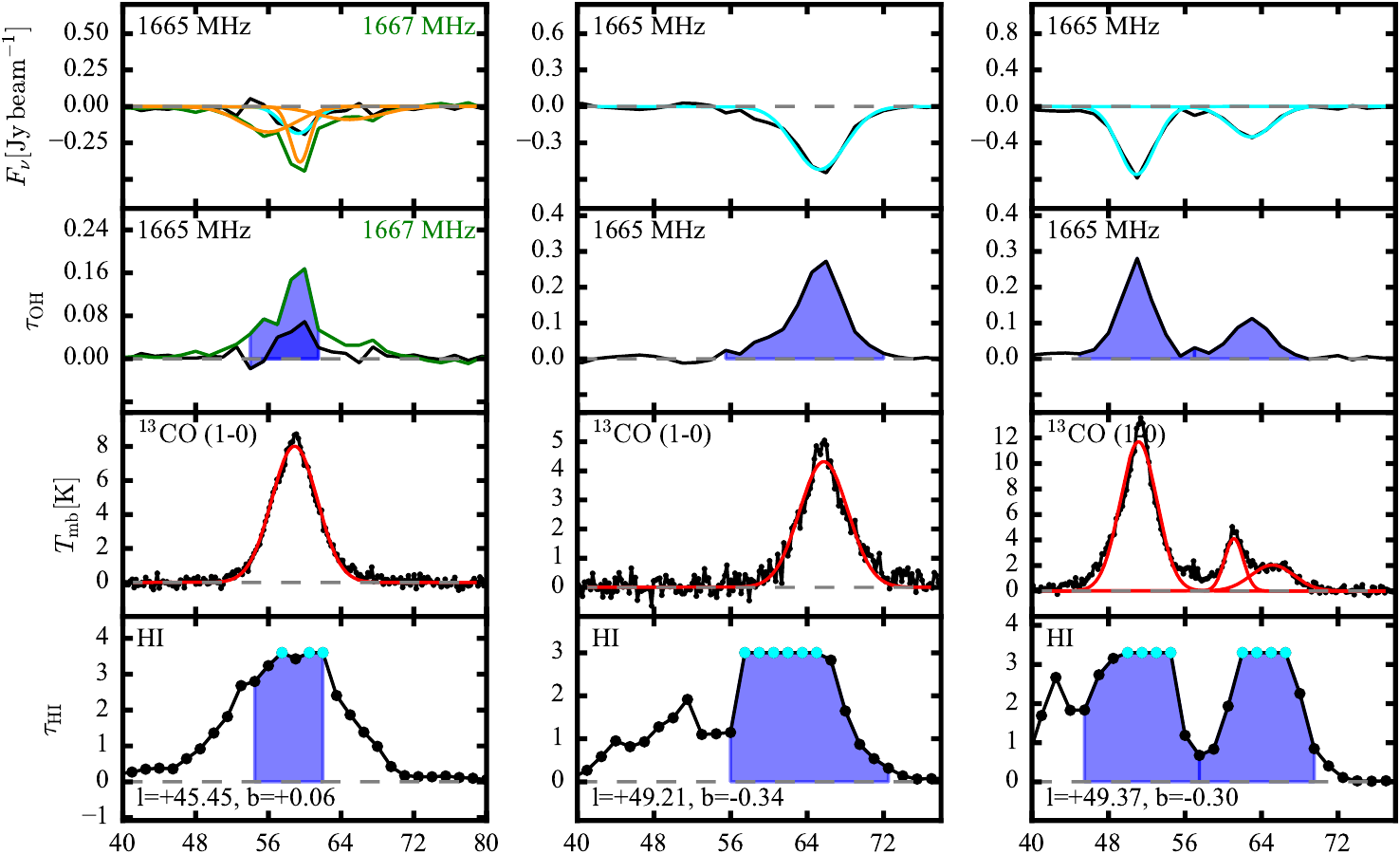}
\includegraphics[width=.95\textwidth]{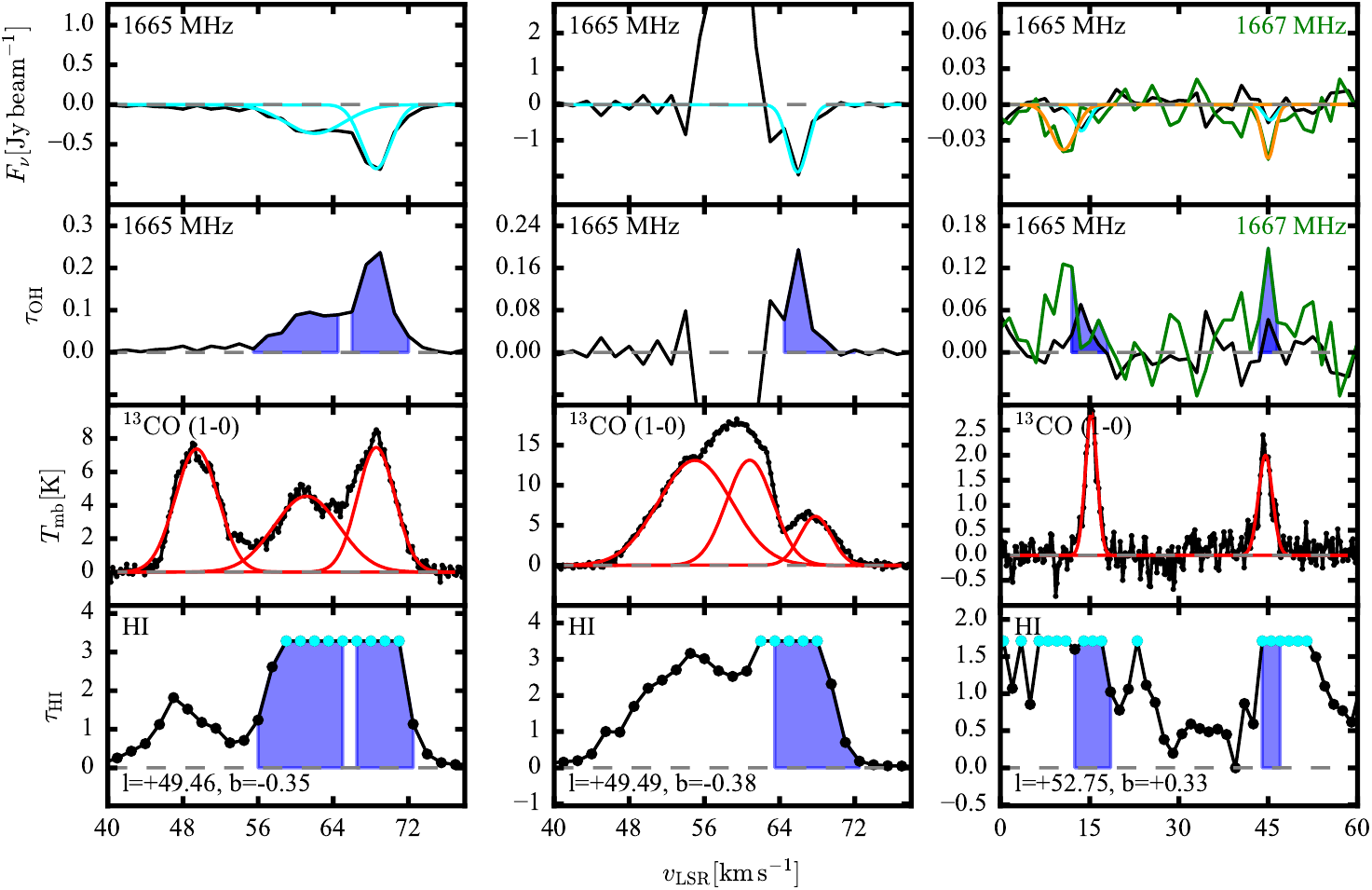}
\caption{As Fig. \ref{fig:plot_positions_00}.}\label{fig:plot_positions_12}\end{figure*}

\begin{figure*} \includegraphics[width=.66\textwidth]{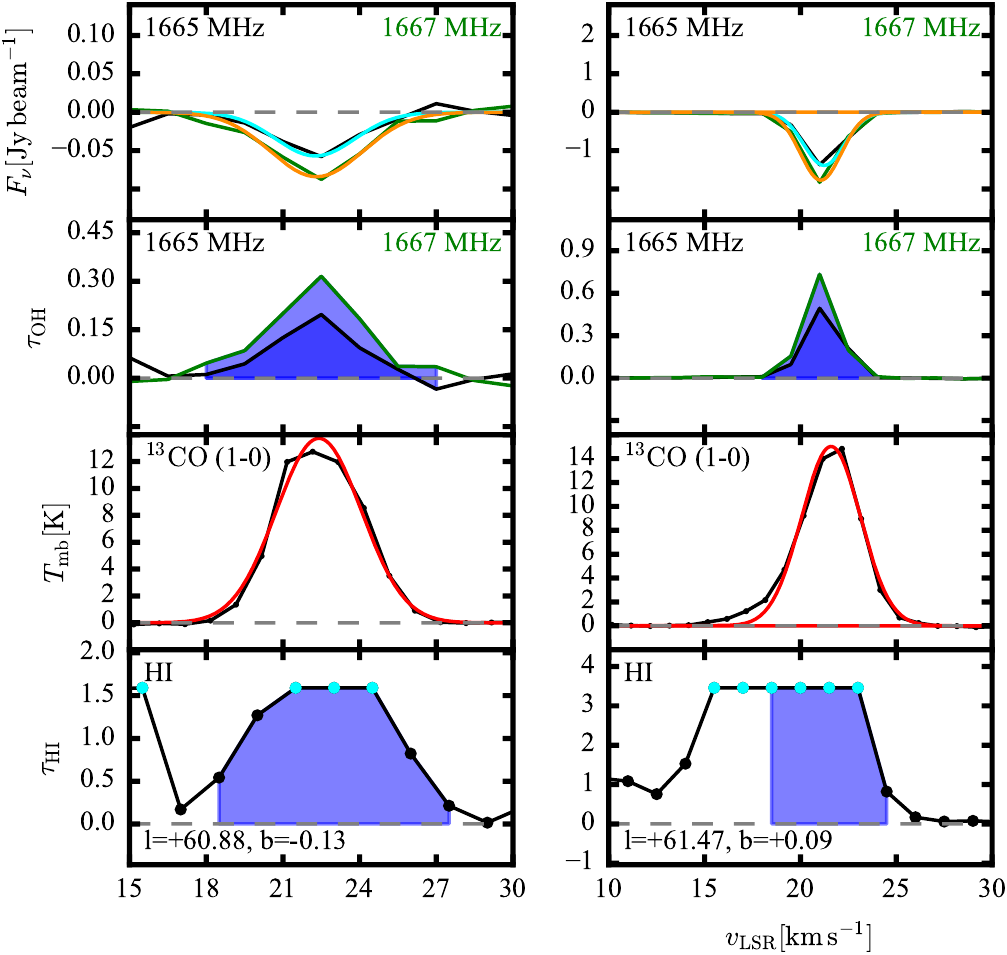}\caption{As Fig. \ref{fig:plot_positions_00}.}\label{fig:plot_positions_14}\end{figure*}

\section{OH absorption towards W43}
Moment zero map of the optical depth of the 1667{\us}MHz transition for different velocity intervals towards W43. 
\begin{figure*}
 \centering
 \includegraphics[width=0.99\textwidth]{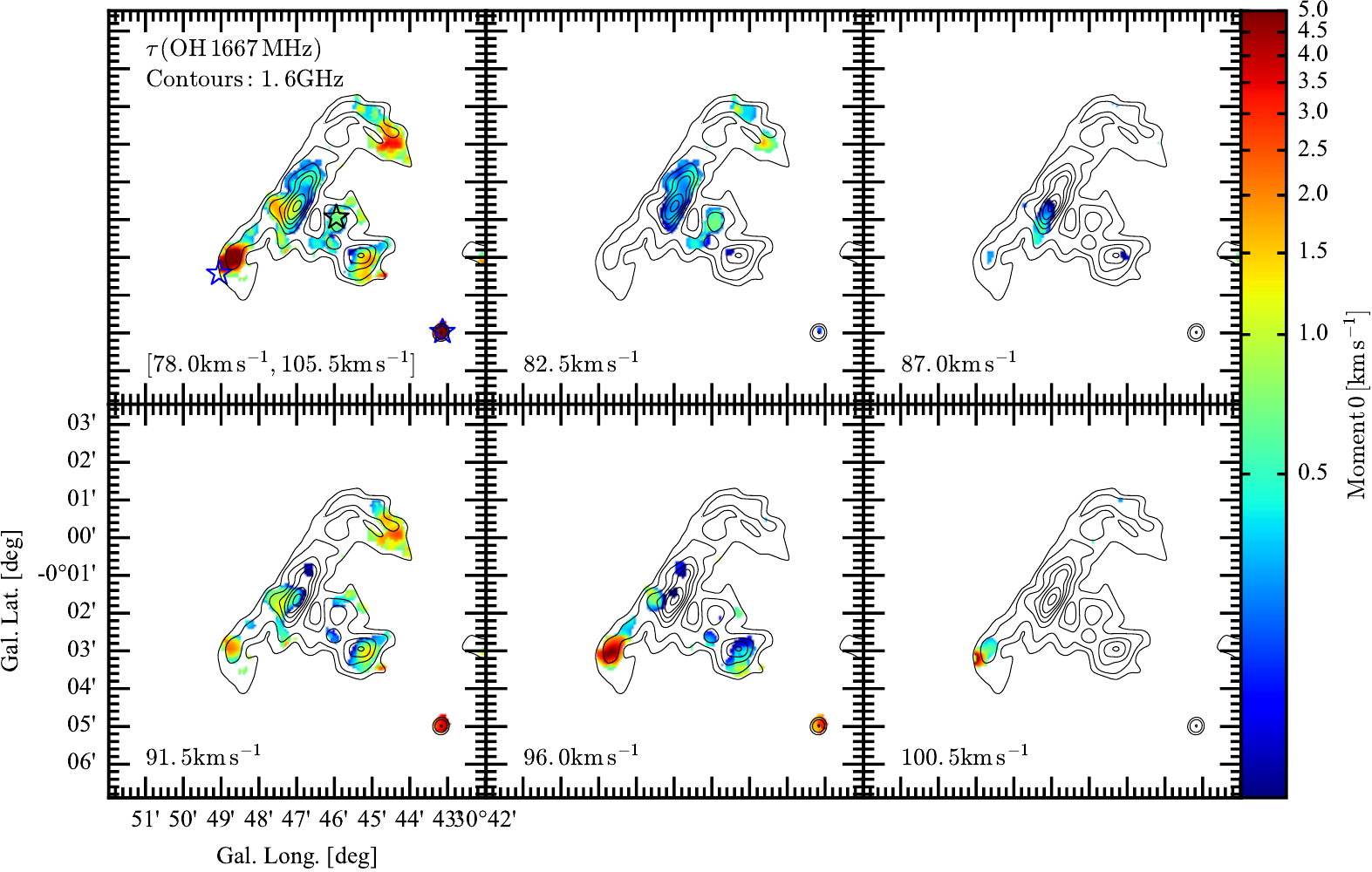}
 \caption{Integrated optical depth of the \ce{OH} 1667{\us}MHz line in the W43 star-forming 
 region. In the top-left panel, $\tau$ is integrated over the same velocity range as in Fig.~\ref{fig:plot_moment_map_w43_ratio_moments}. 
  The other panels show $\tau$ around the indicated velocities after integrating over three channels of 1.5\us\kms\ width. For each pixel, only channels that are detected 
 at a 3-$\sigma$ level or higher contribute to the integrated $\tau$-map. 
 The optical depth map is overlayed with contours of the 18~cm continuum emission 
 (black, in levels of 0.1, 0.2, 0.4, 0.6, 0.8, 1.0, 1.25, 1.5 and 1.75\us\jyb). 
 Symbols as in Fig.~\ref{fig:plot_moment_map_w43_ratio_moments}.}
 \label{fig:plot_channel_map_w43_tau_1667}
\end{figure*}

\end{appendix}
\end{document}